\documentclass[reprint,superscriptaddress,amsmath,amssymb,aps,floatfix]{revtex4-1}
 
\usepackage{graphicx}
\usepackage{dcolumn}
\usepackage{bm}
\usepackage{url} 

\begin{document}


\title[Energy Shift of dns-Si by SiO$_2$ vs. Si$_3$N$_4$]{Electronic Structure Shift of Deep Nanoscale Silicon by SiO$_2$- vs. Si$_3$N$_4$-Embedding as Alternative to Impurity Doping} 

\author{Dirk K{\"o}nig}
\email{solidstatedirk@gmail.com}
\affiliation{Integrated Materials Design Centre (IMDC) Node, University of New South Wales, NSW 2052, Australia}
\altaffiliation[Also at ]{Institute of Semiconductor Electronics (IHT), RWTH Aachen University, 52074 Aachen, Germany}
\author{No{\"e}l Wilck}
\affiliation{Institute of Semiconductor Electronics (IHT), RWTH Aachen University, 52074, Germany}
\author{Daniel Hiller}
\affiliation{Research School of Engineering, The Australian National University, ACT 2601, Australia}
\author{Birger Berghoff}
\affiliation{Institute of Semiconductor Electronics (IHT), RWTH Aachen University, 52074, Germany}
\author{Alexander Meledin}
\affiliation{Ernst-Ruska Centre for Microscopy and Spectroscopy with Electrons, RWTH Aachen University, 52074, Germany}
\author{Giovanni Di Santo}
\author{Luca Petaccia}
\affiliation{Elettra Sincrotrone Trieste, Strada Statale 14 km 163.5, 34149 Trieste, Italy}
\author{Joachim Mayer}
\affiliation{Ernst-Ruska Centre for Microscopy and Spectroscopy with Electrons, RWTH Aachen University, 52074, Germany}
\author{Sean Smith}
\affiliation{Department of Applied Mathematics, Research School of Physics and Engineering, The Australian National University, ACT2601, Australia}
\altaffiliation[Also at ]{Integrated Materials Design Lab (IMDL), The Australian National University, ACT 2601, Australia}
\author{Joachim Knoch}
\affiliation{Institute of Semiconductor Electronics (IHT), RWTH Aachen University, 52074, Germany}

\begin{abstract}
Conventional impurity doping of deep nanoscale silicon (dns-Si) 
used in ultra large scale integration (ULSI) faces serious 
challenges below the 14 nm technology node. We report on a new fundamental effect in theory and experiment, namely the electronic structure of dns-Si experiencing energy offsets of ca. 1 eV as a function of SiO$_2$- vs. Si$_3$N$_4$-embedding with a few monolayers (MLs). An interface charge transfer (ICT) from dns-Si specific to the anion type of the dielectric is at the core of this effect and arguably nested in  quantum-chemical properties of oxygen (O) and nitrogen (N) vs. Si. We investigate the size up to which this energy offset defines the electronic structure of dns-Si by density functional theory (DFT), considering interface orientation, embedding layer thickness, and approximants featuring two Si nanocrystals (NCs); one embedded in SiO$_2$ and the other in Si$_3$N$_4$. Working with synchrotron ultraviolet photoelectron spectroscopy (UPS), we use SiO$_2$- vs. Si$_3$N$_4$-embedded Si nanowells (NWells) to obtain their energy of the top valence band states. These results confirm our theoretical findings and gauge an analytic model for projecting maximum dns-Si sizes for NCs, nanowires (NWires) and NWells where the energy offset reaches full scale, yielding to a clear preference for electrons or holes as 
majority carriers in dns-Si. Our findings can replace impurity doping for n/p-type dns-Si as used in ultra-low power electronics and ULSI, eliminating dopant-related issues such as inelastic carrier scattering, thermal ionization, clustering, out-diffusion and defect generation. As far as majority carrier preference is concerned, the elimination of those issues effectively shifts the lower size limit of Si-based ULSI devices to the crystalization limit of Si of ca. 1.5 nm and enables them to work also under cryogenic conditions.
\end{abstract}

\maketitle
\section{\label{intro}Introduction}
Impurity doping of Si has been a prerequisite for Si-based electronics for about 70 years \cite{Pear49}. Over the last decade, impurity doping gradually became a major issue in ultra-large scale integration (ULSI) as fin/nanowire (NWire) device features approached the characteristic lengths of dopant out-diffusion, clustering and inactivation \cite{Koel13}. Considerable broadening of dopant profiles from drain/source regions into gate areas persists even when using self-regulatory plasma doping combined with rapid spike annealing \cite{Kamb13}.
To add, required ULSI transistor functionality \cite{Duff18} and emerging applications of Si-nanocrystals (NCs) \cite{Heit05} unveiled additional doping issues such as self-purification \cite{Dalp06,Steg09}, dopant ionization failing at room temperature \cite{Chan08,Koe15} and dopant-associated defect states \cite{Pere12,Koe15,Hill17a,Hill18b}.

In the late 1970s, modulation doping of III-V semiconductor combinations such as GaAs/AlAs was discovered \cite{Din78}. Lately, this concept was applied successfully to Si by acceptor doping of adjacent SiO$_2$ \cite{Koe17a,Hill18a,Koe18b} and proposed for donor-doping Al$\rm{_x}$Ga$\rm{_{1-x}}$N barriers from adjacent Si-rich Si$_3$N$_4$ during Si-NC formation anneal \cite{Koe13}.

Ideally, majority carrier preference for electrons/holes and thus n/p-type conductivity would not require doping if the electronic structure of deep nanoscale Si (dns-Si) could be shifted over energy as per n/p-type section of an electronic device. Such an energy offset $\Delta E$ would avoid all dopant-related issues mentioned above, leading to lower inelastic carrier scattering rates and higher carrier mobilities which allow for decreased heat loss and bias voltages in ULSI. Together with directed self-assembly of block copolymers as new lithography approach to dns-Si structures \cite{Liu18}, such properties enable Si-FET technology to work at yet smaller structure sizes potentially enabling Moore's Law to approach the Si-crystallization limit of ca. 1.5\,nm \cite{Schu94}. 
In a recent paper \cite{Koe18a} we were able to demonstrate in theory and experiment that $\Delta E$ exists in dns-Si when embedding one part of the Si nanovolume in SiO$_2$ and another part in Si$_3$N$_4$. However, a detailed investigation of the impact originating from interface bond densities per square, thickness of embedding dielectric and sample size on the magnitude of the effect is missing. The latter is very important for the applicability of the effect in real devices. As an effect induced via the interface of dns-Si, the extension of the p/n-junction is on the order of ca. 15 {\AA}, allowing for a leap in device miniaturization. Moreover, we deliver a detailed phenomenologial description of the fundamental quantum-chemical origin of $\Delta E$ which is underpinned by experimental data and DFT results.

In our work presented here, we consider SiO$_2$- and Si$_3$N$_4$-embedding of Si-NCs as a function of shape, size, interface orientation and thickness of embedding dielectric by using density functional theory (DFT) to calculate their electronic structure. As the next step, we compute DFT approximants containing two NCs, one embedded in SiO$_2$ and the other in Si$_3$N$_4$, to verify the $\Delta E$ we found for single Si-NCs within one system, accounting for interactions of both Si-NCs. From there, we present experimental data on Si-nanowells (NWells) embedded in SiO$_2$ vs. Si$_3$N$_4$, namely information on the ionization energy of the valence band edge of the Si-NWells by long-term synchrotron ultraviolet photoelectron spectroscopy (UPS)\cite{Berg64a,Berg64b,Rein05,Shig14,Koe18a,Koe18aa}. With this wealth of theoretical and experimental data, we propose a model explaining the interface impact of oxygen (O) and nitrogen (N) onto dns-Si, the associated characteristic impact length and other prominent features observed in the electronic structure of dns-Si. The massive $\Delta E$ observed in theory and experiment presents a new fundamental effect which can replace doping by forcing dns-Si into p- or n-type carrier preference as a function of embedding dielectric. Thereby, advantages of fully depleted field effect transistor (FET) ULSI devices can be combined with carrier preferences hitherto only achieved by impurity doping.

\section{\label{meth}Methods}

\subsection{\label{DFT-Meth}Density Functional Theory Calculations}
Hybrid density functional theory (h-DFT) calculations were carried out in real space with a molecular orbital basis set (MO-BS) and both Hartree-Fock (HF) and h-DFT methods as described below, employing the {\sc Gaussian}09 program suite \cite{G09}. Initially, the MO-BS wavefunction ensemble was tested and optimized for stability with respect to describing the energy minimum of the approximant (variational principle; $stable=opt$) with the HF method using a 3-21G MO-BS \cite{Gor82} (HF/3-21G). This MO wavefunction ensemble was then used for the structural optimization of the approximant to arrive at its most stable configuration (maximum integral over all bond energies), again following the HF/3-21G route. Using these optimized geometries, their electronic structure was calculated again by testing and optimizing the MO-BS wavefunction ensemble with the B3LYP hybrid DF \cite{Beck88,Lee88} and the 6-31G(d) MO-BS \cite{Fra82} (B3LYP/6-31G(d)\,). Root mean square (RMS) and peak force convergence limits were 8 meV/{\AA} and 12 meV/{\AA}, respectively.  Tight convergence criteria were applied to the self-consistent field routine. Ultrafine integration grids were used throughout. During all calculations, no symmetry constraints were applied to MOs. Extensive accuracy evaluations can be found elsewhere \cite{Koe08,Koe14,Koe18aa}. We note that in real-space DFT calculations, exact calibration to an absolute energy scale (the vacuum level $E_{\mathrm{vac}}$) is known to be ambiguous \cite{Riss11}. Nevertheless, relative changes of energy values such as $\Delta E$ or fundamental energy gaps $E_{\rm{gap}}$ between approximants with different interface termination are accurate within the same computation route, leaving a constant energy shift to all states with respect to $E_{\mathrm{vac}}$ as the only uncertainty. Approximants and MOs were visualized with 
{\sc Gview\,5} \cite{GV5}. Electronic density of states (DOS) were calculated from MO eigenenergies, applying a Gaussian broadening of 0.2 eV. The sizes given for NCs $d_{\rm{NC}}$ were calculated as the product of atomic volume of Si and the number of Si atoms forming the NC, assuming a spherical shape. This approach allows for a direct comparison of results between approximants of different shape.

\subsection{\label{Prep-Meth}Sample Preparation}
Samples comprising a Si$_3$N$_4$-embedded NWell were fabricated by plasma enhanced chemical vapour deposition (PECVD) using SiH$_4+$NH$_3+$N$_2$ for Si$_3$N$_4$ buffer and top layers (thickness 5.5 and 1.5 nm, respectively) and SiH$_4+$ Ar for amorphous Si \cite{Hill14}. The Si$_3$N$_4$ spacer layer served to suppress excited electrons from the Si wafer to interfere with electrons from the Si-NWell during UPS. The substrates, i. e. n-type Si wafers (Sb doping, $5$ to $15\times10^{-3}\,\Omega$cm) of (111)-surface orientation, were wet-chemically cleaned. After deposition the wafers were annealed in a quartz tube furnace for 1 min at $1100\,^{\circ}$C in pure N$_2$ ambient to induce Si-crystallization. Subsequently, samples were H$_2$-passivated at $450\,^{\circ}$C for 1 hour. 

Samples comprising a SiO$_2$-embedded NWell were dry-etched with an isotropic SF$_6$/O$_2$ plasma to thin down a 85 nm thick top c-Si layer of an Si-on-insulator (SOI) wafer with 200 nm buried SiO$_2$ (BOX), arriving at ca. 2.7 to 6.5 nm thick c-Si on top. The back-etch was followed by oxidation in 68 wt-\% HNO$_3$ at 120 $^{\circ}$C, resulting in a 1.1 to 5.0 nm Si-NWell with 1.4 nm SiO$_2$ capping. 

Si reference samples were processed by etching a 001-Si wafer (Sb-doped n-type, 
$5$ to $15\times10^{-3}\ \Omega$cm) 
in 1 wt-\% buffered hydrofluoric acid and immediate sample mounting under a N$_2$-shower with swift loading into the ultra-high vacuum (UHV) annealing chamber.

\begin{figure}[h!]
\includegraphics[width=8.6cm,keepaspectratio]{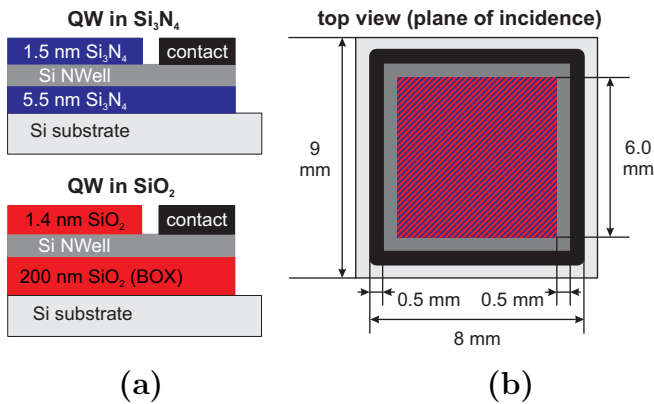}
\begin{center}
{\bf\large(a)\hspace{4.3cm}(b)}
\vspace*{-0.1cm}
\end{center}
\caption{\label{fig01} Cross section view of sample layout of Si NWells embedded in Si$_3$N$_4$ or SiO$_2$ for synchrotron UPS measurements, showing the edge region of samples {\bf(a)}.  Top view on the plane of incidence with mesa structure inside the contact frame{\bf(b)}. Hatched area stands for respective dielectric.} 
\end{figure}

All NWell samples were contacted via a lateral metal contact frame on the front surface which was processed by photolithographical structuring, wet-chemical  etching in 50 \% HF plus 0.1 \% HNO$_3$ for opening the top Si$_3$N$_4$ or in 1 wt-\% hydrofluoric acid for opening the top SiO$_2$ layer and thermal evaporation of 300 nm Al, followed by a lift-off in dimethyl-sulfoxide ($\rm{H_3C\!-\!(S\!=\!O)\!-\!CH_3}$). The Si reference samples were contacted directly on their front surface. Fig. \ref{fig01} shows the structure of the mesa samples.

\subsection{\label{Char-Meth} Characterization}
Ultraviolet photoelectron spectroscopy (UPS) measurements were carried out at the BaDElPh beamline \cite{Peta09} at the Elettra Synchrotron in Trieste, Italy,  see \cite{Koe18a} for details. All samples were subject to a UHV anneal for 90 min at 500 K to desorb water and air-related species from the sample surface prior to measurements. Single scans of spectra were recorded over 12 h per NWell sample and sub-sequently added up for eliminating white noise.
Scans for the Si-reference sample were recorded over 2\,h and sub-sequently added up \cite{Koe18a}. All samples of beamtime 1 were exited with photon energies of $h\nu=8.7$ eV and a photon flux of $2\times 10^{12}$ s$^{-1}$. Samples of beamtime 2 were excited with $h\nu=8.0$ eV and a photon flux of $2\times 10^{12}$ s$^{-1}$, yielding to slightly better signal-to-noise ratio (SNR) due to an increased inelastic mean free path $\lambda_{\rm{imfp}}$ of escaping electrons. The incident angle of the UV beam onto the sample was 50$^{\circ}$ with respect to the sample surface normal, excited electrons were collected with an electron analyzer along the normal vector of the sample surface. Energy calibration of the UPS was realized using a tantalum (Ta) stripe in electrical contact to the sample as a work function reference. Further UPS-data of SiO$_2$ and Si$_3$N$_4$ reference samples as well as on UPS signal normalization and background substraction are available in \cite{Koe18a,Koe18aa} and in Appendix Sec. \ref{Apx-UPS-Data-Eval}.

All samples for transmission electron microscopy (TEM) investigation were capped with a protective 100 nm thick SiO$_2$-layer to facilitate the preparation of cross sections by the focused ion beam (FIB) technique using a FEI Strata FIB 205 workstation. Some samples were further thinned by means of a Fischione NanoMill. TEM analysis of the cross sections was performed at a FEI Tecnai F20 TEM operated at 200 kV at the Central Facility for Electron Microscopy, RWTH Aachen University, and at the spherical aberration corrected FEI Titan 80-300 TEM operated at 300 kV at Ernst Ruska-Centre, Forschungszentrum J\"ulich \cite{TITAN}. 

In addition, Si-NWell thicknesses were measured by ellipsometry. The thickness of Si-NWells in Si$_3$N$_4$ and in SiO$_2$ were measured using a Woollam M-2000 ellipsometer and an ACCURION nanofilm ep4se ellipsometer, respectively. All thickness measurements were confirmed by values obtained from TEM. Yet, samples were given a standard deviation in NWell thickness of $\pm 1$ monolayer (ML) due to the large extension of the measurement spot during UPS as compared to NWell thickness.

\section{\label{DFT-Res}DFT Results}

\subsection{\label{DFT-Res-1NC}Single Si-NCs coated with SiO$_2$ vs. Si$_3$N$_4$}
\begin{figure}[h!]
\includegraphics[width=4.25cm,keepaspectratio]{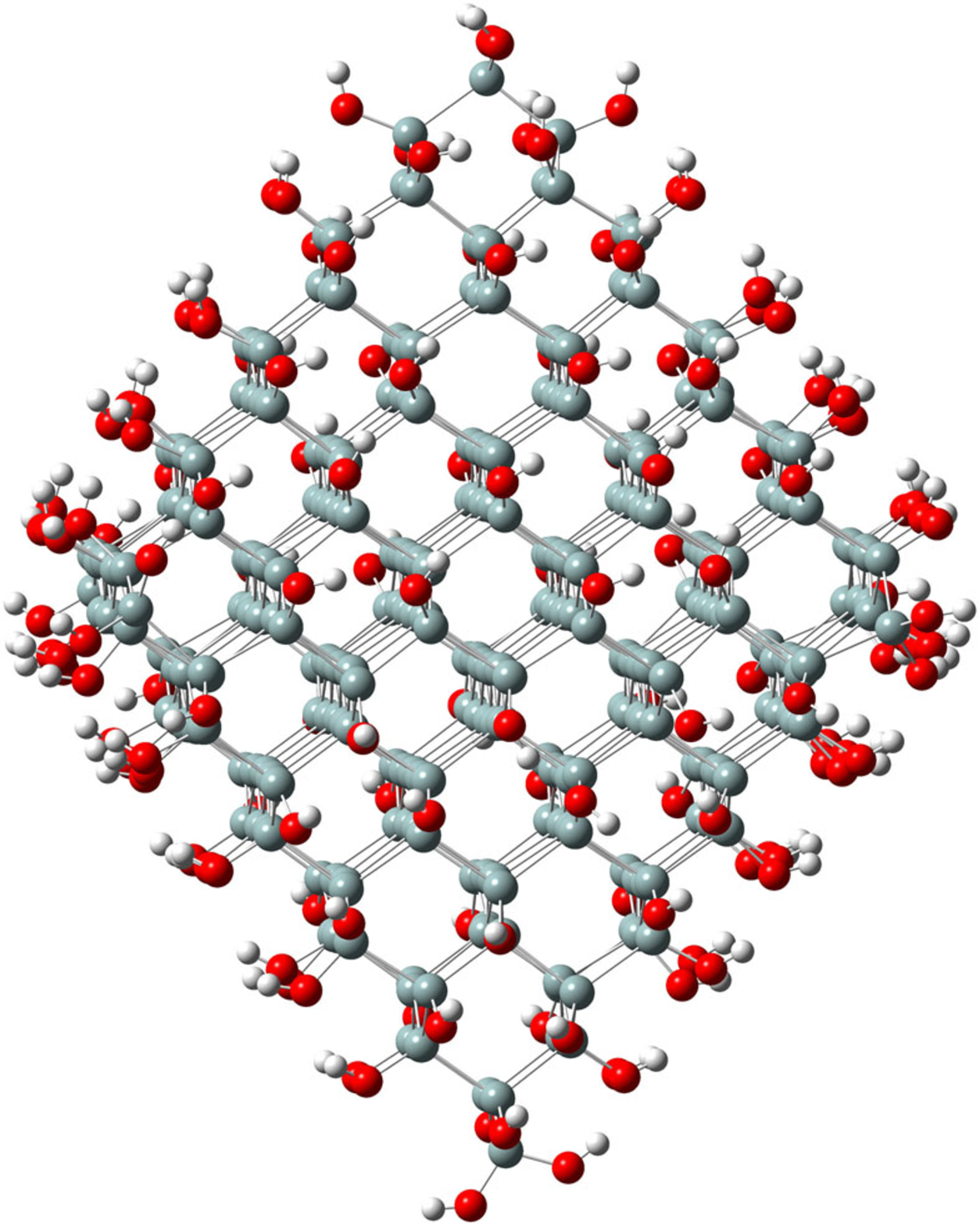}
\includegraphics[width=4.25cm,keepaspectratio]{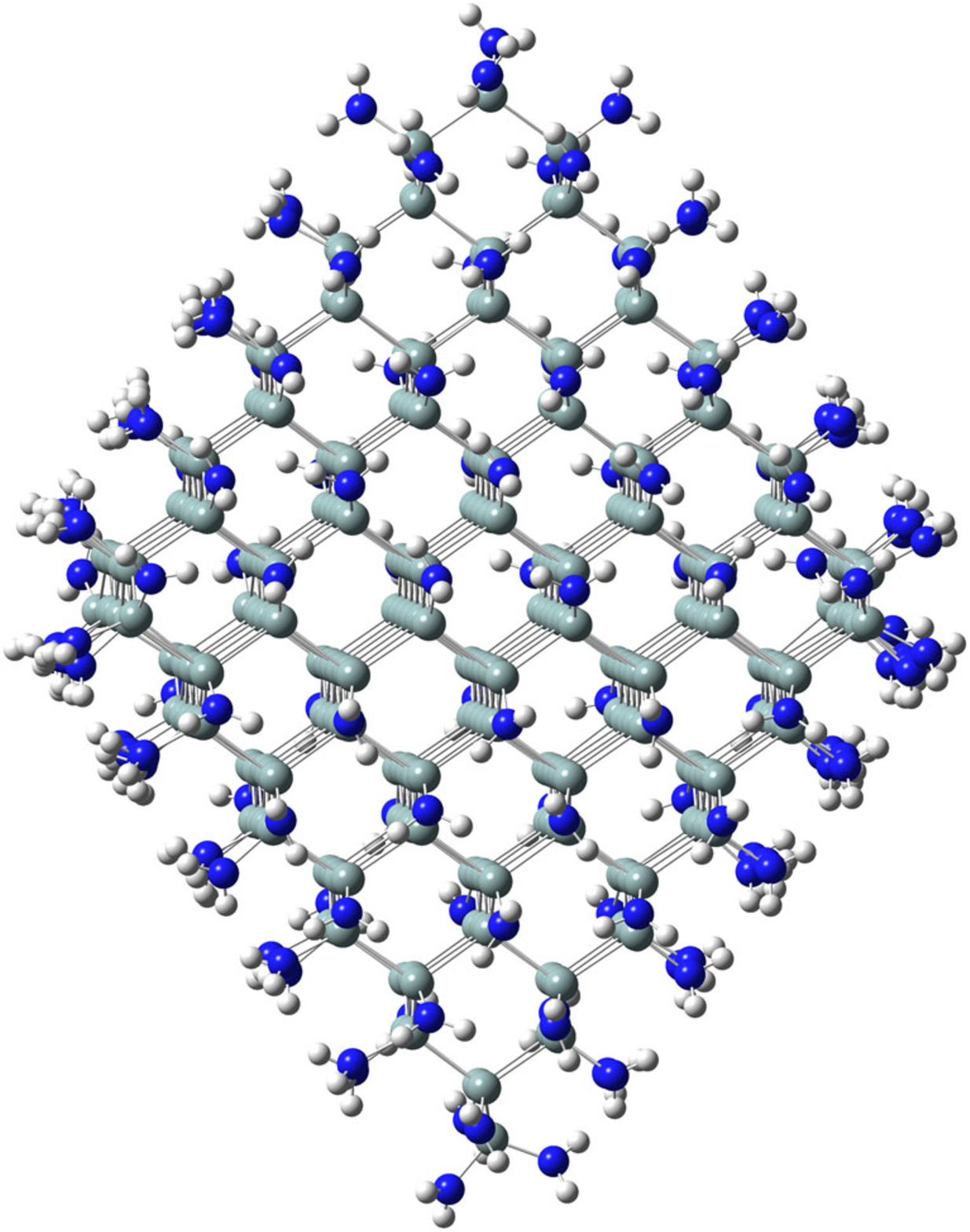}
\begin{center}
\vspace*{-0.2cm}
{\bf\large(a)}
\end{center}
\includegraphics[width=4.25cm,keepaspectratio]{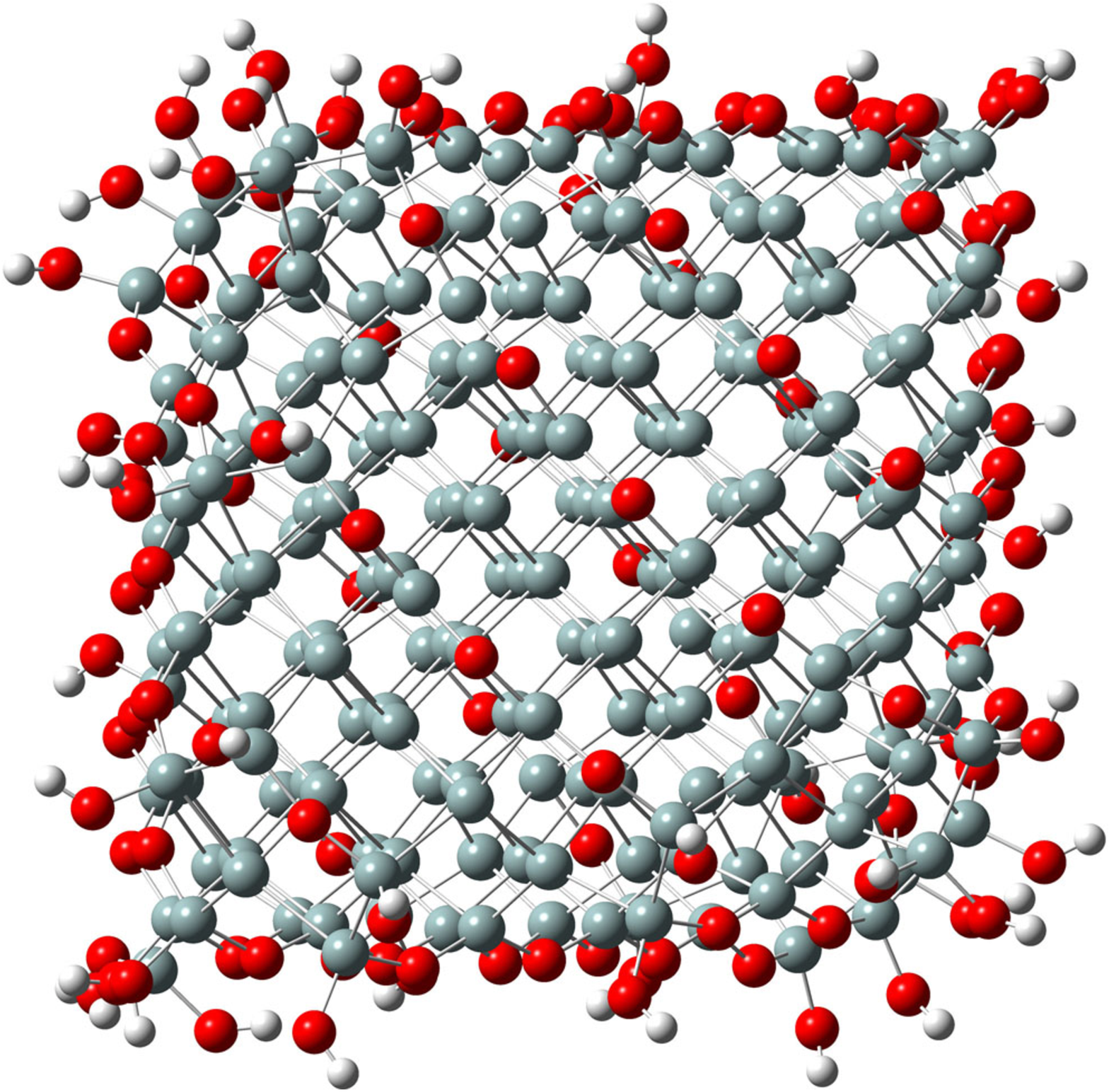}
\includegraphics[width=4.25cm,keepaspectratio]{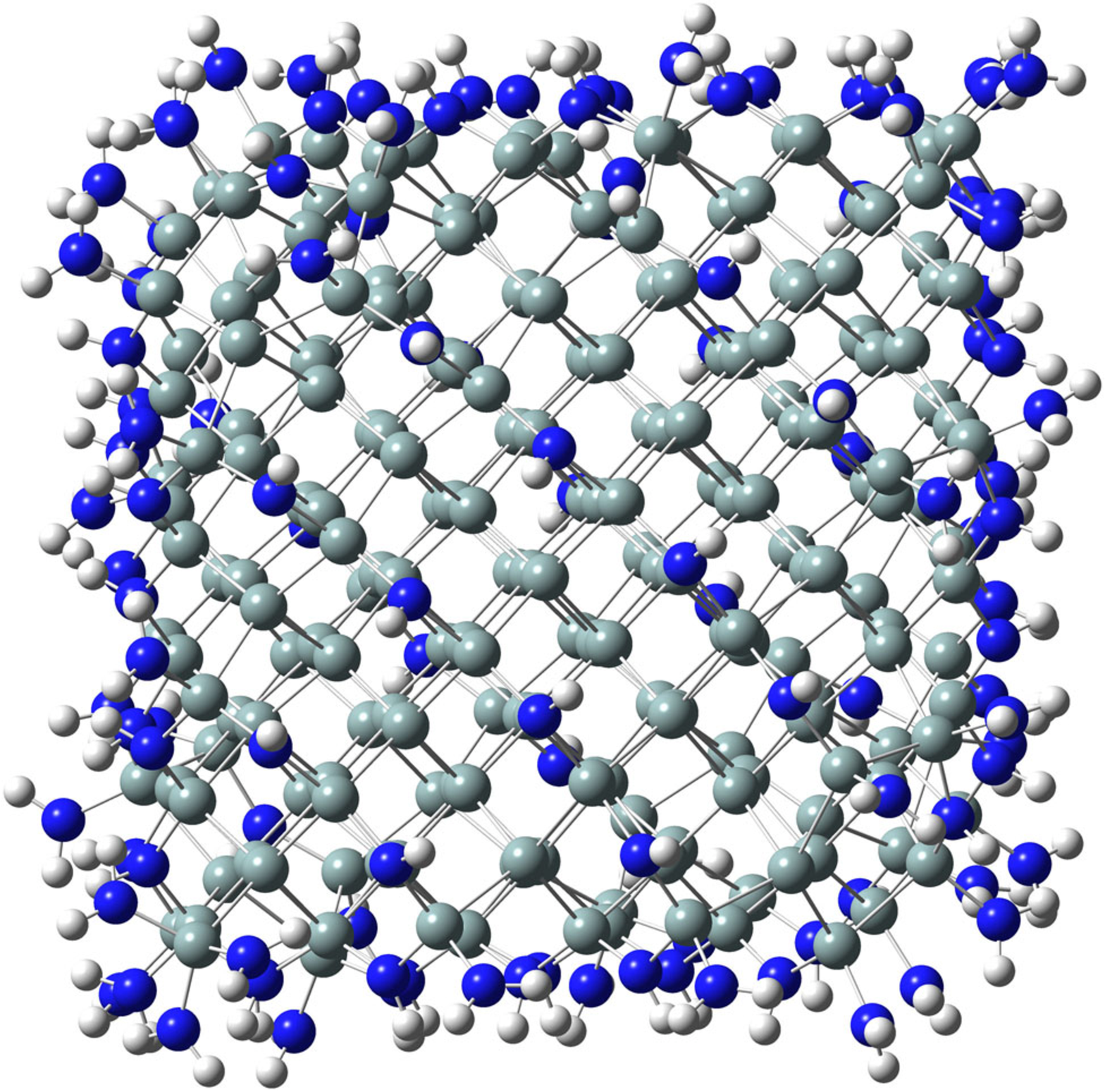}
\begin{center}
\vspace*{-0.2cm}
{\bf\large(b)}
\vspace*{-0.1cm}
\end{center}
\includegraphics[width=4.25cm,keepaspectratio]{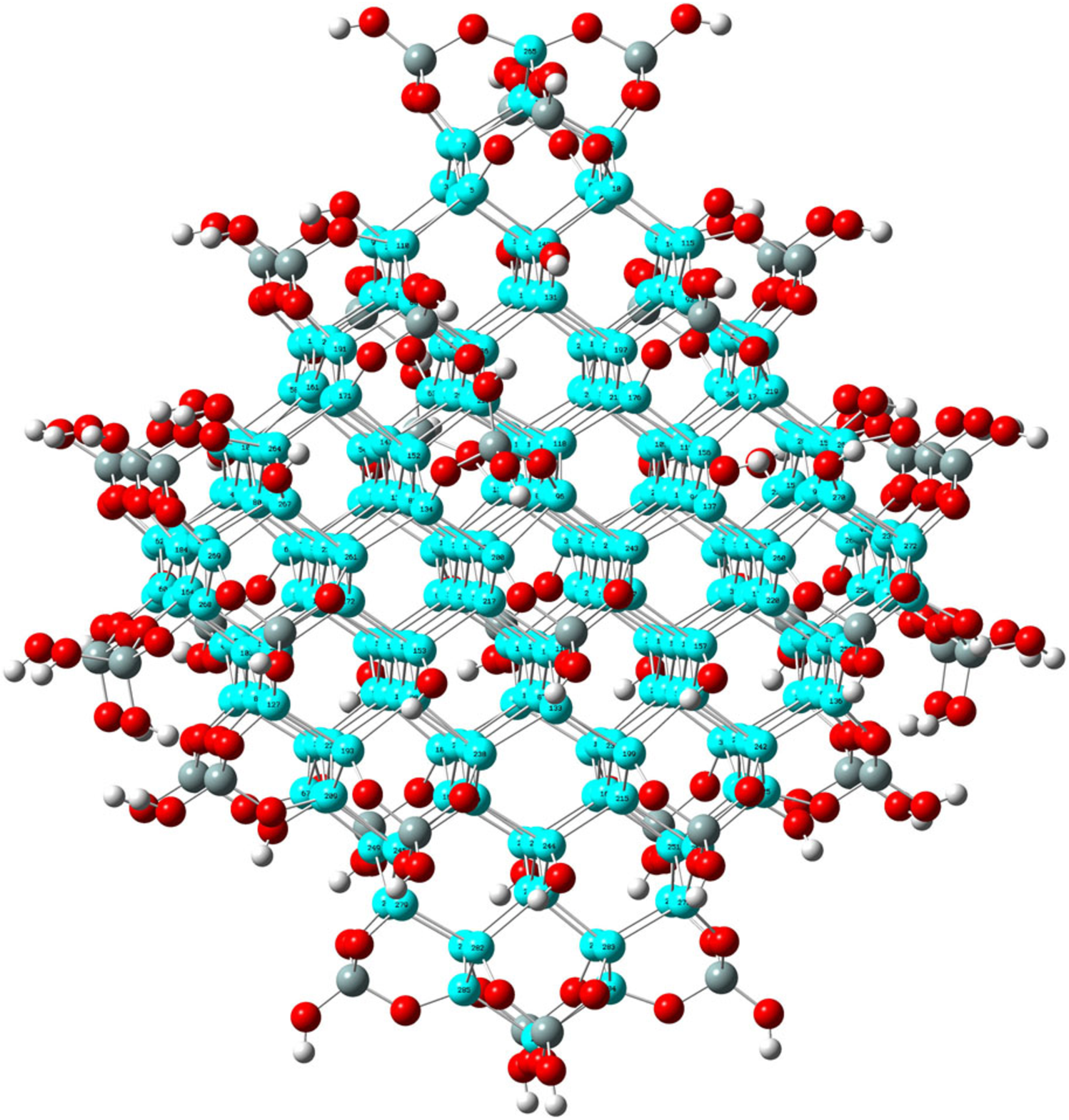}
\includegraphics[width=4.25cm,keepaspectratio]{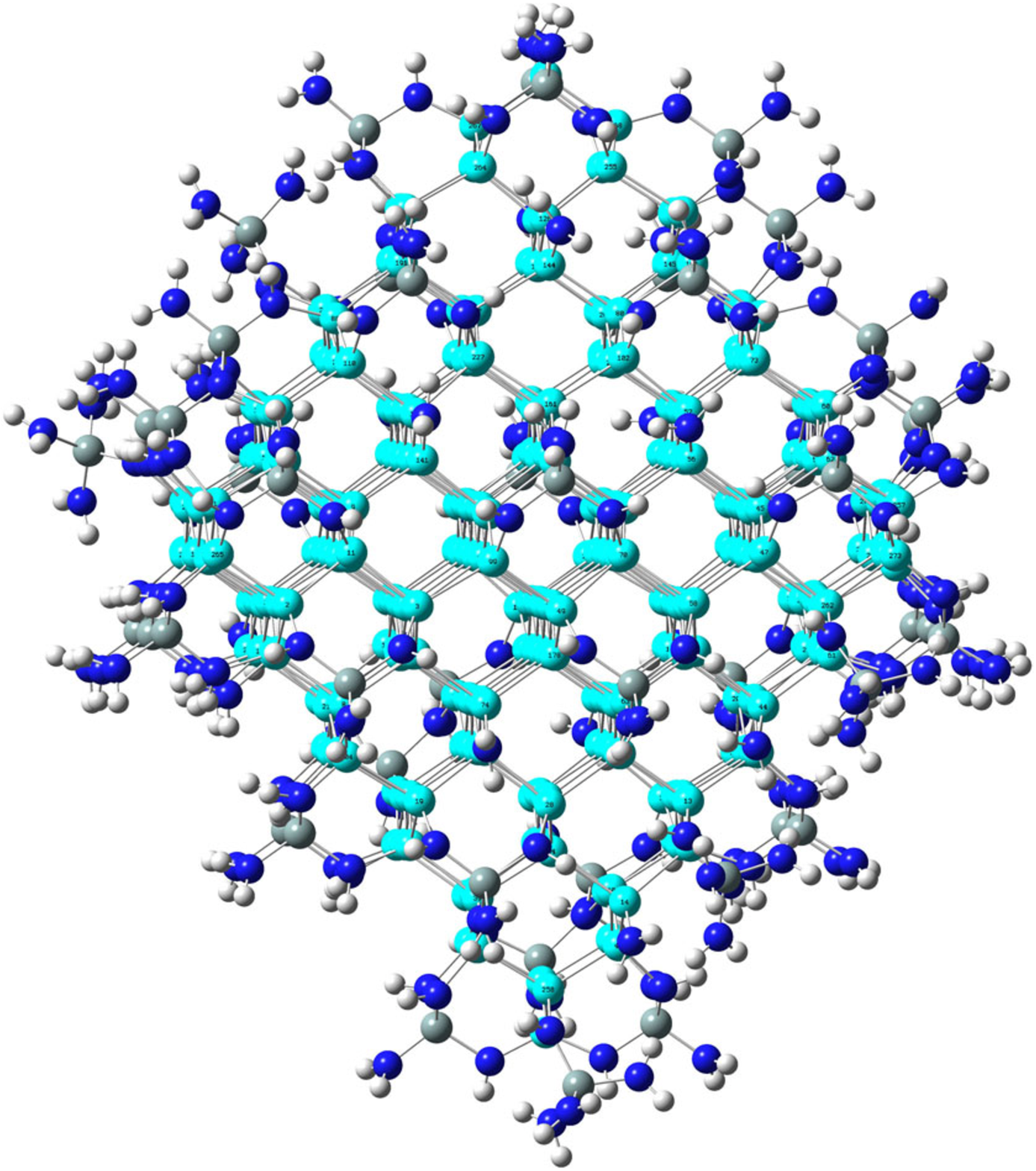}
\begin{center}
\vspace*{-0.2cm}
{\bf\large(c)}
\vspace*{-0.1cm}
\end{center}
\caption{\label{fig02} {\bf Approximants to explore the impact of interface modification on electronic structure of Si-NCs.} Octahedral $\langle111\rangle$-faceted Si-NCs (Si$_{286}$X$_{144}$, X$=$OH, NH$_2$) of 22 {\AA} size {\bf(a)}. Cubic $\langle001\rangle$-faceted Si-NCs (Si$_{216}(>$XH$_{-1})_{75}$X$_{48},\,>$XH$_{-1}=\ >$O/NH) of 20.2 {\AA} size {\bf(b)}. Same Si-NCs as shown in (a), embedded in 1.5 ML SiO$_2$ or Si$_3$N$_4$. NC atoms are highlighted in cyan {\bf(c)}. Atoms colors: Si is grey, O is red, N is blue and H is white.} 
\end{figure}

With the interface charge transfer (ICT) at the core of $\Delta E$, we investigate the nature of the interface and the interface-to-volume ratio of dns-Si as key parameters defined by system size. The latter can be expressed by the ratio of interface bonds to the number of Si-NC atoms $N_{\rm{IF}}/N_{\rm{NC}}$ \cite{Koe16}, whereby this ratio can be applied to other dns-Si volumes such as NWires or fins \cite{Koe19a}. Apart from its termination with OH- vs. NH$_2$-groups, the nature of the interface is defined by its orientation with characteristic bond densities per square. We choose the  $\langle111\rangle$- and $\langle001\rangle$-orientations with their interface bond densities per cm$^2$ of $\Box N_{111}=7.8\times 10^{14}$ cm$^{-2}$ and $\Box N_{001}=13.6\times 10^{14}$ cm$^{-2}$ \cite{Hes93}, respectively, for two reasons. First, the technologically most relevant interface orientations of dns-Si are $\langle001\rangle$, $\langle110\rangle$ and $\langle111\rangle$. Second,  
$\Box N_{110}=9.6\times 10^{14}$ cm$^{-2}$ for $\langle110\rangle$-orientations of Si interfaces \cite{Hes93} is close to $\Box N_{111}$ whereby a significant difference of its impact onto the ICT is not given. We calculated octahedral approximants with exclusive $\langle111\rangle$-orientations up to $d_{\rm{NC}}=25.9$ {\AA} -- Si$_{455}$X$_{196}$  with X$=$ OH, NH$_2$, see Fig. \ref{fig02}a -- and cubic approximants with exclusive $\langle001\rangle$-orientations and associated bridge bonds ($>$) of NH- and O-groups up to $d_{\rm{NC}}=20.2$ {\AA} -- Si$_{216}$($>$XH$_{-1}$)$_{75}$X$_{48}$ with XH$_{-1}=$ $>$O/$>$NH, see Fig. \ref{fig02}b. Computations of bigger cubic NCs such as Si$_{512}$($>$XH$_{-1}$)$_{75}$X$_{48}$ were not successful due to excessive strain induced by $>$NH and in particular $>$O disintegrating the cubic shape. 

Another important parameter is the thickness of the embedding dielectric. 
We calculated octahedral Si-NCs with exclusive $\langle111\rangle$-orientations as above, but covered in 1.5 ML SiO$_2$ or Si$_3$N$_4$ up to Si$_{286}$ NCs ($d_{\rm{NC}}=22.2$ {\AA}). Thereby, we can directly compare the electronic structure of both approximant cohorts. Fig \ref{fig02}c shows such approximants where the NCs have been highlighted.
 
\begin{figure}[h!]
\includegraphics[width=8.6cm,keepaspectratio]{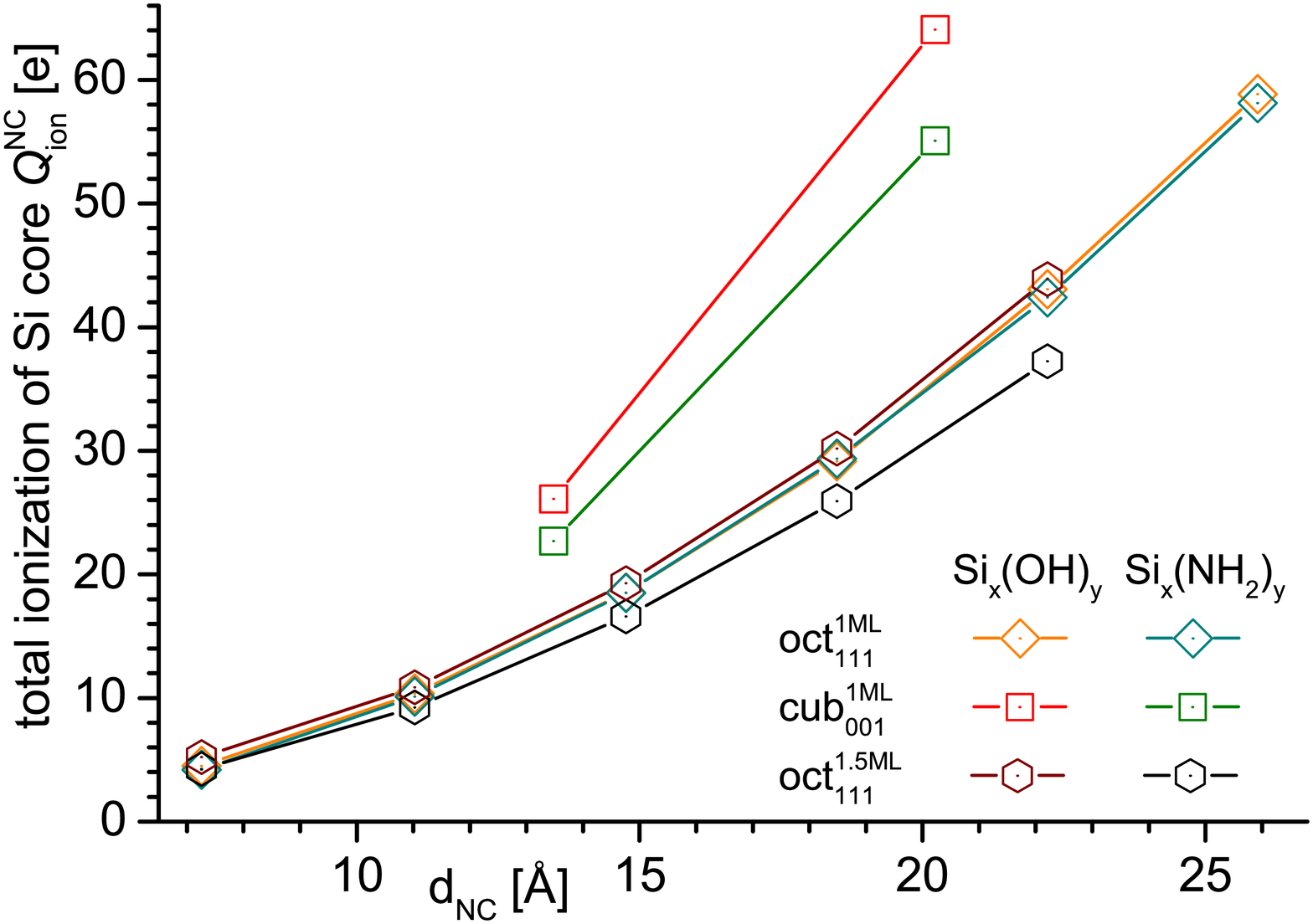}
\caption{\label{fig03}
Total ionization of Si-NCs of NC classes shown in Fig. \ref{fig02}. Terminations with OH (NH$_2$) are considered to present 1ML of SiO$_2$ (Si$_3$N$_4$).}  
\end{figure}

We now focus on the ionization of Si-NCs, see Fig. \ref{fig03}. It provides a blueprint of the strength of the ICT which is closely linked to $\Delta E$. All total NC ionizations $Q_{\rm{ion}}^{\rm{NC}}$ over $d_{\rm{NC}}$ show a quadratic dependence which clearly points to a surface effect \cite{Koe14,Koe16}. In general, O-terminated species have a slightly higher ionization although the difference in ionicity of the bond Si---O with ca. 53 \% ionic vs. Si---N with ca. 35 \% is significant \cite{HolWi95,Koe08}. The minute difference for OH-/NH$_2$-terminated octahedral NCs shows that the ICT is saturated for $d_{\rm{NC}}\leq 25.9$ {\AA}. As we will see in Sec. \ref{Mechnsm}, this value is merely limited by tractable DFT system size and surpassed roughly 3-fold when feeding experimental results into an analytic impact model of $\Delta E$ \cite{Koe14}. Accordingly, the embedding of Si-NCs into 1.5 ML SiO$_2$ does only yield to a slightly higher $Q_{\rm{ion}}^{\rm{NC}}$, whereby ionization drops notably for coating in 1.5 ML Si$_3$N$_4$. The origin of the latter is due to the positive electron affinity $X$ of N which deflects the valence electron wavefunctions originating from Si-NC atoms back into the NC, resulting in a delocalization of these wavefunctions, see Sec. \ref{Mechnsm} for details. Due to MO hybridization, the unoccupied states follow suit. Considering $Q_{\rm{ion}}^{\rm{NC}}$ as a function of interface orientation, it becomes apparent that $\langle001\rangle$-faceted interfaces with their increased bond density yield a stronger ICT. A direct comparison of $N_{\rm{IF}}/N_{\rm{NC}}$ between  $\langle001\rangle$-cubic vs. $\langle111\rangle$-octahedral NCs cannot be accomplished due to their different $d_{\rm{NC}}$. As a good estimate, we can compare the average value of the Si$_{165}$X$_{100}$ and Si$_{286}$X$_{144}$ NCs of $N_{\rm{IF}}/N_{\rm{NC}}=0.555$ with their average size of $d_{\rm{NC}}=20.4$ {\AA} to $N_{\rm{IF}}/N_{\rm{NC}}=0.917$ of the $\langle001\rangle$-cubic approximant Si$_{216}$($>$XH$_{-1}$)$_{75}$X$_{48}$ with $d_{\rm{NC}}=20.2$ {\AA} \cite{Koe16}. We thus arrive at $^5\!/_3$ of the number of interface bonds per NC atom when going from  $\langle111\rangle$-octahedral to  $\langle001\rangle$-cubic NCs. Accordingly, $\langle001\rangle$-cubic NCs are significantly more ionized than $\langle111\rangle$-octahedral NCs which can be seen in Fig. \ref{fig03}.

\begin{figure}[h!]
\includegraphics[width=8.6cm,keepaspectratio]{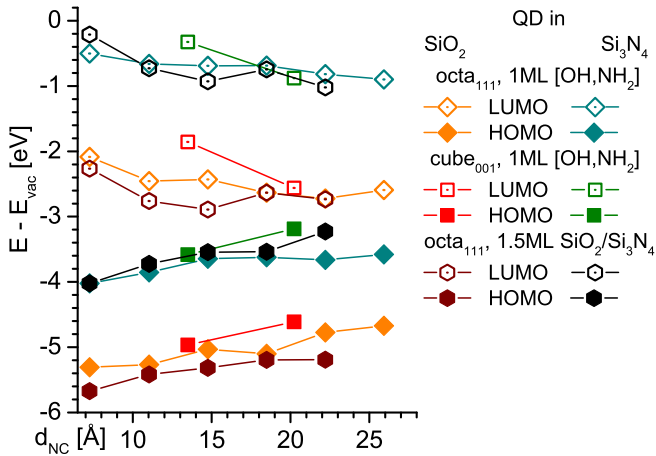}
\caption{\label{fig04} HOMO and LUMO energies relative to $E_{\rm{vac}}$ of NC classes shown in Fig. \ref{fig02}. Terminations with OH (NH$_2$) are considered to present 1ML of SiO$_2$ (Si$_3$N$_4$).} 
\end{figure}

Alternations of the ICT as function of interface orientation and thickness of embedding dielectric result in a modification of the NC electronic structure, namely energies of the highest occupied MO (HOMO) $E_{\rm{HOMO}}$ and of the lowest unoccupied MO (LUMO) $E_{\rm{LUMO}}$, see Fig. \ref{fig04}. 
The increased ICT due to a higher ratio $N_{\rm{IF}}/N_{\rm{NC}}$ for $\langle001\rangle$-cubic NCs or by increasing the dielectric embedding from 1 to 1.5 ML results in a more pronounced shift with respect to the vacuum level $E_{\rm{vac}}$. For N-terminated NCs, a more pronounced shift towards $E_{\rm{vac}}$ is clearly visible for the $\langle111\rangle$-octahedral NCs embedded in 1.5 ML Si$_3$N$_4$ and in particular for the $\langle001\rangle$-cubic NCs. The results for O-terminated $\langle111\rangle$-octahedral NCs show the same trend with more pronounced shifts further below $E_{\rm{vac}}$. The cubic NCs Si$_{64}$($>$O)$_{27}$(OH)$_{30}$ ($d_{\rm{NC}}=13.5$ {\AA}) and Si$_{216}$($>$O)$_{75}$(OH)$_{48}$ do not follow the trend which appears surprising on a first glance. When looking at the latter cubic Si-NC (see Fig. \ref{fig02}b), it becomes apparent that considerable strain deforms the NC cube. For $\langle111\rangle$-octahedral NCs, strain has only a minor influence on their electronic structure even with embedding  -- see Fig \ref{fig02} and \cite{Koe18a}. However, with many $>$O bond configurations on $\langle001\rangle$ interfaces, strain may become more influential on the electronic structure. Such strain becomes evident when comparing bond lengths and angles as listed in Table \ref{tab_1}. The bond length of O---Si bonds in ultrathin SiO$_2$ at the NC is ca. 7.4 \% longer when compared to SiO$_2$, while the value for Si---N bonds is only stretched by ca. 3.8 \% with respect to Si$_3$N$_4$ \cite{Koe18aa}. A similar situation exists with bond angles of Si atoms when going from the NC center to its interface. Together with a bond energy of $E_{\rm{bond}}\mbox{(Si---O)}\approx 4.72$ eV vs. $E_{\rm{bond}}\mbox{(Si---N)}\approx 3.89$ eV \cite{HolWi95}, the combination of higher $E_{\rm{bond}}$ and shorter bond lengths shows that $>$O bond configurations to Si are capable of generating notably more local strain as opposed to $>$NH bond configurations. With O-terminated $\langle001\rangle$-cubic NCs, it appears that significant strain lowers the binding energy of the system which is reflected in a slight shift towards $E_{\rm{vac}}$, see Fig. \ref{fig04}. It was reported that strain dominates the electronic structure of even $\langle111\rangle$-octahedral Si-NCs embedded in SiO$_2$ \cite{Guer09}, although we cannot confirm such results with own calculations of such Si-NCs in up to 3 ML of SiO$_2$ \cite{Koe18a}. Hence, only massive strain emerges as modifier to the electronic structure, though still exceeded considerably by the impact of embedding dielectric via ICT. Bigger $\langle001\rangle$-cubic approximants such as Si$_{512}$($>$O)$_{147}$(OH)$_{66}$ with $d_{\rm{NC}}=27.0$ {\AA} get partially disintegrated by strain due to $>$O bonds, showing that stress destroys the NC before becoming dominant over the ICT.
\begin{table}[h!]
\caption{\label{tab_1} Bond lengths $d_{\rm{Si-X}}$ (X $=$ O, N) of bridging anion species $>$O or $>$NH, bond angles over associated Si atoms $\angle_{\rm{X-Si-X}}$, and bond angles $\angle_{\rm{Si-Si-Si}}$ over Si--Si--Si in the center of cubic NCs shown for 20.2 {\AA} cubic Si-NCs with exclusively $\langle001\rangle$-oriented interfaces, \emph{cf.} Fig. \ref{fig02}b. For comparison, $d_{\rm{Si-X}}$ bond lengths of SiO$_2$ and Si$_3$N$_4$ approximants are listed. All lengths and angles are given as average values with absolute standard deviations. All approximants had identical parameters in HF/3-21G structural optimization.}
\begin{ruledtabular}
\begin{tabular}{l||c|c}
&Si$_{216}$($>$O)$_{75}$(OH)$_{48}$&SiO$_2$ [Si$_{29}$O$_{76}$H$_{36}$]\\
\hline
$d_{\rm{Si-O}}$\,[{\AA}]$^{\rm{a}}$&$1.732\pm 0.031$&$1.612\pm 0.009$ \cite{Koe18aa}\\
$\angle_{\rm{O-Si-O}}$\,[$^{\circ}$]&$132.86\pm 1.24$&\\
$\angle_{\rm{Si-Si-Si}}$\,[$^{\circ}$]$^{\rm{b}}$&$109.76\pm 0.91$&\\ \hline
&Si$_{216}$($>$NH)$_{75}$(NH$_2$)$_{48}$&Si$_3$N$_4$ [Si$_{40}$N$_{86}$H$_{98}$]\\
\hline
$d_{\rm{Si-N}}$\,[{\AA}]$^{\rm{c}}$&$1.824\pm 0.032$&$1.757\pm 0.016$ \cite{Koe18aa}\\
$\angle_{\rm{N-Si-N}}$\,[$^{\circ}$]&$125.19\pm 1.27$&\\
$\angle_{\rm{Si-Si-Si}}$\,[$^{\circ}$]$^{\rm{b}}$&$109.75\pm 0.96$&\\
\end{tabular}
\end{ruledtabular}
\begin{flushleft}
$^{\rm{a}}$ excluding --OH groups\\
$^{\rm{b}}$ angles from central Si atom up to 2-nn Si\\
$^{\rm{c}}$ excluding --NH$_2$ groups
\end{flushleft}
\end{table} 

In summary, we observed a substantial energy offset of the electronic structure of Si-NCs terminated with O vs. N, yielding to $\Delta E_{\rm{HOMO}}\approx 1.75$ eV and $\Delta E_{\rm{LUMO}}\approx 1.93$ eV with 1.5 ML coating by SiO$_2$ vs. Si$_3$N$_4$. These values correspond to ca. 75 \% and 55 \%, respectively, of the nominal HOMO-LUMO gap of Si-NCs in the size range of 20 {\AA}. We found that the ICT is dominated by bond densities per square given by interface orientation via the ratio $N_{\rm{IF}}/N_{\rm{NC}}$ and to a lesser degree by the thickness of the embedding dielectric. Interface strain originating from $>$O bonds does have a minor influence on the electronic structure by decreasing binding energies of $\langle001\rangle$-cubic NCs, shifting their electronic structure slightly towards $E_{\rm{vac}}$, opposing the downshift of the electronic structure by SiO$_2$-embedding to some extent.

\subsection{\label{DFT-Res-2NCs} One Approximant Featuring a Si-NC in SiO$_2$ Adjacent to a Si-NC in Si$_3$N$_4$}
An approximant consisting of two Si-NCs, with the first (second) NC embedded in SiO$_2$ (Si$_3$N$_4$), presents the ultimate DFT system to test the energy offset $\Delta E$ and to gain further insight into its working principle. We calculated four approximants which mainly differ in the size of embedded Si-NCs: $2\times$Si$_{10}$ (7.2 {\AA} each) \cite{Koe18a}, $2\times$Si$_{35}$ (11.0 {\AA} each), $2\times$Si$_{84}$ (14.8 {\AA} each) and $2\times$Si$_{165}$ (18.5 {\AA} each). The latter two NC sizes overlap with experimental observations, thus providing a solid ground for Sec. \ref{Mechnsm} where we estimate the size range up to which dns-Si systems are dominated by $\Delta E$. 

\begin{figure*}[h!]
\vspace*{20.5cm}
 \begin{picture}(0,0)
  \includegraphics{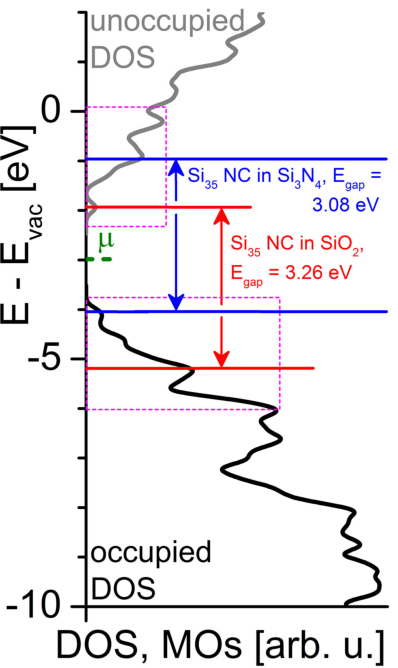}
  \put(-126,504){{\large\bf(a)}}
 \end{picture}
 \begin{picture}(0,0)
 \includegraphics{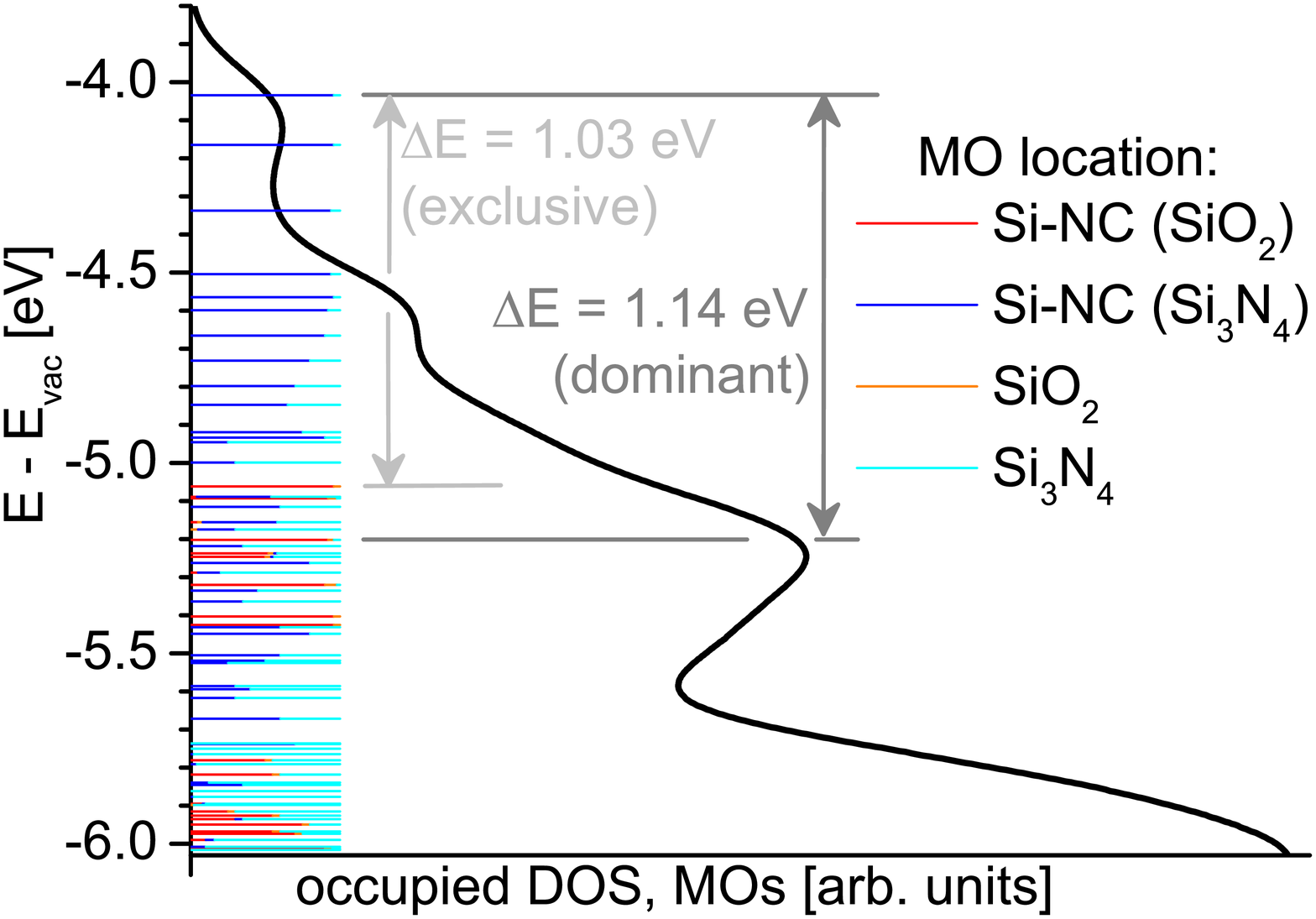}
 \end{picture}
 \begin{picture}(0,0)
 \includegraphics{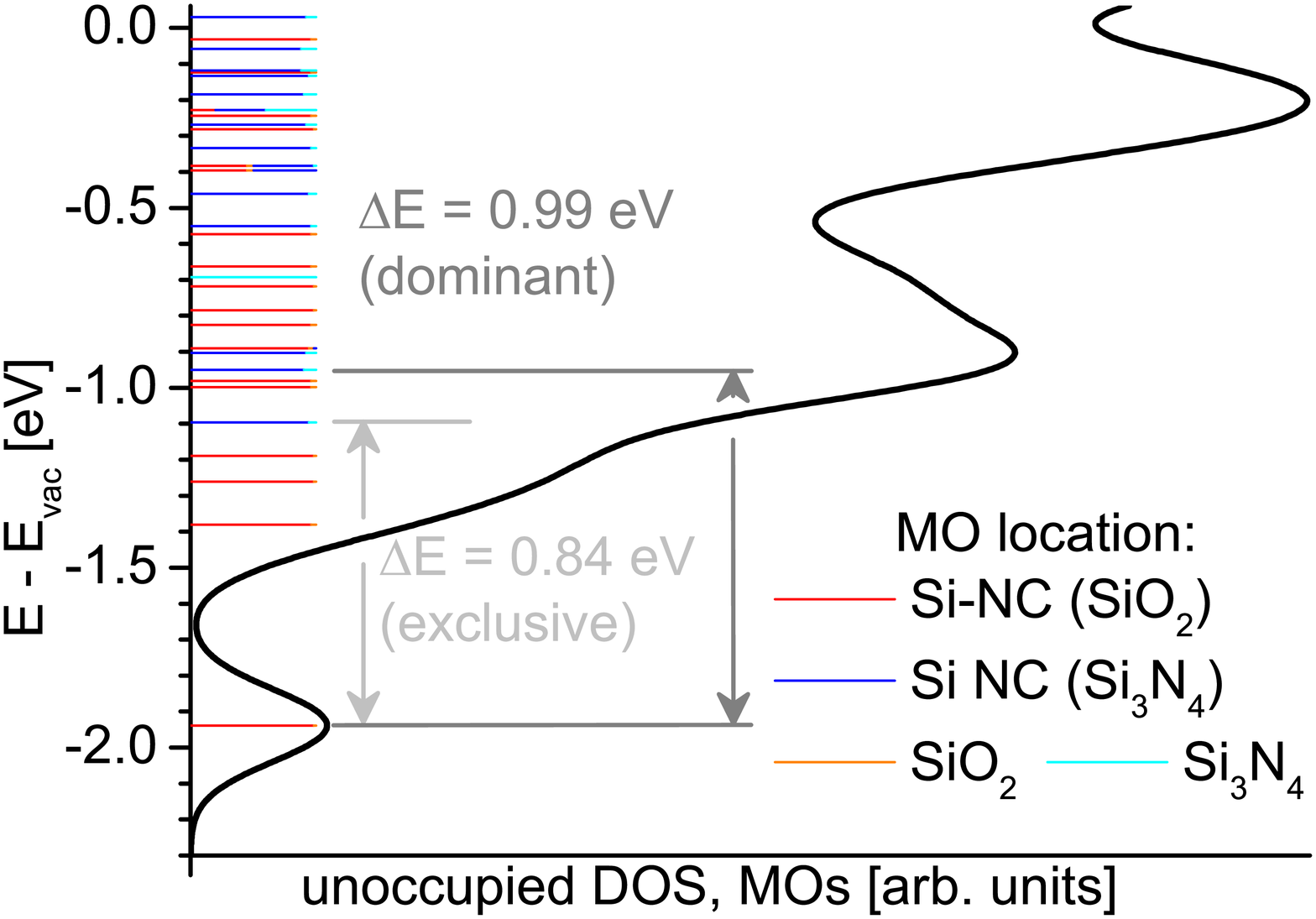}
 \end{picture}
 \begin{picture}(0,0)
 \includegraphics{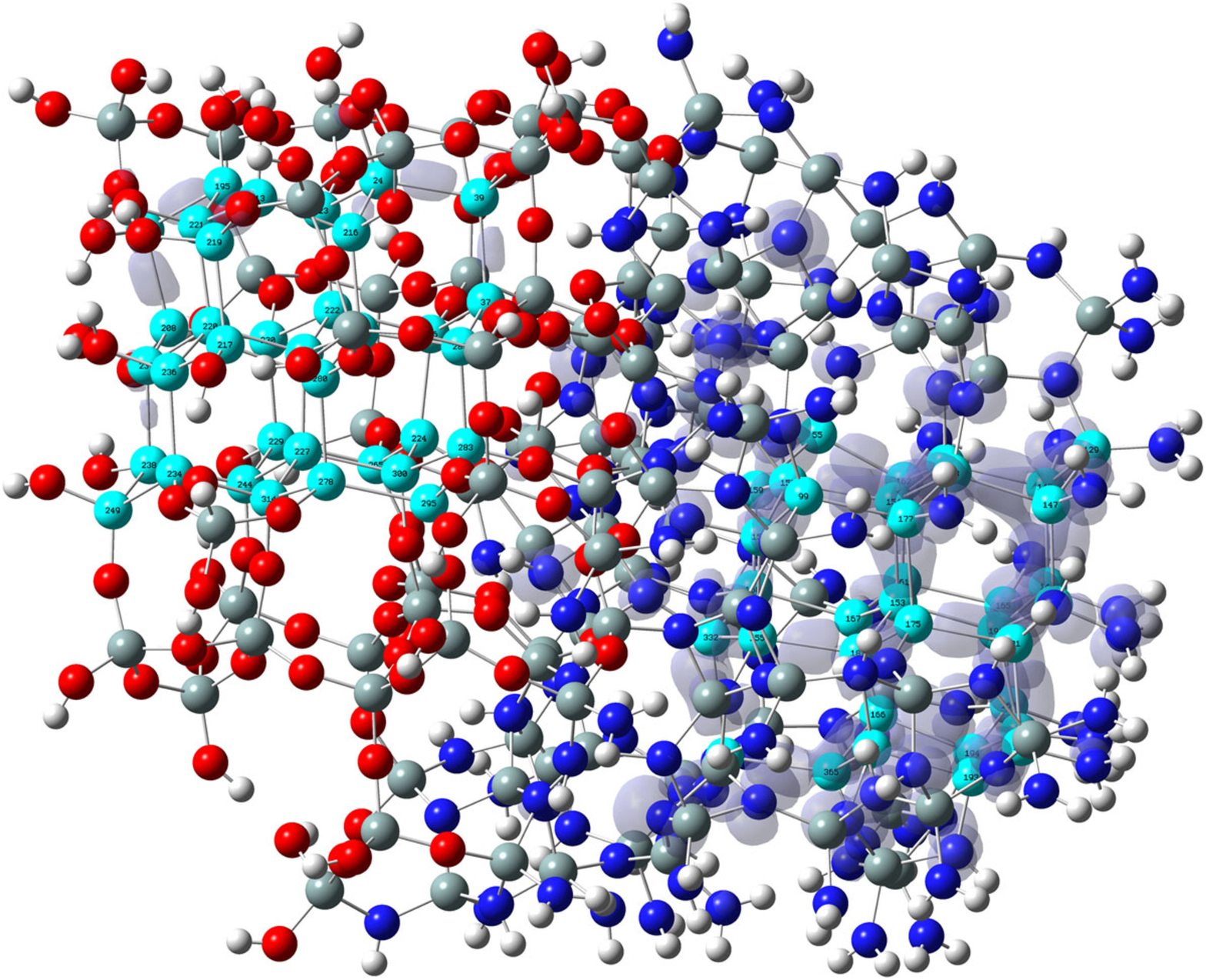}
 \put(245,504){{\large\bf(c)}}
 \end{picture}
  \begin{picture}(0,0)
  \includegraphics{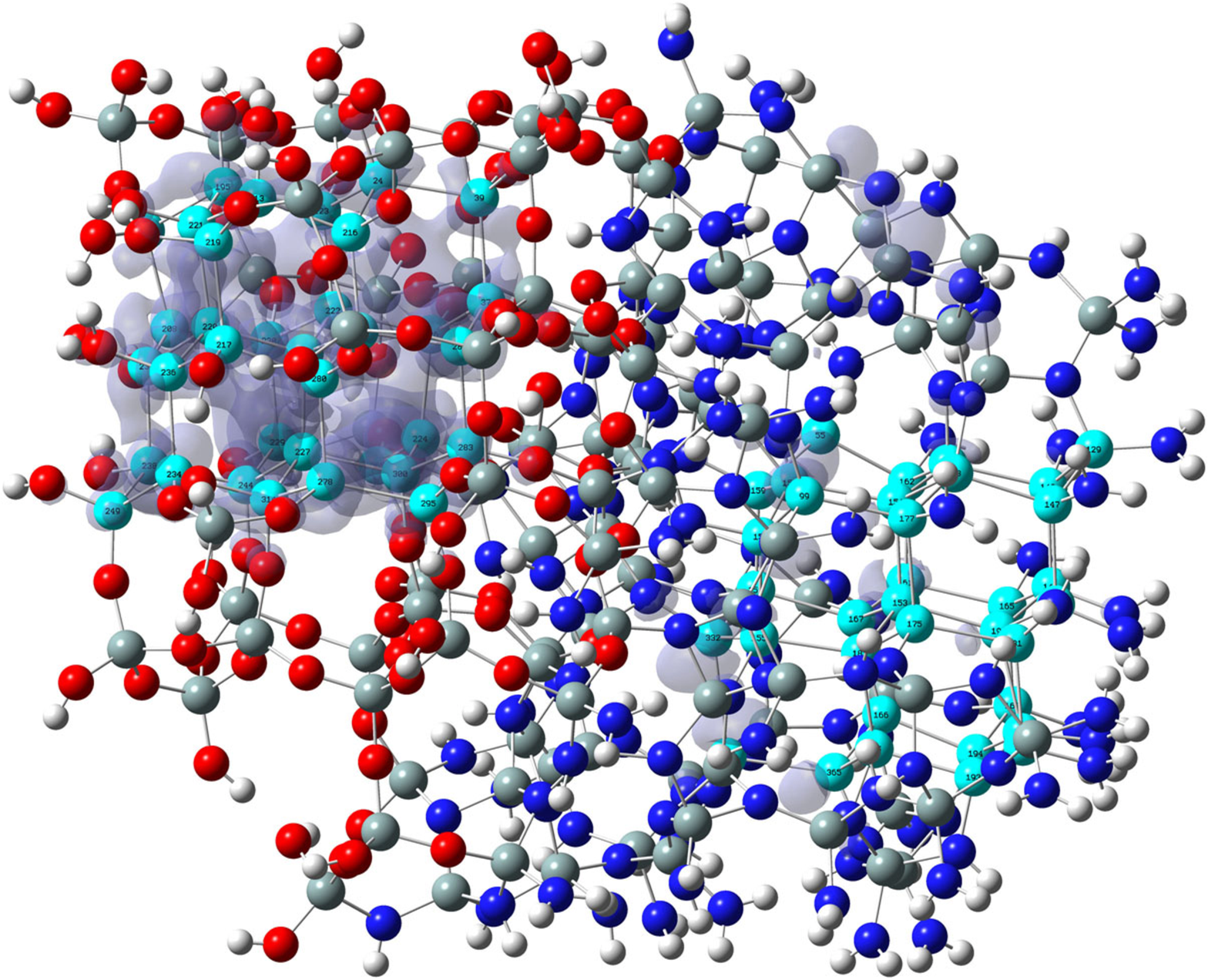}
  \put(41.7,504){{\large\bf(b)}}
 \end{picture} 
 \begin{picture}(0,0)
 \includegraphics{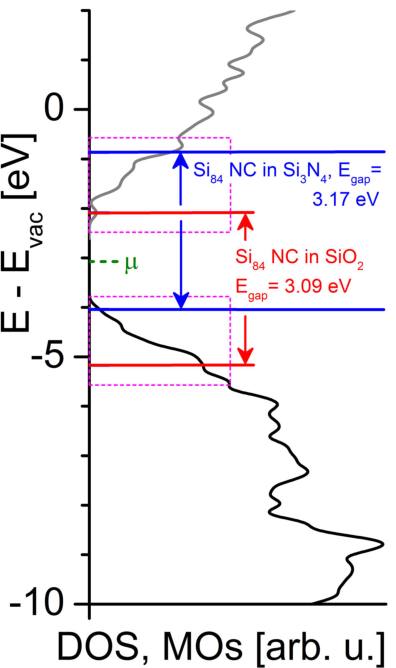}
  \put(-141.5,307){{\large\bf(d)}}
 \end{picture}
 \begin{picture}(0,0)
 \includegraphics{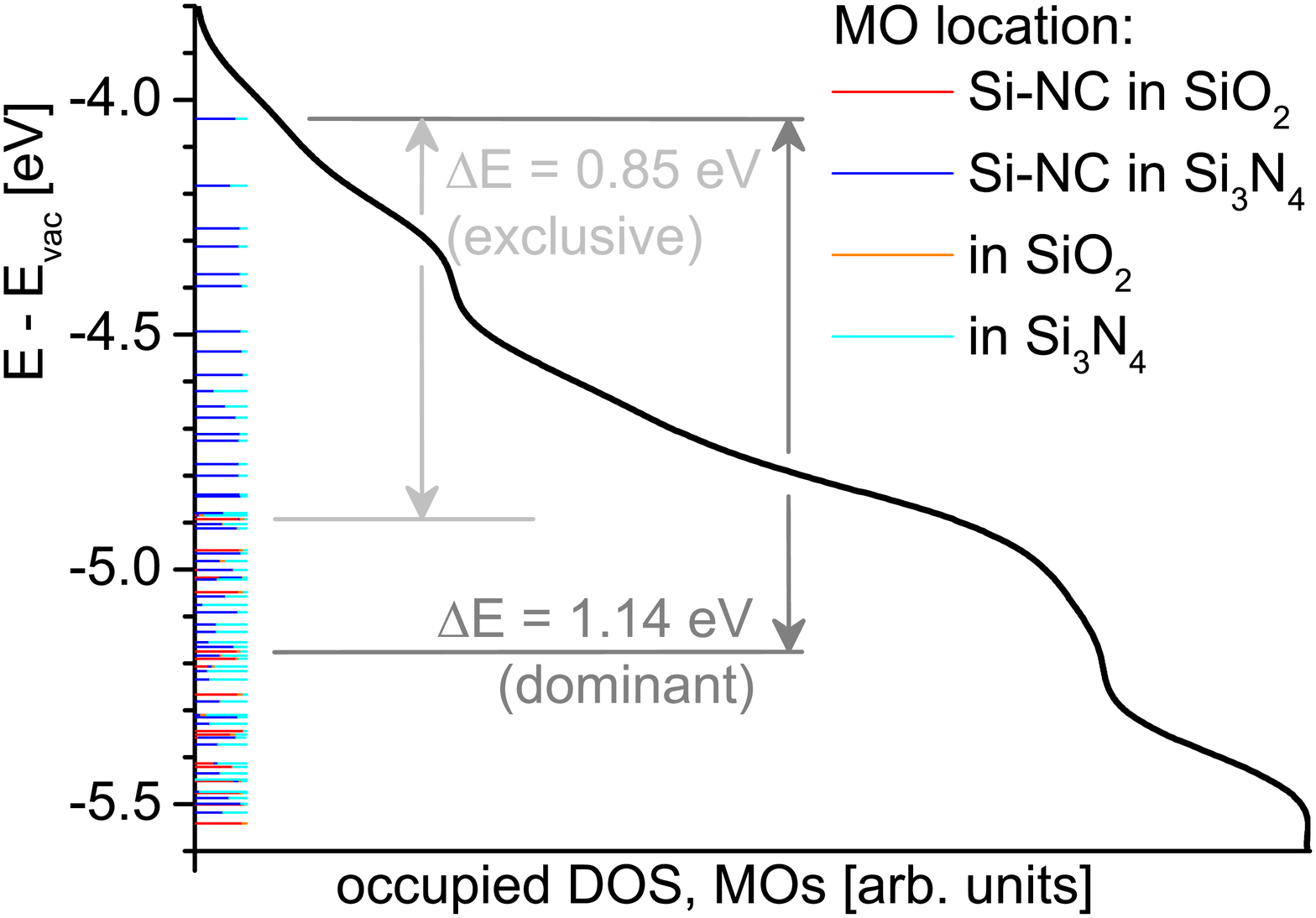}
 \end{picture}
 \begin{picture}(0,0)
 \includegraphics{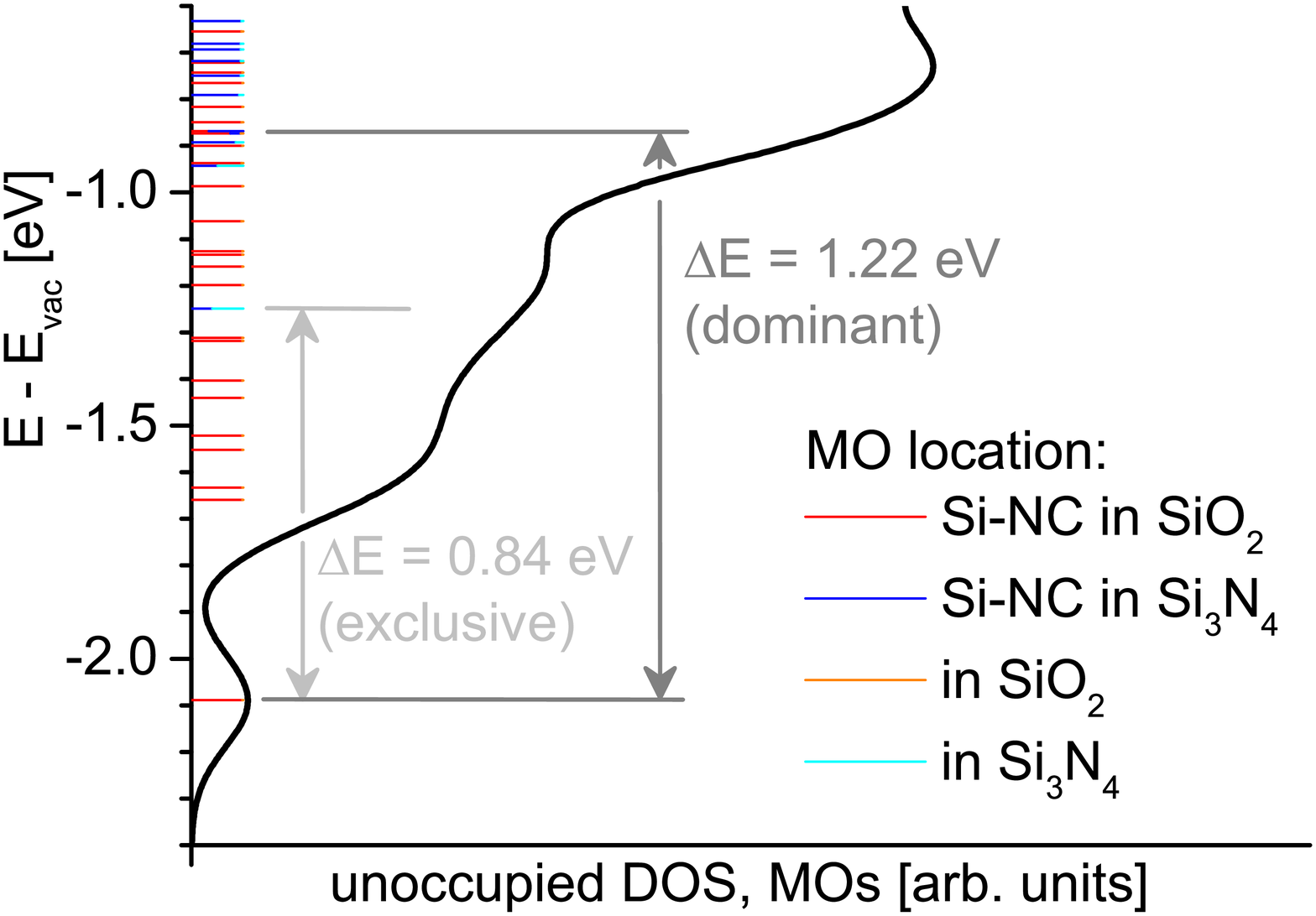}
 \end{picture}
 \begin{picture}(0,0)
 \includegraphics{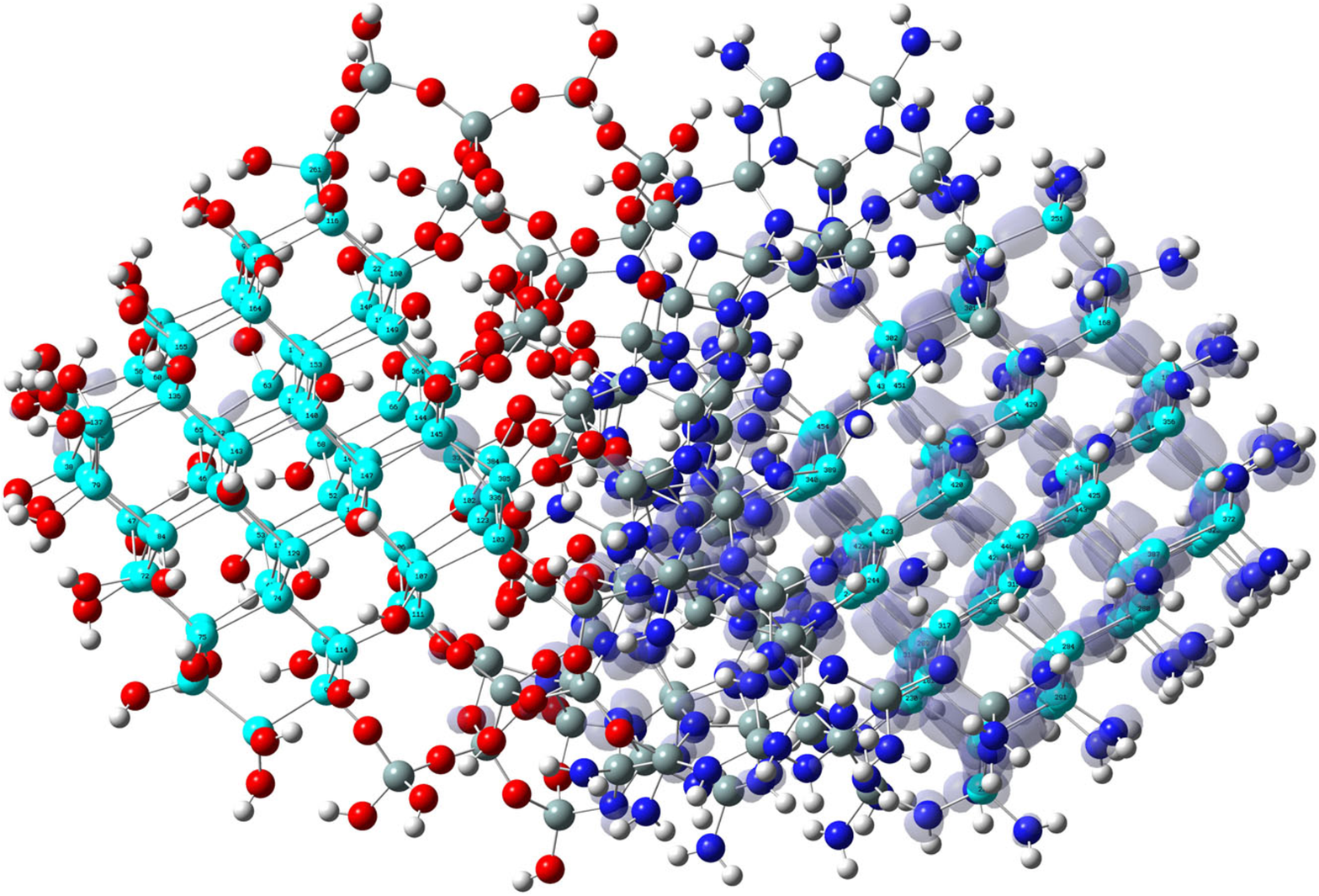}
  \put(29,307){{\large\bf(e)}}
 \end{picture}
 \begin{picture}(0,0)
 \includegraphics{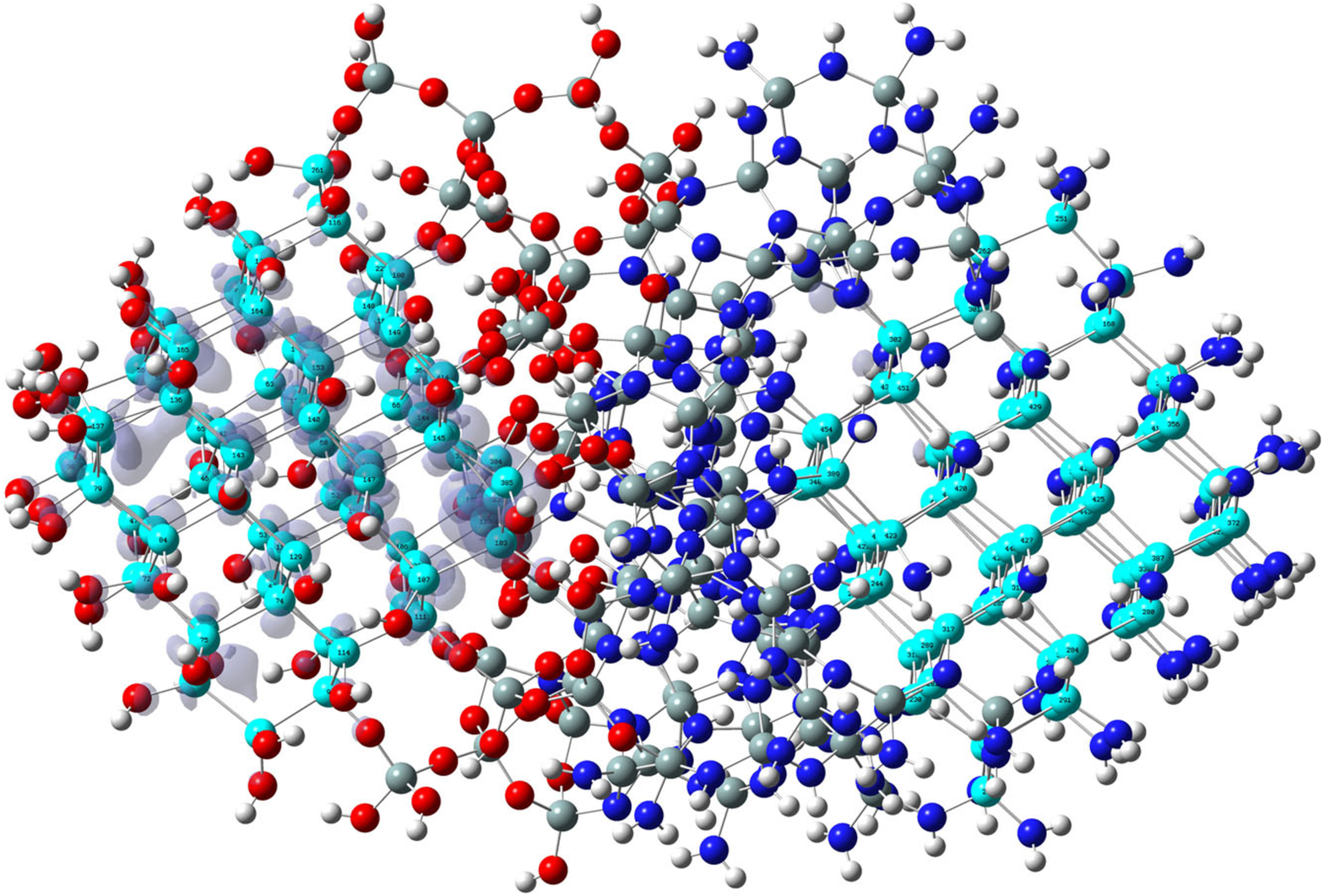}
 \put(226.5,307){{\large\bf(f)}}
 \end{picture} 
 \begin{picture}(0,0)
 \includegraphics{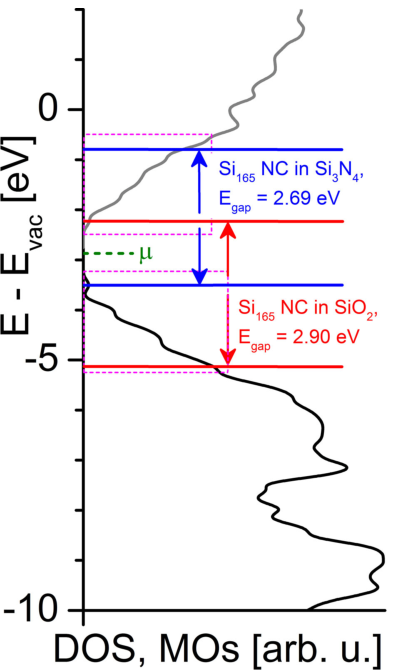}
  \put(-151,110){{\large\bf(g)}}
 \end{picture}
 \begin{picture}(0,0)
 \includegraphics{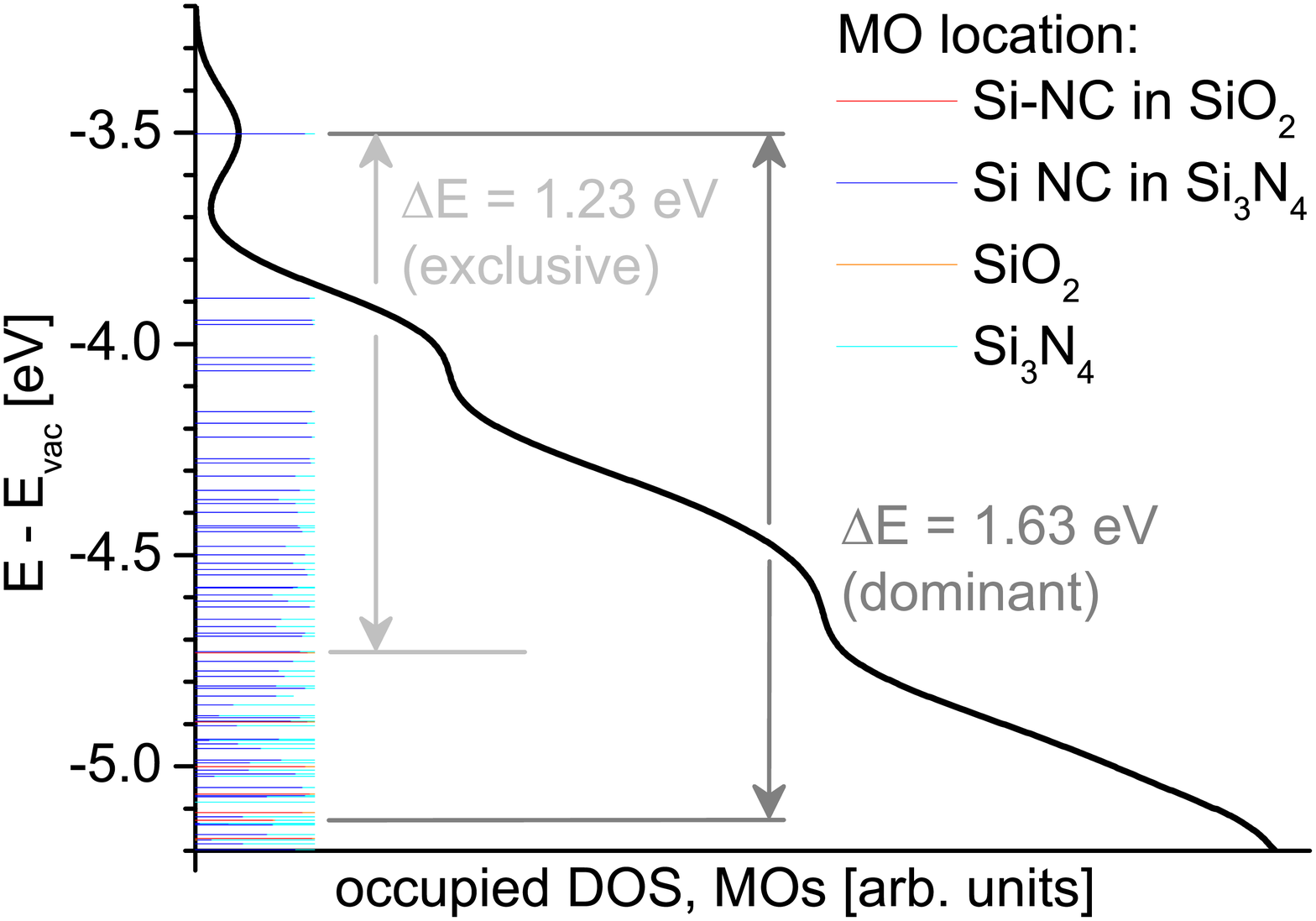}
 \end{picture}
 \begin{picture}(0,0)
 \includegraphics{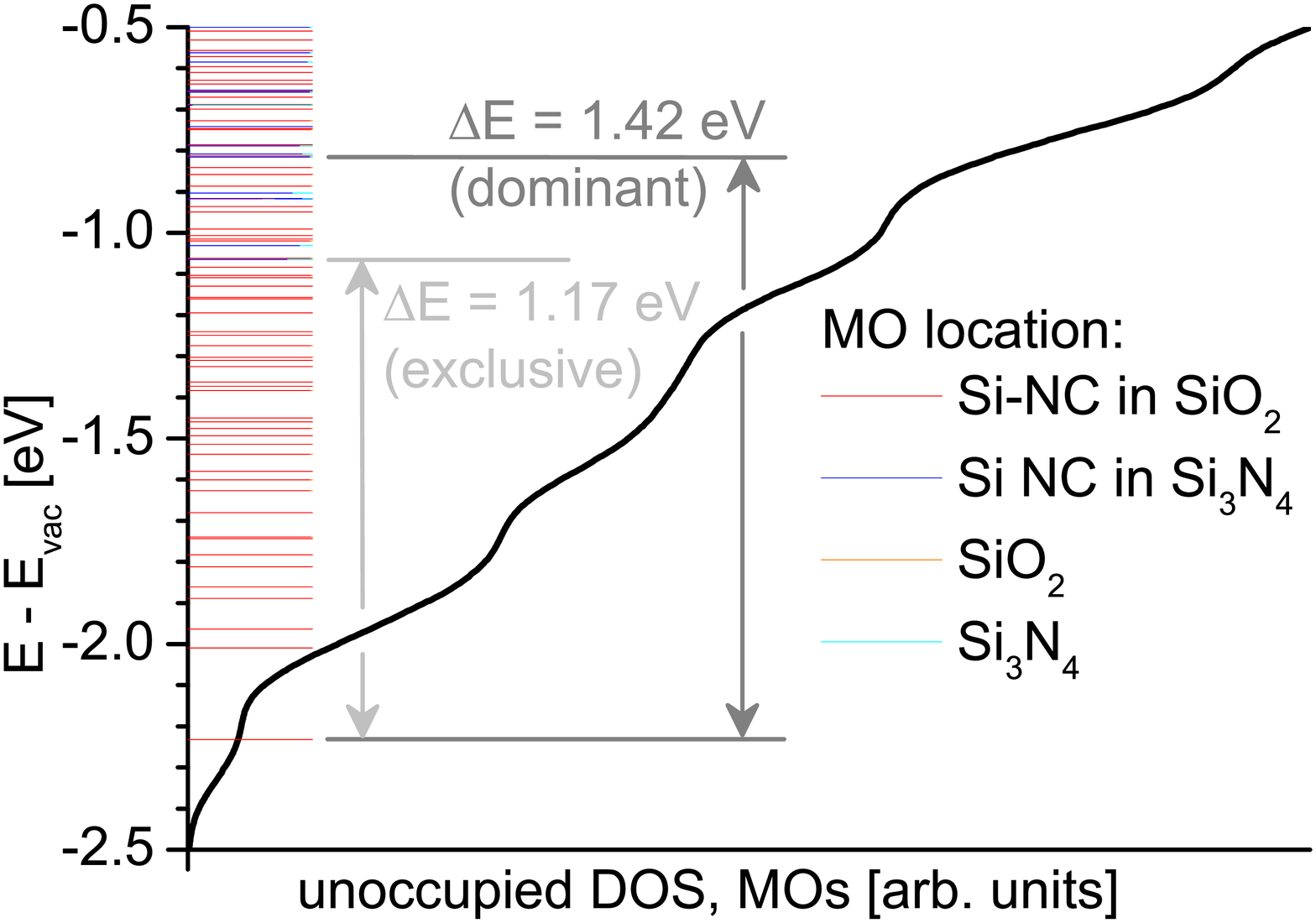}
 \end{picture}
 \begin{picture}(0,0)
 \includegraphics{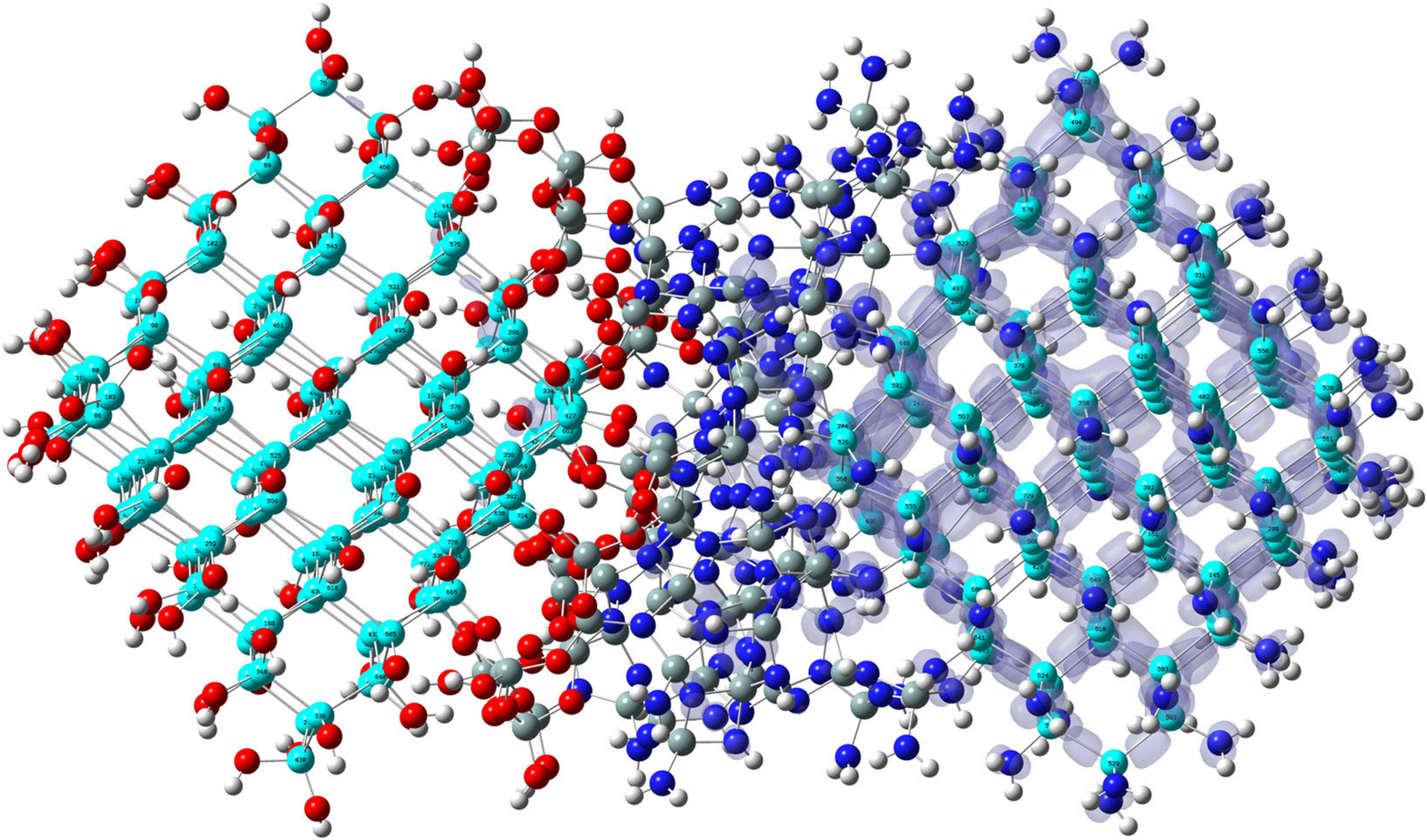}
  \put(13,110){{\large\bf(h)}}
 \end{picture}
 \begin{picture}(0,0)
 \includegraphics{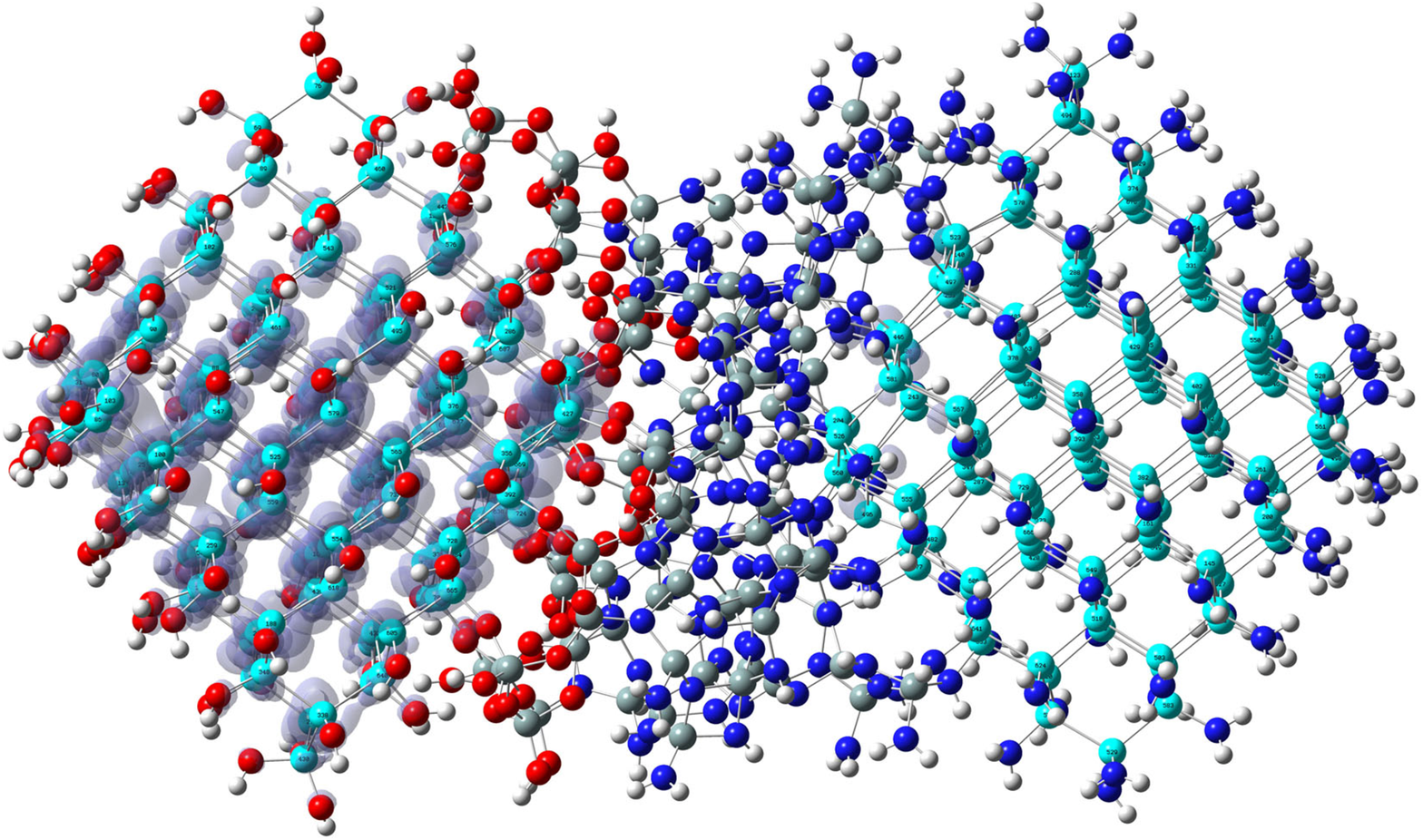}
  \put(211.3,110){{\large\bf(i)}}
 \end{picture} 
\caption{\label{fig05}{\bf DFT results of two embedded Si-NCs in SiO$_2$/Si$_3$N$_4$ within one approximant.} 
DOS of $2\times$Si$_{35}$ NCs, of 11.0 {\AA} size {\bf(a)}. The green dashed line shows the chemical potential $\mu$ of the approximant. Detailed DOS of frontier OMOs contained in magenta rectangle, showing their localization, together with iso-density plot of OMOs [$4.6\times 10^{-3}\,e/a_{\rm{B,0}}$] within energy window where NC in Si$_3$N$_4$ dominates electronic structure {\bf(b)}. Detailed DOS of frontier UMOs contained in magenta rectangle, showing their localization, together with iso-density plot of UMOs [$3.2\times 10^{-3}\,e/a_{\rm{B,0}}$] within energy window where NC in SiO$_2$ dominates electronic structure {\bf(c)}. All iso-density plots have the same value per frontier OMOs/UMOs. Graphs {\bf(d)}-{\bf(f)} show results for $2\times$Si$_{84}$ NCs of 14.8 {\AA} size, graphs {\bf(g)}-{\bf(i)} for $2\times$Si$_{165}$ NCs of 18.5 {\AA} size. For atom colors see Fig. \ref{fig02}.}
\end{figure*}

\begin{figure}
\includegraphics[width=8.6cm,keepaspectratio]{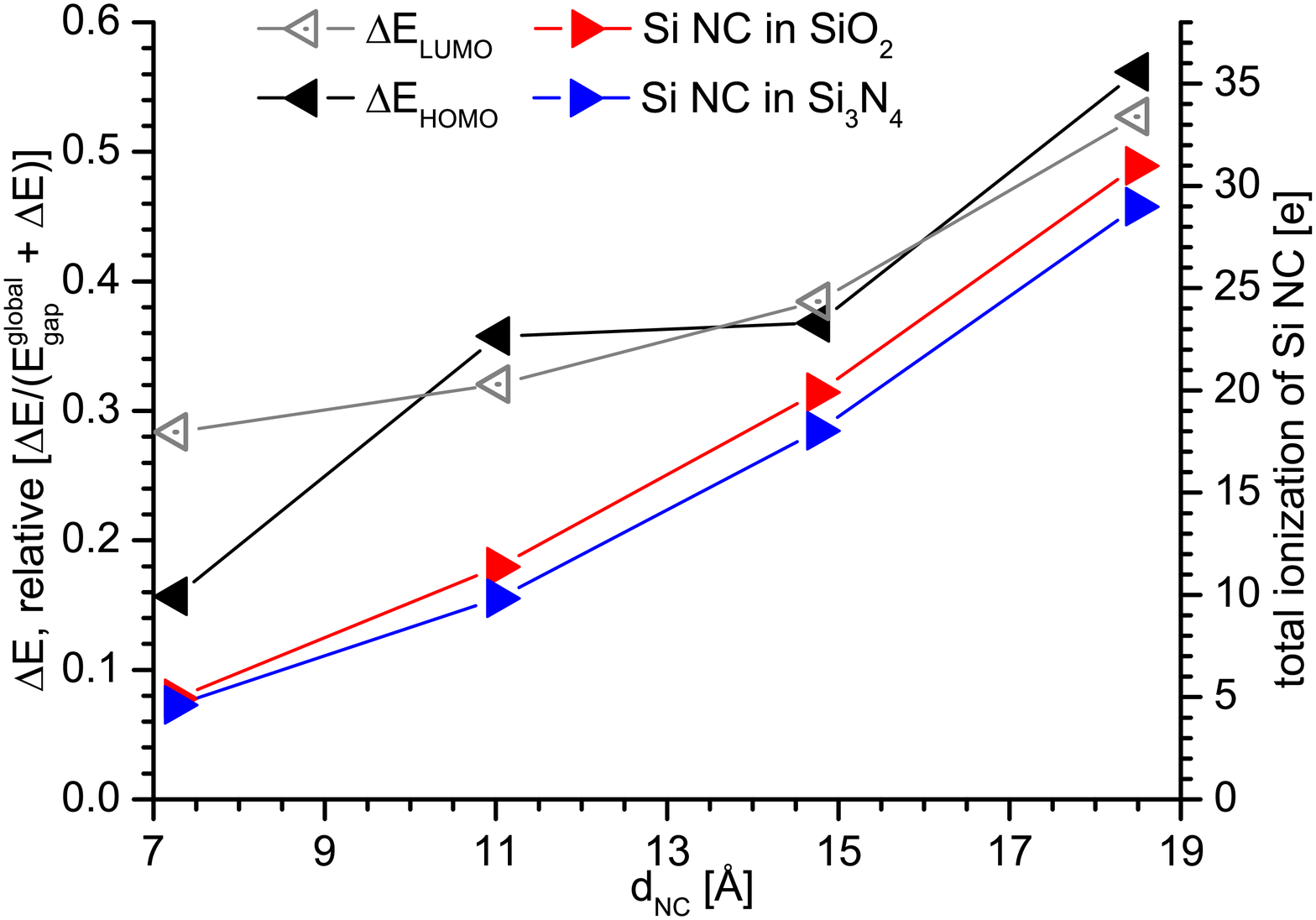}
\caption{\label{fig06} Relative change in OMO and UMO energy offsets due to SiO$_2$- vs. Si$_3$N$_4$-embedding as function of NC size, referring to global HOMO-LUMO gap of Si-NCs in SiO$_2$ vs. Si$_3$N$_4$, see Fig. \ref{fig05}. Colored symbols show total Si-NC ionization as function of NC size and embedding and refer to right scale.} 
\end{figure}

Fig. \ref{fig05} shows the results for the $2\times$Si$_{35}$, $2\times$Si$_{84}$ and $2\times$Si$_{165}$ approximants, with respective DOS ranges enlarged to show the localization of the electronic states within the approximants. Underneath these DOS ranges, iso-density plots of the partial DOS located dominantly within one of the NCs are shown as a function of their embedding. 
We can clearly see two effects: $\Delta E_{\rm{HOMO}}$ and $\Delta E_{\rm{LUMO}}$ increase with NC separation growing with NC size \footnote{There are 2 ML, 3 ML, 4 ML and 4.5 ML of SiO$_2$ plus Si$_3$N$_4$ for the $2\times$Si$_{10}$-, $2\times$Si$_{35}$-, $2\times$Si$_{84}$- and $2\times$Si$_{165}$-approximants separating the NCs. Increasing inter-NC layer thickness of the combined dielectrics or thickness of the dielectric at outer NC facets is not possible at present due to intractability of DFT computations for $>1,500$ heavy atoms with the level of theory we use.}. The global energy gap of the entire approximant $E_{\rm{gap}}^{\rm{global}}$ decreases with NC size. Both effects lead to a superlinear increase in n/p-type preference for charge carriers with increasing NC size, \emph{cf.} Fig. \ref{fig06}, and in general with the size of dns-Si systems up to a certain limit, see Sec. \ref{Mechnsm}. We note that $\Delta E_{\rm{HOMO}}$ and $\Delta E_{\rm{LUMO}}$ of the $2\times$Si$_n$ approximants ($n=10,\,35,\,84,\,165$) do not reach the values obtained for individual NCs in Sec. \ref{DFT-Res-1NC}. The proximity of SiO$_2$ to Si$_3$N$_4$ results in a transient region where the electron-localizing nature of O cancels out with the electron-delocalizing nature of N \cite{Koe18a}. Since the combined thickness of SiO$_2$ and Si$_3$N$_4$ between NCs gets thicker with increasing NC size, this effect diminishes to yield to increased $\Delta E_{\rm{HOMO}}$ and $\Delta E_{\rm{LUMO}}$.

The total NC ionization in Fig. \ref{fig06} only shows minor deviations from the ionization of $\langle111\rangle$-octahedral N Cs with OH/NH$_2$-termination, see Fig. \ref{fig03}, confirming that both 18.5 {\AA} Si-NCs (Si$_{165}$) are still within range of ICT saturation. These minor deviations are instructive to understand the impact of O vs. N onto dns-Si, see Sec. \ref{Mechnsm}. 

For all NC sizes, N-terminated Si-NCs slightly increase their ionization from coverage with all-around 1.5 ML Si$_3$N$_4$ via inclusion into 1 to 3 ML Si$_3$N$_4$ in $2\times$Si$_n$ approximants up to all-around 1ML Si$_3$N$_4$ coverage. As pointed out in Sec. \ref{DFT-Res-1NC}, this increase with decreasing thickness of Si$_3$N$_4$-embedding originates from the positive electron affinity $X$ of N. 
The 1 to 3 ML Si$_3$N$_4$ coverage takes an intermediate position between full coverages with 1 and 1.5 ML Si$_3$N$_4$. In other words, the 1 ML Si$_3$N$_4$ coverage of the outer NC interface in $2\times$Si$_n$ approximants does not delocate and/or deflect valence electron wavefunctions originating from Si-NC atoms enough to provide a full upshift of energy levels. Arguably, and backed by experimental data (Sec. \ref{Expr-UPS}), the electronic states of Si$_3$N$_4$-embedded NCs in the $2\times$Si$_n$ approximants could be located nearer $E_{\rm{vac}}$ if a complete multi-ML embedding in Si$_3$N$_4$  with $>1,500$ heavy atoms could be computed. 

For Si-NCs in SiO$_2$, NC ionization increases from 1 ML via 1.5 ML SiO$_2$ to embedding in 1 to 3ML SiO$_2$ in $2\times$Si$_n$ approximants. Here, the strong localization of valence electrons from Si-NCs at O drives the ionization process and thus the shift of valence electron wavefunctions of the Si-NC to higher binding energies. We thus can presume an average 2 ML embedding in SiO$_2$ for the  $2\times$Si$_n$ approximants, yielding 1, 1.5 and 2 ML for the increasing ionization of Si-NCs by SiO$_2$, clearly correlating with the number of O atoms per Si-NC atom.

To recapitalize, $\Delta E$ also exists in the complete system manifested by 
$2\times$Si$_n$ approximants, one embedded each in SiO$_2$ and Si$_3$N$_4$. 
The strength of $\Delta E$ increases with NC separation given by the combined inter-NC thickness of SiO$_2$ and Si$_3$N$_4$, though not quite reaching the values for individual NCs in SiO$_2$ vs. Si$_3$N$_4$. As a likely cause we identified the transition range between SiO$_2$ and Si$_3$N$_4$ where the nature of both anions -- O and N -- tend to cancel each others impact on the electronic structure.
Apart from proving the full scale of $\Delta E$ in a complete system, the $2\times$Si$_n$ approximants investigated in this section yield to a more detailed evaluations of NC ionization and $\Delta E$.  

\section{\label{Expr-UPS}Experimental Results: Synchrotron UPS}

\begin{figure}[h!]
\includegraphics[width=8.6cm,keepaspectratio]{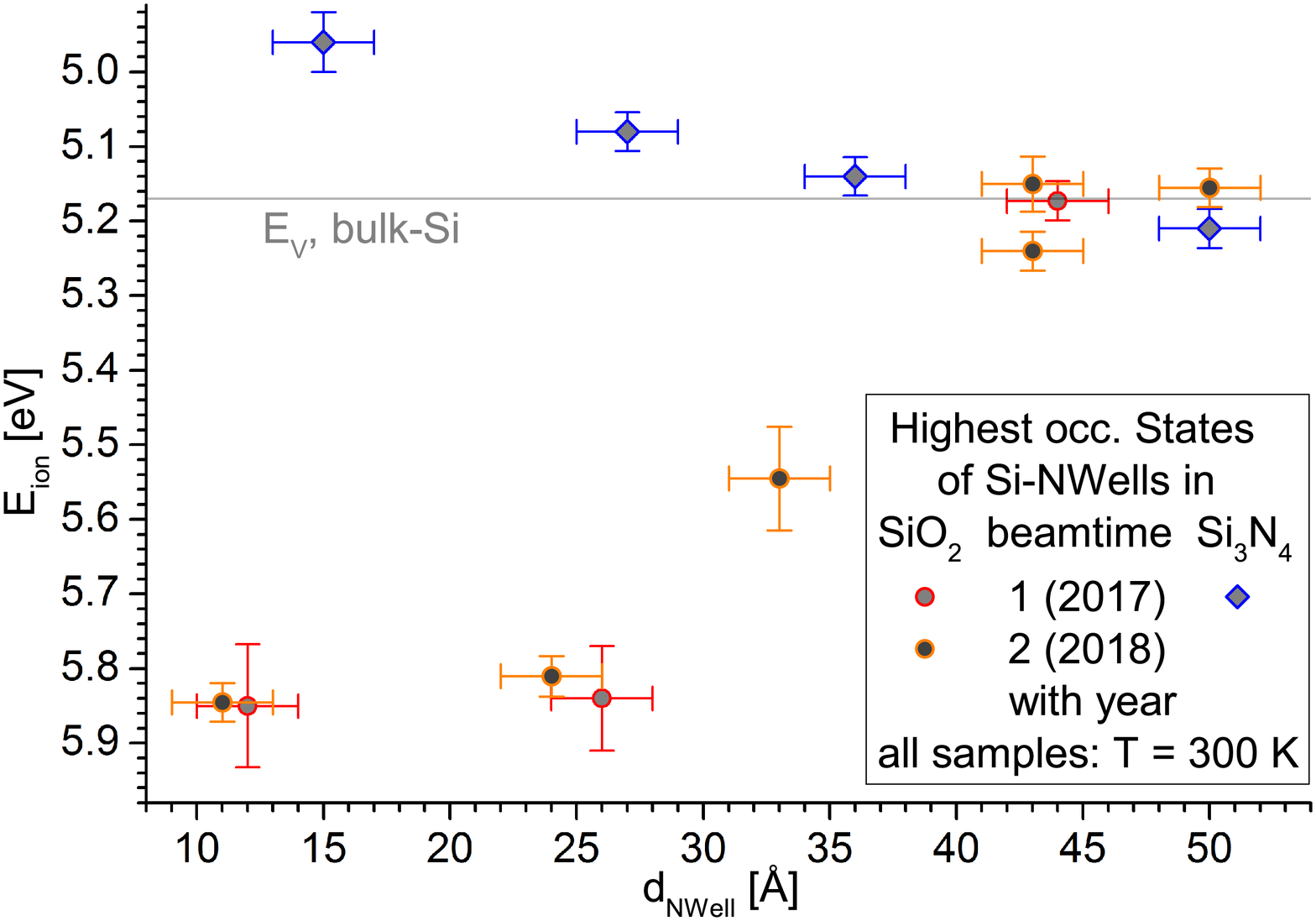}
\caption{\label{fig07} Ionization energies $E_{\rm{ion}}$, representing $E_{\rm{V}}$ of Si-NWells embedded in SiO$_2$ or in Si$_3$N$_4$ as a function of $d_{\rm{NWell}}$. Each point represents a different sample. Valence band edge of bulk-Si is shown as reference. Error bars refer to standard deviations in thickness and in energy, see Sec. \ref{Apx-UPS-Data-Eval} and \cite{Koe18aa}).} 
\end{figure}

After an initial UPS measurements confirming the effect of $\Delta E$ for Si-NWells in SiO$_2$ vs. Si$_3$N$_4$ \cite{Koe18a}, we characterized 10 SiO$_2$- and 4 Si$_3$N$_4$-embedded NWells with thicknesses ranging from 11 to 50 {\AA}. Since we work with NWells and thus quasi-2D band structures, we replace $E_{\rm{HOMO}}$ and $E_{\rm{LUMO}}$ with the energy of the highest occupied \emph{viz.} top valence band states $E_{\rm{V}}$ and of the lowest unoccupied states \emph{viz.} bottom conduction band $E_{\rm{C}}$, respectively, to account for the limits on continuous electronic DOS. 
Details on deriving the ionization energy $E_{\rm{ion}}=E_{\rm{V}}$ of such valence band states and respective standard deviations are treated in Appendix Sec. \ref{Apx-UPS-Data-Eval} and in \cite{Koe18aa}. Fig. \ref{fig07} shows $E_{\rm{V}}$ values as a function of NWell thickness $d_{\rm{NWell}}$ for SiO$_2$- and Si$_3$N$_4$-embedding. 

For $d_{\rm{NWell}}=11$ to 15 {\AA}, we obtained $E_{\rm{V}}\approx E_{\rm{vac}}-5.85$ eV for NWells in SiO$_2$ and $E_{\rm{V}}\approx E_{\rm{vac}}-4.96$ eV for NWells in Si$_3$N$_4$. For SiO$_2$-embedding, a minute change to $E_{\rm{V}}\approx E_{\rm{vac}}-5.82$ eV occurs for $d_{\rm{NWell}}=25\pm 1$ {\AA}. Changes are more notable for Si$_3$N$_4$-embedding where we obtained $E_{\rm{V}}\approx E_{\rm{vac}}-5.08$ eV for $d_{\rm{NWell}}=27$ {\AA}. We also note that $E_{\rm{V}}(d_{\rm{NWell}})$ changes rather continuously as compared to NWells embedded in SiO$_2$ which is subject to discussion in Sec. \ref{Mechnsm}. In the range of $d_{\rm{NWell}}\approx 30$ to 40 {\AA}, $E_{\rm{V}}$ of Si-NWells in SiO$_2$ and Si$_3$N$_4$ change to approach the common value $E_{\rm{V}}=E_{\rm{vac}}-5.17$ eV as known for bulk-Si. Expressing these values in terms of  $\Delta E_{\rm{V}}$, we get $\Delta E_{\rm{V}}(d_{\rm{NWell}}=11\ \mbox{to}\ 20\ \rm{\AA, SiO_2})=-0.68$ eV and $\Delta E_{\rm{V}}(d_{\rm{NWell}}=11\ \mbox{to}\ 20\ \rm{\AA, Si_3N_4})=+0.21$ eV, decreasing to $\Delta E_{\rm{V}}(d_{\rm{NWell}}=25\pm 1\ \rm{\AA, SiO_2})=-0.65$ eV and $\Delta E_{\rm{V}}(d_{\rm{NWell}}=25\pm 1\ \rm{\AA, Si_3N_4})=+0.09$ eV. The difference in $E_{\rm{V}}$ for Si-NWells embedded in SiO$_2$ vs. Si$_3$N$_4$ thus undergoes a minor decrease from 0.89 eV for $d_{\rm{NWell}}=11\ \mbox{to}\ 15$ {\AA} to 0.74 eV for $d_{\rm{NWell}}=25\pm 1$ {\AA}. Since Si-NWells have significantly lower fundamental gaps as compared to Si-NCs (\emph{cf.} Figs. \ref{fig04} and \ref{fig05}), a pronounced n/p-type charge carrier preference arises from the energy offsets above for $d_{\rm{NWell}}\leq 28$ {\AA}. 

We are planning to measure the conduction band states of NWell samples along with UPS characterizations on the same samples, motivated by Zimina \emph{et al.} \cite{Zim06} who characterized $E_{\rm{C}}$ of Si-NCs embedded in SiO$_2$ for $d_{\rm{NC}}=16$ to 40 {\AA} by soft X-ray emission spectroscopy (SXES). A \emph{downshift of} $E_{\rm{C}}$ up to 0.2 eV \emph{below the conduction band edge} $E_{\rm{C}}$ \emph{of bulk-Si} was obtained \cite{Zim06}. Conventional quantum physics strongly suggests $E_{\rm{C}}$ to be located a few 100 meV above $E_{\rm{C}}$ of bulk-Si which already includes contributions from excitonic binding and dielectric screening. The same work also noted that the interface of Si-NCs is not abrupt but rather gradual, with sub-oxide shells in a thickness range of 3 to 5 {\AA}. This value is likely underestimated since similar superlattices of Si-NCs with sub-oxide shells characterized by atom probe tomography were found to have sub-oxide shells of ca. 8 {\AA} \cite{Gnas14,Koe15}. Such thicker sub-oxide shells will attenuate the ICT and thus $\Delta E$ due to the sub-oxide not having the full $\Delta E $ impact onto dns-Si and working as an electrostatic buffer between the Si-NC and SiO$_2$ at the same time. Since our NWell samples have plane interfaces and were not processed by excess Si segregation from a sub-oxide \cite{Heit05}, the interface thickness is on the order of 5 {\AA} as occurring in conventional c-Si oxidation \cite{Afa01}. Hence, it appears that higher energy offsets $\Delta E_{\rm{C}}$ and consequently $\Delta E_{\rm{V}}$ exist for NWells as compared to Si-NCs formed by segregation from Si-rich dielectrics \cite{Heit05,Hill14}.

To sum up, $E_{\rm{V}}=E_{\rm{vac}}-5.17\ \mbox{eV}=E_{\rm{V}}(\mbox{bulk-Si})$ for NWells with $d_{\rm{NWell}}=40$ to 50 {\AA} in both, SiO$_2$ and Si$_3$N$_4$. We take this value as the common baseline from which to count $\Delta E_{\rm{V}}$ with decreasing $d_{\rm{NWell}}$ as per dielectric. For $d_{\rm{NWell}}=30$ to 40 {\AA}, a transition range exists with increasing $\Delta E_{\rm{V}}$ between NWells in SiO$_2$ vs. Si$_3$N$_4$. At $d_{\rm{NWell}}=25\pm 1$ {\AA}, the respective $E_{\rm{V}}$ are fully developed with $\Delta E_{\rm{V}}\approx 0.74$ eV, still increasing to $\approx0.9$ eV for $d_{\rm{NWell}}=11$ to 15 {\AA}. Thus, a strong n/p-type carrier preference exists for dns-Si in SiO$_2$/Si$_3$N$_4$ which can induce a p/n junction for $d_{\rm{NWell}}\leq 28$ {\AA}, \emph{cf.} Fig. \ref{fig05}g-i. Published characterisation data on Si-NCs prepared by segregation anneal from Si-rich SiO$_2$ \cite{Zim06} show \emph{negative} $\Delta E_{\rm{C}}$, resulting in  $E_{\rm{C}}\approx E_C(\mbox{bulk-Si})-0.2$ eV. As electronic quality and abruptness of the interface of Si-NWells is notably superior to segregated Si-NCs having warped/multi-faceted interfaces, the negative $\Delta E_{\rm{C}}$ of NWells should be more pronounced in SXES measurements. 

\section{\label{Mechnsm}Proposed Mechanism of Energy Offset}
With the quantitative theoretical and experimental data presented above, we propose a qualitative model for the energy offsets $\Delta E_{\rm{V}}$ and $\Delta E_{\rm{C}}$ induced by embedding in SiO$_2$ vs. Si$_3$N$_4$. From there, we determine the impact length of $\Delta E_{\rm{V}}$ and $\Delta E_{\rm{C}}$ in dns-Si systems. 

\subsection{\label{Mech-basicPrinc}Basic Principle}
A brief look at some element-specific quantum-chemical parameters of O, N and Si as listed in Table \ref{tab_2} is very helpful to explain $\Delta E$.
\squeezetable
\begin{table}
\caption{\label{tab_2}Fundamental properties of N, O and Si: Ionization energy ($E_{\mathrm{ion}}$), electron affinity ($X$), electronegativity (EN), ensuing ionicity of bond (IOB) to Si and experimental values of characteristic bond lengths \cite{HolWi95}.}
 \renewcommand{\baselinestretch}{1.2}\small\normalsize
 \begin{tabular}{c|cccccl}
  \hline
  $\in$ &\ $E_{\mathrm{ion}}$\footnote{refers to first valence electron}\ &\ $X$\ &\ EN\footnote{values after Allred \& Rochow}\ &\ IOB &\multicolumn{2}{c}{\ $d_{\mathrm{bond}}$ to Si}\\
  &\ [eV]\ &\ [eV]\ &&\ [\%]\ &\multicolumn{2}{c}{\ [{\AA}]\ }\\
  \hline
  N & 14.53 & $\mathbf{+}$0.07 & 3.07 & 36 & 0.1743 & (Si$_3$N$_4$)\\
  O & 13.36 & $\mathbf{-}$1.46 & 3.50 & 54 & 0.1626 & (SiO$_2$)\\
  Si & 8.15 & $\mathbf{-}$2.08 & 1.74 & 0 & 0.2387 & (bulk Si)\footnote{ with unit cell length of 0.5431 nm \cite{CoData14}}\\
  \hline
 \end{tabular}\\
\end{table}

\begin{figure}[h!]
\includegraphics[width=8.6cm,keepaspectratio]{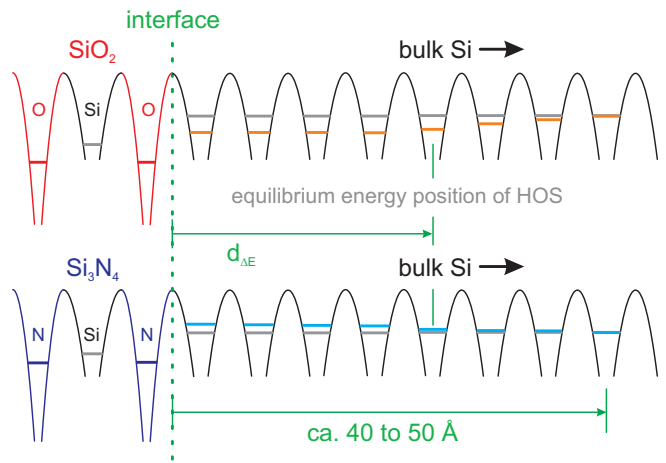}
\caption{\label{fig08} Principle of how the energy offset in dns-Si is induced on the quantum scale. Top graph shows atomic potential at SiO$_2$/Si interface, bottom graph shows the situation at Si$_3$N$_4$/Si interface. Gray valence levels refer to equilibrium positions in bulk-Si. The impact length of the energy offset $d_{\Delta E}$ defines the distance from the interface where $\Delta E$ leaves the saturation regime as obtained from UPS measurements. For further explanations see text.} 
\end{figure}

For O, the explanation of $\Delta E$ is fairly straightforward. With a high negative $X$ and an ionicity of bond (IOB) to Si of ca. 53 \%, O localizes electronic charge from dns-Si, thereby increasing the binding energy of such electrons which present the valence states of dns-Si with the lowest binding energy. As a consequence, $E_{\rm{V}}$(dns-Si) is shifted further below $E_{\rm{vac}}$. Since Si atoms have a reasonably high $E_{\rm{ion}}$ which counteracts electron removal, the ICT feeding into electron localization at O decreases from its saturation value (maximum positive ionization) after a few MLs of the dns-Si system. Eventually, the impact of the ICT due to O at the dns-Si interface vanishes, whereby $E_{\rm{V}}\mbox{(dns-Si)}\rightarrow E_{\rm{V}}\mbox{(bulk-Si)}$. The top graph of Fig. \ref{fig08} illustrates this behavior.

The situation with N requires some broader considerations of its properties. Out of all chemical elements which are known to expose anionic properties (main groups IV to VII of the periodic table), N is the only element which possesses a \emph{positive} electron affinity $X$, meaning that turning N into a negative ion (a Lewis acid \cite{HolWi95}) requires a significant activation energy. Therefore, N is very reluctant to accept negative charge as from dns-Si, which becomes even more prominent when compared to O. This fact is well known from some basic chemical reactions such as creating ammonia (NH$_3$) from the elements,
\begin{equation}
\rm{N_2}\,+\,3\rm{H_2}\ \stackrel{650\,^{\circ}\rm{C},\, 300\,\rm{bar,\,Fe\  contact}}{\longrightarrow}\ 2\rm{NH_3}\ +92.3\ \rm{kJ/mol}\ , \nonumber
\end{equation}
requiring a catalytic iron contact at 300 bar and 650 $^{\circ}$C to trigger above reaction known as Haber-Bosch synthesis \cite{HolWi95}. In contrast, O reacts violently with H at room temperature at 1 bar (oxyhydrogen reaction) \cite{HolWi95}:
\begin{equation}
\rm{O}_2\,+\,2\rm{H_2}\ \longrightarrow 2\rm{H_2O}\ +572.0\ \rm{kJ/mol}\ .\nonumber
\end{equation}
Considering Si, oxidation of Si surfaces to SiO$_2$ in elemental O$_2$ proceeds readily at room temperature and is promoted by the presence of H$_2$O. In contrast, the nitridation reaction requires Si surfaces to be offered atomic/ionic N in a plasma several 100 $^{\circ}$C above room temperature to achieve the reaction.

This uncommon behavior of N is at the heart of the energy offset: While N delocalizes electrons of a dns-Si system at the top of its valence band, it does not localize such electrons, but rather deflects them back into the dns-Si and adjacent Si$_3$N$_4$. This delocalization results in a lower binding energy of these valence electrons. From Fig. \ref{fig07} we also see that the delocalizing impact of N onto $E_{\rm{V}}(\mbox{dns-Si})$ does not show a strict saturation for $d_{\rm{NWell}}\leq 25$ {\AA} as is the case for SiO$_2$-embedding. This behavior has to be seen in the context of electron delocalization which is more sensitive to environmental changes such as an increase of the number of NWell Si atoms per N atom in Si$_3$N$_4$. We can therefore state that $\Delta E$ generated by Si$_3$N$_4$-embedding is not as robust as $\Delta E$ due to SiO$_2$. 

The partial deflection of delocalized electrons into the first couple of ML of Si$_3$N$_4$ explains the stronger shift of $E_{\rm{HOMO}}$ and $E_{\rm{LUMO}}$ towards $E_{\rm{vac}}$ for $\langle111\rangle$-octahedral Si-NCs in 1.5 ML vs. 1 ML Si$_3$N$_4$, \emph{cf.} Fig. \ref{fig04}. The same consequence occurs when going from $\langle111\rangle$-octahedral to $\langle001\rangle$-cubic Si-NCs, where the ratio $N_{\rm{IF}}/N_{\rm{NC}}$ increases significantly. To add, such behavior can clearly be seen for the frontier OMOs at approximants featuring a Si-NC in Si$_3$N$_4$ plus a Si-NC in SiO$_2$, see Sec. \ref{DFT-Res-2NCs} and Fig. \ref{fig05}b, e and h. The resulting shift of $E_{\rm{V}}\mbox{(dns-Si)}$ towards $E_{\rm{vac}}$ within the first few MLs of dns-Si is shown in the bottom graph of Fig. \ref{fig08}.

Fig. \ref{fig07} shows that $|\Delta E_{\rm{V}}\rm{(SiO_2)}|\approx 5\pm 2\times$$|\Delta E_{\rm{V}}\rm{(Si_3N_4)}|$. This finding correlates qualitatively with the IOB of N vs. O to Si. Other factors as anion-to-Si ratio, positive $X$(N) vs. negative $X$(O), bond length and packing fraction in the dielectric have an influence on this ratio as well.

\begin{figure}[h!]
\includegraphics[width=8.0cm,keepaspectratio]{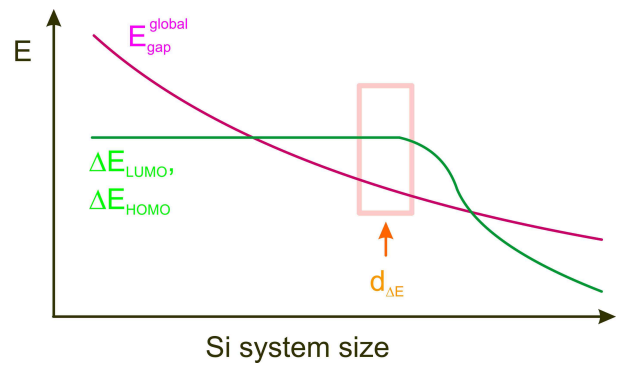}
\caption{\label{fig09}Qualitative scheme to explain the dependence of energy offset and global energy gap as function of dns-Si system size. The technological optimum for maximum n/p-type carrier preference is shown by the semi-transparent pink frame, defining $d_{\Delta E}$ as per explanation in text.} 
\end{figure}

Since we intent to use the $\Delta E$ for undoped p/n junctions, we consider it useful to define its impact length $d_{\Delta E}$ as the distance from the dns-Si interface to the point where $E_{\rm{V}}$ comes out of saturation. This definition of $d_{\Delta E}$ provides the optimum impact for undoped p/n-junctions since the global energy gap of the dns-Si system decreases with its increasing size, see Fig. \ref{fig09}.

\subsection{\label{Mech-lengthDeltaE}Calculation of Impact Length of $\Delta E$ in dns-Si}
\begin{figure}[h!]
\includegraphics[width=8.6cm,keepaspectratio]{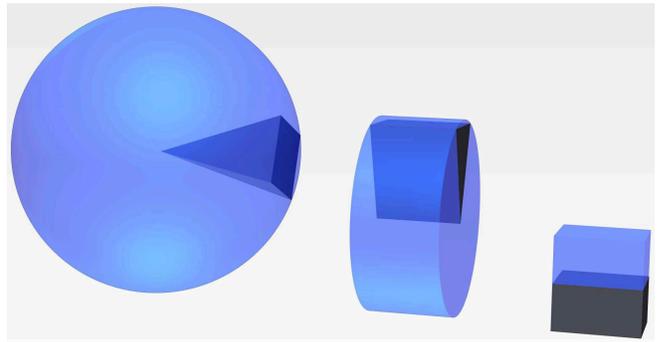}
\caption{\label{fig10} Illustration of volume elements $dV$ to describe the volume of NCs (left), NWires (center), and NWells (right). For NWires and NCs having warped surfaces, the Riemann sum over all $dV(i)$ converges against the volume integral of the dns-Si system for $i\,\rightarrow\infty$, resulting in the base areas of $dV$ to converge to zero. Scaling of system sizes matches $dV$ shape with constant volume and is thus 3:2:1 from left to right.} 
\end{figure}

From the measurements of $E_{\rm{V}}(d_{\rm{NWell}})$ by synchrotron UPS, we derived $d_{\rm{NWell}}\approx 28$ {\AA} as the saturation limit beyond which $\Delta E $ decreases significantly. NWells have two adjacent interfaces which immediatley yields $d_{\Delta E}(\mbox{bulk-Si})=\!\,^1\!/_2\times d_{\rm{NWell}}\approx 14$ {\AA}. This value presents the limit for the semi-infinite case, \emph{id est} an interface to bulk-Si. The volume element defining $d_{\Delta E}(\mbox{bulk-Si})$ is a cubicle of height $d_{\Delta E}(\mbox{bulk-Si})$, its width and length is defined by the smallest periodic unit in the respective direction. For NWells, we thus get $d_{\Delta E}(\mbox{Si-NWell})=2\times d_{\Delta E}(\mbox{bulk-Si})$, arriving at the value obtained by synchrotron UPS, see right graph in Fig \ref{fig10}.

As an approximation, NWires and NCs have warped interfaces which do not allow for a straightforward derivation of $dV$ as is the case for NWells and bulk-Si.
Therefore, we derive the gauge for these dns-Si systems by simple geometric arguments of volume integration via Riemann's sums \cite{OxMathUsrGde04}. For NWires, we cut out a slab which presents the smallest periodic unit along the NWire axis. This slab is then filled with volume elements $dV(i)$, presenting the Riemann sum $\sum_i\,dV$ of the slab volume. For NWires, these $dV(i)$ are wedges, see center graph in Fig. \ref{fig10}. With $i\rightarrow\infty$, we achieve the transition from the Riemann sum to the volume integral, resulting in an infinite number of wedges with vanishing base area. With equal base areas, the volume of a wedge equals the volume of a cubicle if the height of the wedge is twice the height of the cubicle. Consequently, the impact length of $\Delta E$ doubles when going from NWells to NWires, resulting in $d_{\Delta E}(\mbox{Si-NWire})=2\times d_{\Delta E}(\mbox{Si-NWell})$. For NCs, it is straightforward that we get pyramids as volume elements $dV$, see left graph in Fig. \ref{fig10}. With the same base area, a pyramid has the same volume as a cubicle if its height is three times the height of the cubicle. Accordingly, we obtain $d_{\Delta E}(\mbox{Si-NC})=2\times d_{\Delta E}(\mbox{Si-NWire})$. Summarizing all $dV$ definitions, we get $d_{\Delta E}$ scaling with bulk-Si\,:\,NWell\,:\,NWire\,:\,NC like 1\,:\,2\,:\,4\,:\,6. With $d_{\Delta E}(\mbox{Si-NWell})\approx 28$ {\AA}, we get $d_{\Delta E}(\mbox{Si-NWire})\approx 56$ {\AA} and $d_{\Delta E}(\mbox{Si-NC})\approx 84$ {\AA}.

In real systems, the rise in $d_{\Delta E}$ with decreasing dimensionality of the dns-Si system should be sublinear, originating from a higher defect density due to warped interfaces which has to be controlled by $\Delta E$ as well. We can account for this effect by exploiting the fact that fundamental gaps $E_{\rm{gap}}$ of Si-NCs are significantly bigger for Si$_3$N$_4$-embedding vs. SiO$_2$-embedding if such NCs are small enough \cite{Koe08,Koe14}. Experimental data showed $E_{\rm{gap}}$ of both embeddings to converge at $d_{\rm{NC}}\approx 56$ {\AA} \cite{Ehrh13}. Assuming a reduction factor per reduced dimension $DF$ and using $d_{\Delta E}(\mbox{Si-NC}) = DF^2\times 4\times d_{\Delta E}(\mbox{Si-NWell})$, we arrive at $DF\approx\!\,^1/_{\!\sqrt{2}}$, yielding $d_{\Delta E}(\mbox{Si-NWire})\approx 40$ {\AA} and $d_{\Delta E}(\mbox{Si-NC})\approx 56$ {\AA} below which $\Delta E$ is saturated.


%

\section{\label{WrapUp} Conclusions}
We presented a detailed quantitative study of considerable energy offsets of frontier electronic states as a new fundamental effect arising from deep nanoscale (dns-) Si volumes when embedded/coated in SiO$_2$ vs. Si$_3$N$_4$. 

Using DFT, we investigated single Si-NCs as a function of interface orientation and thickness of embedding SiO$_2$ vs. Si$_3$N$_4$ up to $d_{\rm{NC}}=26$ {\AA}. We found $E_{\rm{HOMO}}$ and $E_{\rm{LUMO}}$ to have little dependence on NC size in the range considered, showing a strong offset due to 1.5 ML SiO$_2$- vs. Si$_3$N$_4$-embedding of $\Delta E_{\rm{HOMO}}\approx 1.75$ eV and $\Delta E_{\rm{LUMO}}\approx 1.93$ eV. These offsets push $E_{\rm{HOMO}}$ and $E_{\rm{LUMO}}$ of SiO$_2$-embedded Si-NCs further below $E_{\rm{vac}}$, while shifting $E_{\rm{HOMO}}$ and $E_{\rm{LUMO}}$ of Si$_3$N$_4$-embedded Si-NCs towards $E_{\rm{vac}}$. Extending the embedding in SiO$_2$/Si$_3$N$_4$ beyond 1 ML and in particular increasing the ratio of interface bonds per NC atom $N_{\rm{IF}}/N_{\rm{NC}}$ by going from $\langle111\rangle$-octahedral to $\langle001\rangle$-cubic NCs increased $\Delta E$ further. 
As ultimate theoretical test, we computed approximants featuring two Si-NCs of up to 18.5 {\AA} size, one each embedded in SiO$_2$ and Si$_3$N$_4$, respectively, confirming $\Delta E$ within one system. 

We characterized 14 samples, each featuring one Si-NWell of $d_{\rm{NWell}}=11$ to 50 {\AA} embedded in SiO$_2$ or in Si$_3$N$_4$, by long-term synchrotron UPS to measure the energy of the top valence band states $E_{\rm{V}}$. For $d_{\rm{NWell}}\alt 28$ {\AA}, we obtained $\Delta E_{\rm{V}}\approx 0.9$ to 0.74 eV for SiO$_2$- vs. Si$_3$N$_4$-embedding. After a transition range at $d_{\rm{NWell}}\approx 30$ to 40 {\AA}, Si-NWells in both embeddings converge to $E_{\rm{V}}=E_{\rm{V}}(\mbox{bulk-Si})$. Zimina \emph{et al.} \cite{Zim06} found the energy of the lowest unoccupied states at $E_{\rm{C}}\approx E_{\rm{C}}\mbox{(bulk-Si)}-0.2$ eV for SiO$_2$-embedded Si-NCs, whereby attractive interactions like exciton binding energy and screening were already included in this value. This finding is consistent with our results on $E_{\rm{V}}$ of thin Si-NWells in SiO$_2$.

Investigating the quantum chemistry of O and N with respect to Si, we proposed a model for $\Delta E$. The uncommon situation of N as the only anionic element with an electron affinity $X>0$ eV and a yet fairly high electronegativity EN inducing a strong polar bond to Si is one cornerstone of the newly discovered effect. While N delocalizes valence electrons from dns-Si, it does \emph{not} localize such electrons but deflects these back into dns-Si or surrounding Si$_3$N$_4$ as also evident from frontier OMO density plots in DFT calculations spreading well into Si$_3$N$_4$. As a consequence, such delocalized NWell valence band states have a lower binding energy which is equivalent to an  upshift of  $E_{\rm{V}}(\rm{NWell})$ towards $E_{\rm{vac}}$. As expected, the strong conventional anionic nature of O results in strong localization of valence electrons from dns-Si at O in SiO$_2$ and an associated downshift of $E_{\rm{V}}(\rm{NWell})$ below $E_{\rm{vac}}$.

With a straightforward geometrical Riemann model, we calculated the impact length $l_{\Delta E}$ up to which $\Delta E$ governs the electronic structure of dns-Si to scale 1\,:\,2\,:\,4\,:\,6 for bulk-Si\,:\,NWell\,:\,NWire\,:\,NC. With $l_{\Delta E}\mbox{(NWell)}\approx 28$ {\AA}, we thus arrived at $l_{\Delta E}\mbox{(NWire)}\approx 56$ {\AA} and $l_{\Delta E}\mbox{(NC)}\approx 84$ {\AA}. The latter two dns-Si structures have warped interfaces which are more prone to defects; experimental data from the literature on the difference on the fundamental gap of Si-NCs embedded in SiO$_2$ vs. Si$_3$N$_4$ yielded $l_{\Delta E}\mbox{(NC)}\approx 56$ {\AA}. We introduced a dimensionality factor $DF$ per reduced dimension (NWells $\stackrel{\scriptscriptstyle \times DF}{\rightarrow}$ NWires $\stackrel{\scriptscriptstyle \times DF}{\rightarrow}$ NCs) of $DF\approx\!\,^1/_{\!\sqrt{2}}$, accounting for an increased number of defects at warped interfaces, yielding $l_{\Delta E}\mbox{(NWire)}\approx 40$ {\AA} and matching $l_{\Delta E}\mbox{(NC)}\approx 56$ {\AA}.

The  $\Delta E$ in dns-Si by coating with or embedding in SiO$_2$ (Si$_3$N$_4$)  induces a strong preference for electrons (holes) and thus to n(p)-type Si. Application to dns-Si would eliminate all issues of impurity doping -- out-diffusion, clustering, self-purification, ionization at room and cryogenic temperatures, inelastic carrier scattering -- and extend device miniaturization potentially down to the minimum size of Si crystallites of ca. 1.5 nm \cite{Schu94} along with much reduced bias voltages and heat loss.

\begin{acknowledgments}
D.~K. wishes to thank J. Rudd for compute cluster administration and acknowledges use of the Abacus compute cluster, IMDC, UNSW, funding by the 2015 UNSW Blue Sky Research Grant and by the 2018 Theodore-von-K{\`a}rm{\`a}n Fellowship of RWTH Aachen University, Germany. D.~K. and D.~H. acknowledge funding by 2012, 2014 and 2016 DAAD-Go8 joint research cooperation schemes. D.~H. acknowledges the Alexander von Humboldt Foundation for a Feodor Lynen Fellowship and the German Research Foundation (DFG) for funding (HI 1779/3-1). The authors thank L. Sancin at Elettra Synchrotron for technical support. N.~W., B.~B. and J.~K.  acknowledge support by the Impulse and Networking Fund of the Helmholtz Association.\\
\end{acknowledgments} 

\begin{appendix}
\section{\label{Apx-UPS-Data-Eval}Evaluation of UPS Data}
In our recent publication, we discussed the calibration of UPS-Scans and data extraction by local fitting of amorphous backgrounds using Si$_3$N$_4$-embedded Si-NWells of 17 and 27 {\AA} thickness in much detail \cite{Koe18aa}.

\begin{figure}[h!]
\includegraphics[width=4.299cm,keepaspectratio]{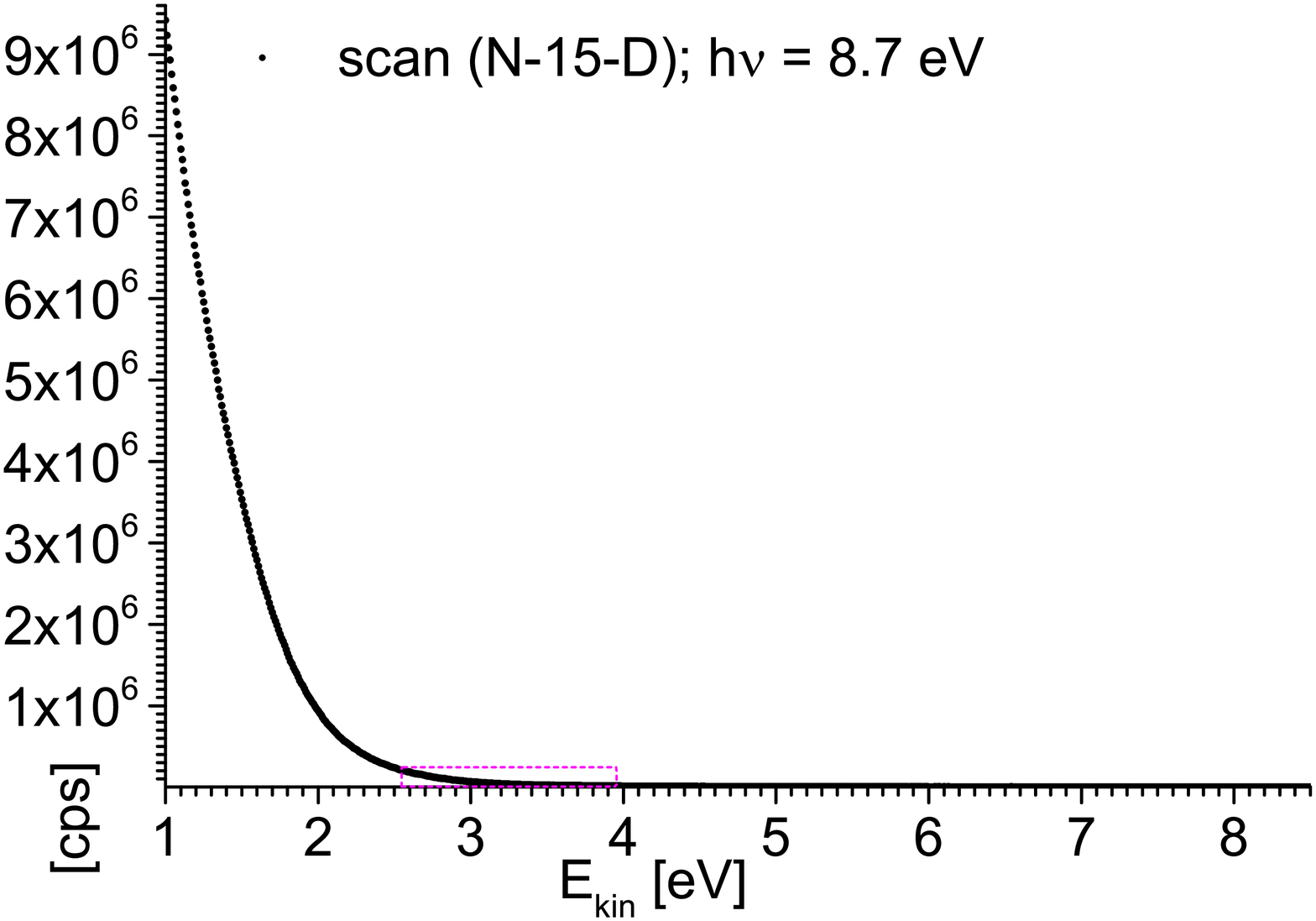}\hfill
\includegraphics[width=4.299cm,keepaspectratio]{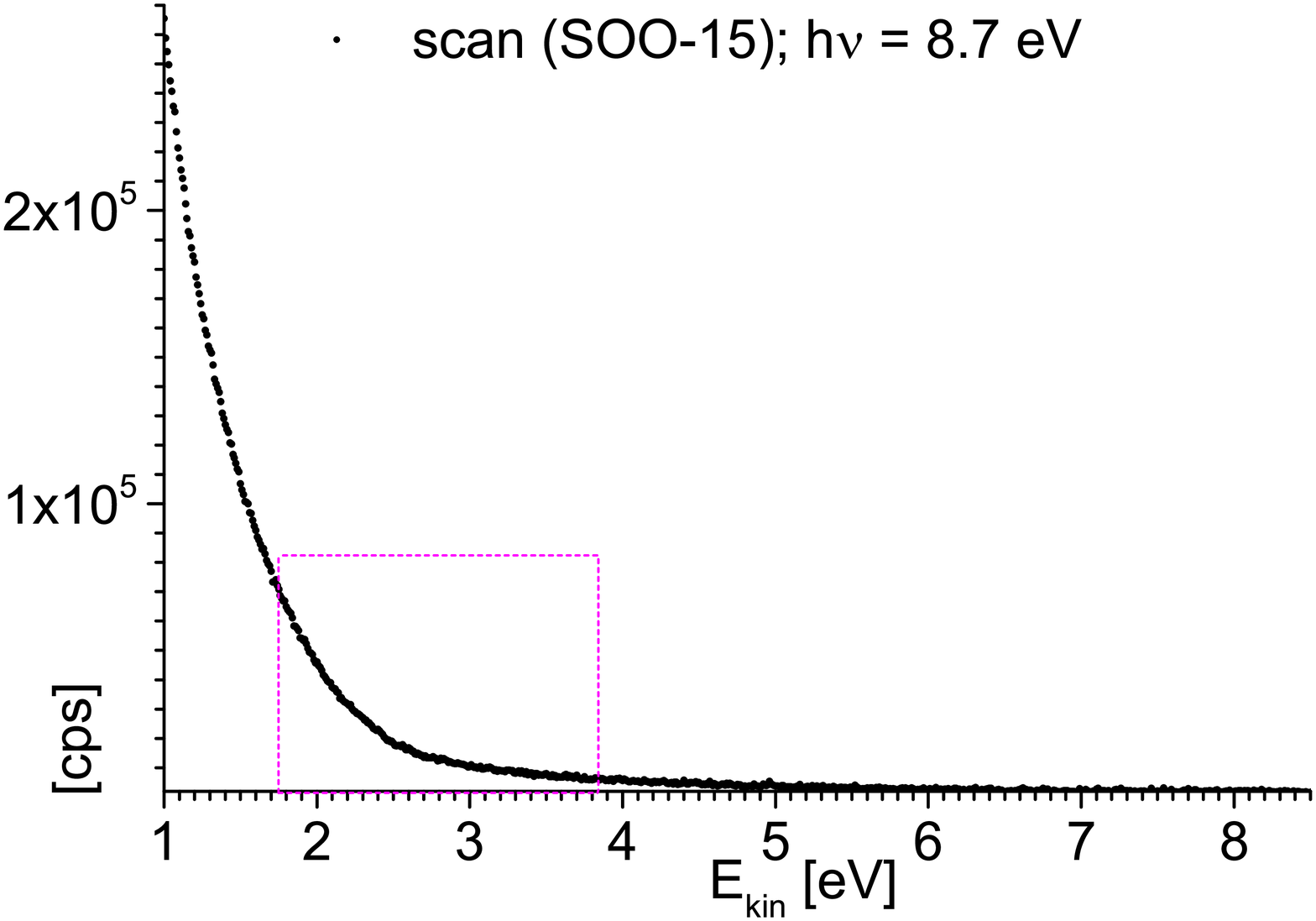}
\begin{center}
\vspace*{-0.25cm}
{\bf(a)\hspace{4.3cm}(b)}\\
\vspace*{-0.5cm}
\end{center}
\caption{\label{fig11} Complete UPS scan of sample N-15-D (a) and SOO-15 (b), shown with respective region of interest marked as magenta frame. Photoexcitation energy was $h\nu=8.7$ eV.} 
\end{figure}

Here, we just focus on one SiO$_2$- and one Si$_3$N$_4$-embedded sample for determining the valence band edge, its standard deviation over energy and associated local background fitting per scan to explain the principle. We further explain several plausibility tests as an additional means to arrive at the correct data. As examples, we pick UPS scans of sample N-15-D, comprising a 15 {\AA} Si-NWell in Si$_3$N$_4$, and of sample SOO-15, comprising a 12 {\AA} Si-NWell in SiO$_2$. Both scans have the lowest signal-to-noise ratio (SNR) against the background due to secondary electrons and SiO$_2$/Si$_3$N$_4$. This circumstance predestines both samples to demonstrate the UPS data filtering algorithm to obtain the valence band edge of buried Si-NWells. Fig. \ref{fig11} shows the global UPS scans of both samples which were measured during beamtime 1 at the BaDElPh beamline at the Elettra Sincrotrone Trieste, Italy, \emph{cf.} Fig. \ref{fig07}. Common parameters are the UV photon energy of $h\nu=8.7$ eV and the background cps value of 1950 which was derived in the $E_{\rm{kin}}$ interval of $[h\nu;\,h\nu-0.5]$ eV = [8.7;8.2] eV. This background cps value derived from the Si reference serves as calibration standard and thus sets the cps intercept for all UPS scans, see \cite{Koe18aa} for details.

\begin{figure}[h!]
\includegraphics[width=4.299cm,keepaspectratio]{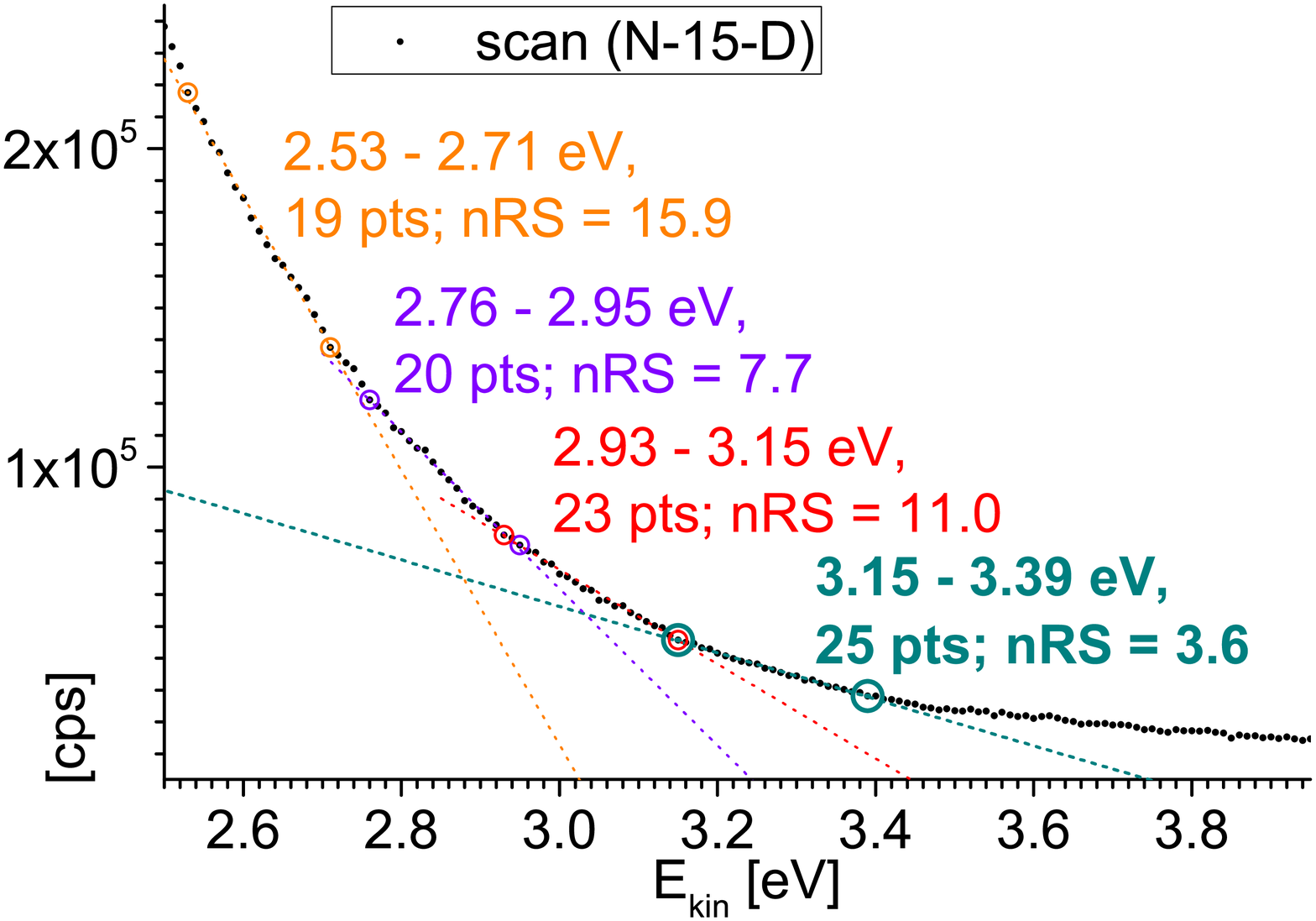}\hfill
\includegraphics[width=4.299cm,keepaspectratio]{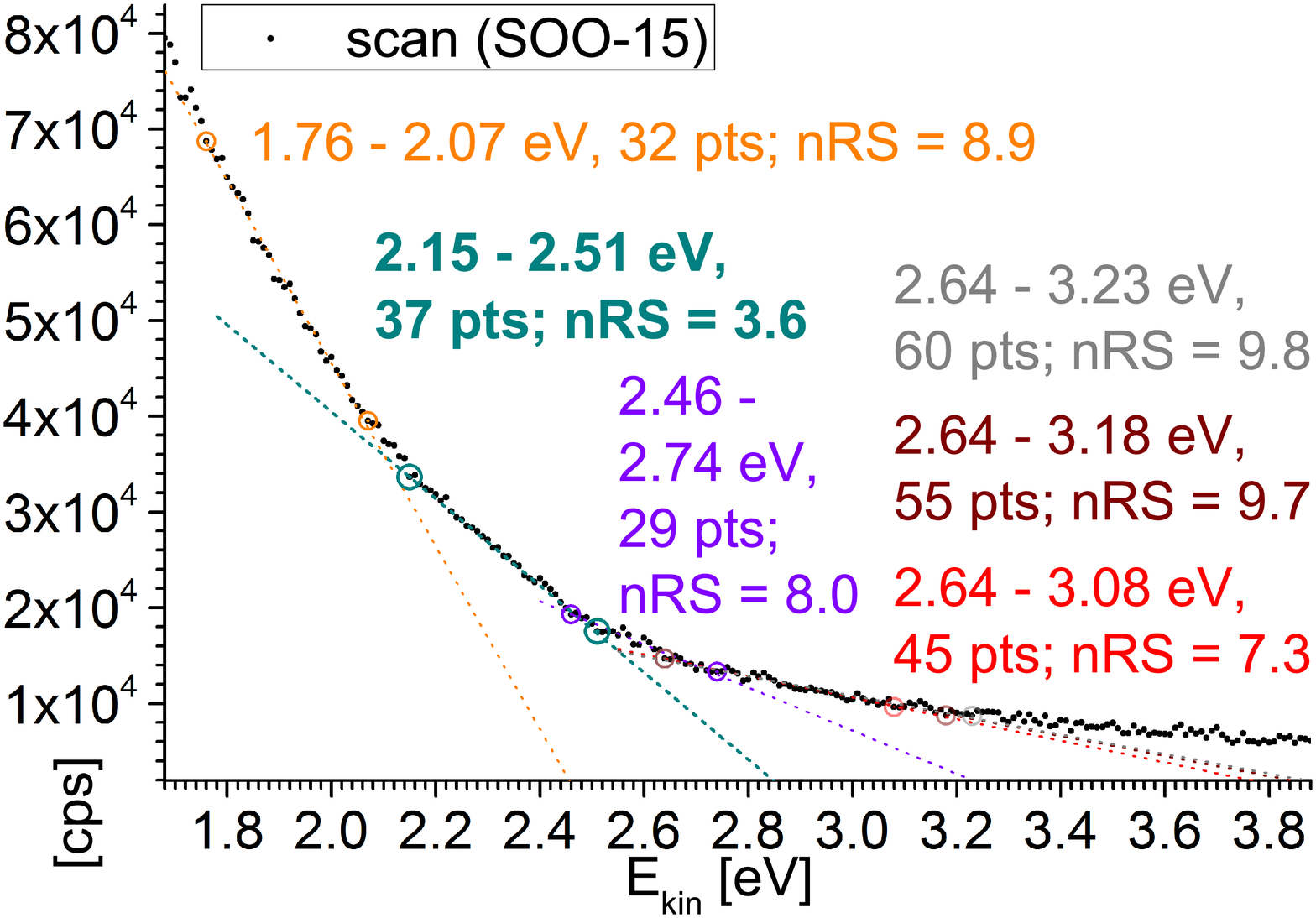}
\begin{center}
\vspace*{-0.25cm}
{\bf(a)\hspace{4.3cm}(b)}\\
\vspace*{-0.5cm}
\end{center}
\caption{\label{fig12} Range of interest of UPS scans of sample N-15-D (a) and SOO-15 (b), shown with all linear region found as per definition in text. Range of $E_{\rm{kin}}$, $N_{\rm{pts}}$ and nRS are shown per linear fit. Bold printed parameters in dark cyan and associated fit are valence band edge of embedded Si-NWell.} 
\end{figure}

Below, we describe the algorithm to determine the valence band edge of a Si-NWell embedded in Si$_3$N$_4$ or SiO$_2$.\\
We begin by checking the UPS scan for regions of reasonably linear values with $\geq 15$ points, equivalent to an $E_{\rm{kin}}$ range of 0.15 eV. The following steps are applied to every linear region found. We assign the number of points forming the linear region to $N_{\rm{pts}}$. We then run a linear fit over the edge region, calculate its sum of residual squares (RS) $\sum[\rm{data}(E_{\rm{kin}})-\rm{fit}(E_{\rm{kin}})]^2$ and its average cps value $\rm{cps}_{\rm{avg}}=1/N_{\rm{pts}}\,\sum \rm{data}(E_{\rm{kin}})$. With these three values, we can calculate the normalized RS (nRS) which takes into account the signal intensity and length of the linear region: $\rm{nRS}=\rm{RS}\,/\,[N_{\rm{pts}}\,\rm{cps}_{\rm{avg}}]$. The linear region with the minimum nRS is the edge of the Si-NWell, subject to passing a number of plausibility tests lined out below. The intercept of the linear fit provides the $E_{\rm{kin}}$ value which yields to the valence band top of the Si-NWell via $E_{\rm{V}}=h\nu - E_{\rm{kin}}(\mbox{intercept})$. Fig. \ref{fig12} shows the range of interest per UPS scan with all linear regions detected.

\begin{figure}[h!]
\includegraphics[width=4.299cm,keepaspectratio]{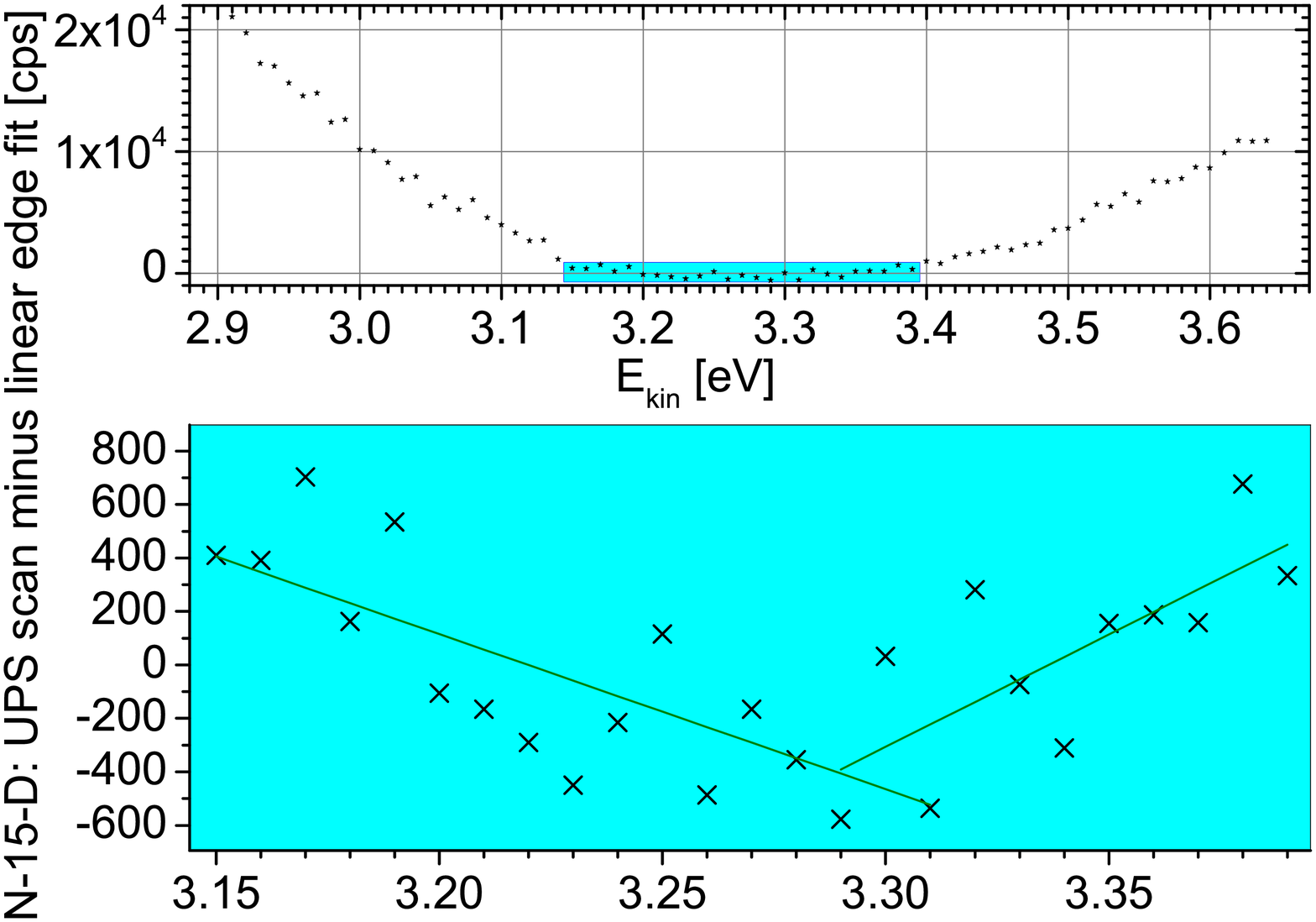}\hfill
\includegraphics[width=4.299cm,keepaspectratio]{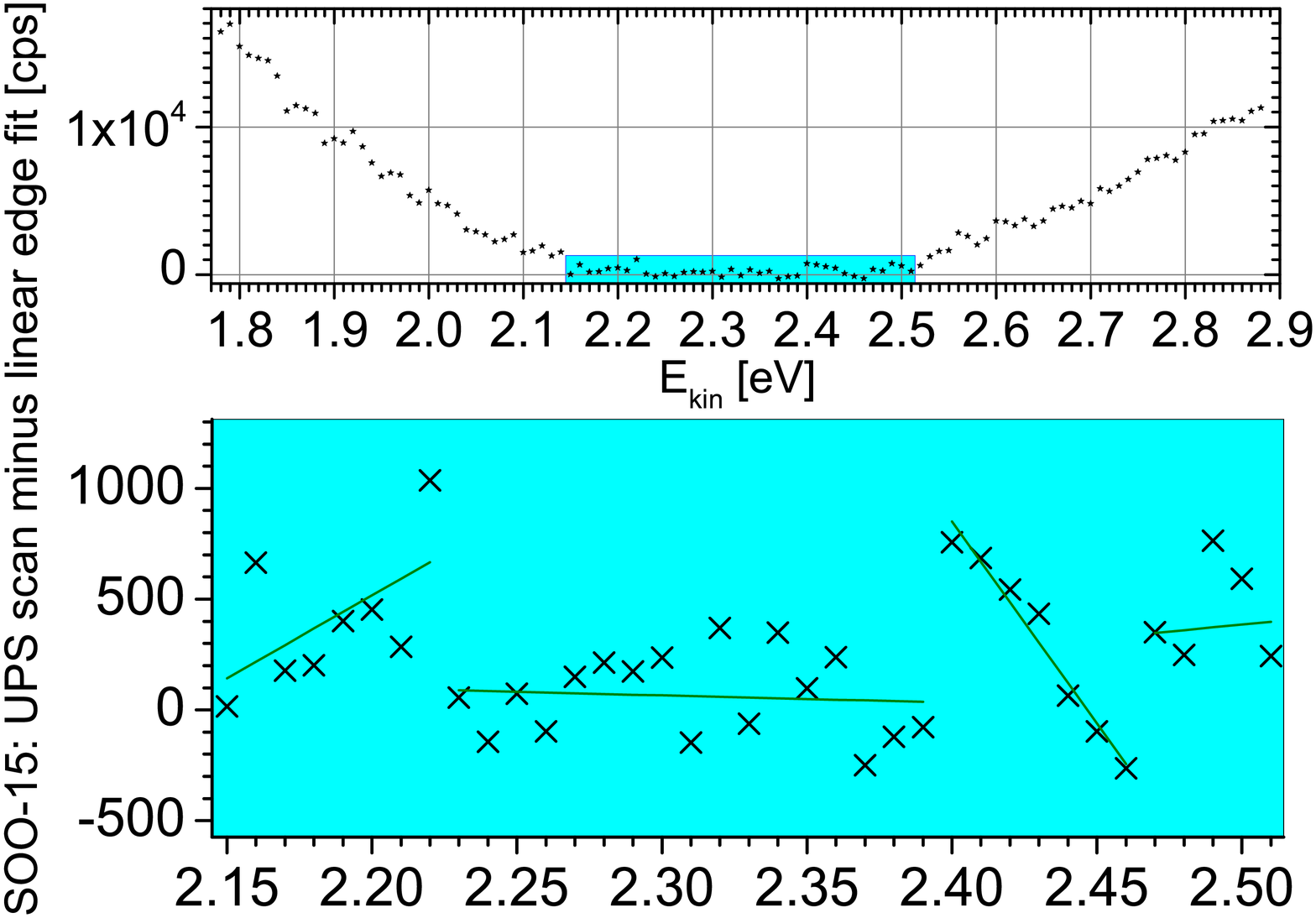}
\begin{center}
\vspace*{-0.25cm}
{\bf(a)\hspace{4.3cm}(b)}\\
\vspace*{-0.5cm}
\end{center}
\caption{\label{fig13} Difference between UPS signal and linear fit to the edge region, shown in $\pm 1$ $E_{\rm{kin}}(\rm{edge})$ interval for sample N-15-D (a) and SOO-15 (b); see text for details. The lower graphs show the edge region, including linear fits to local point groups with associated alignment.  These fits are used for estimating the standard deviation of the valence band edge over $E_{\rm{kin}}$.} 
\end{figure}

\begin{figure}[h!]
\includegraphics[width=4.299cm,keepaspectratio]{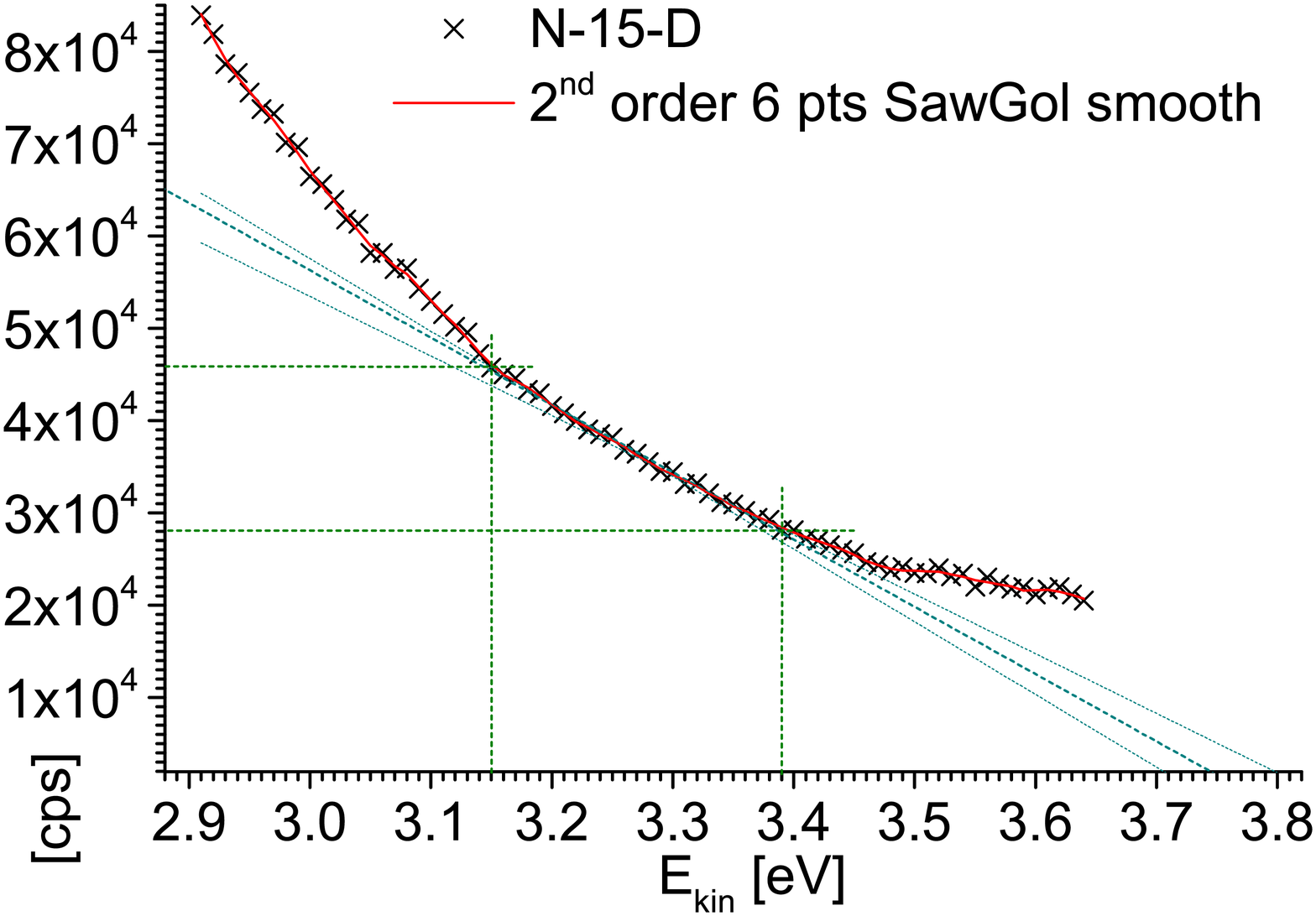}\hfill
\includegraphics[width=4.299cm,keepaspectratio]{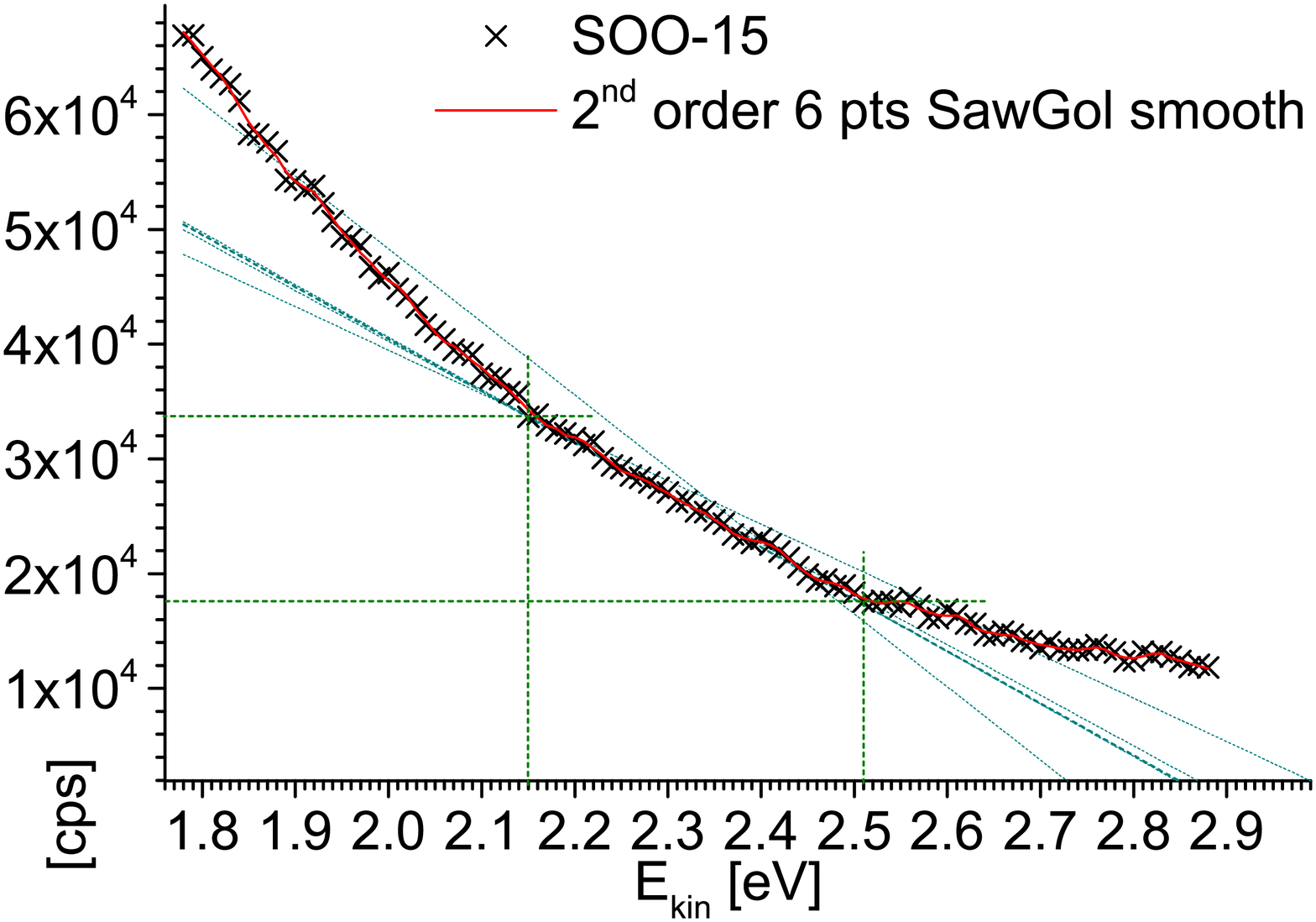}\\
\begin{center}
\vspace*{-0.25cm}
{\bf(a)\hspace{4.3cm}(b)}\\
\vspace*{-0.5cm}
\end{center}
\caption{\label{fig14} UPS scan and its Sawitzky-Golay smoothed version of edge region $\pm 1$ $E_{\rm{kin}}(\rm{edge})$ interval around it, shown for sample N-15-D (a) and SOO-15 (b). Thick dashed lines show valence band edge of Si-NWell, thin dashed lines show local linear fits to point groups showing local alignment within edge region. The latter are used to calculate the standard deviation of the edge over $E_{\rm{kin}}$ and thus $E_{\rm{V}}$. Green dashed lines show cps and $E_{\rm{kin}}$ limits of edge region.} 
\end{figure}

Next, we calculate the standard deviation $\sigma(E_{\rm{kin}})$, using the difference between the UPS signal and its linear fit, \emph{cf.} Fig. \ref{fig13}. Within the edge region, we group points with associated alignment. For these, we calculate local fits and determine their intercepts, \emph{cf.} Fig. \ref{fig14}. 
The resulting $E_{\rm{kin}}$ values are used together with the value of the linear fit to the entire edge region to determine $\sigma(E_{\rm{kin}})=\sqrt{1/n\,\sum_{i=1}^n[\overline{E}_{\rm{kin}}-E_{\rm{kin}}(i)]^2}$ \cite{OxMathUsrGde04} and are shown as error bars of $E_{\rm{ion}}$ in Fig. \ref{fig07}. We note that -- as samples were measured at $T=300$ K -- the minimum error bar for $E_{\rm{kin}},\,E_{\rm{ion}}$ is set to $k_{\rm{B}}T=0.026$ eV.

\begin{figure}[h!]
\includegraphics[width=4.299cm,keepaspectratio]{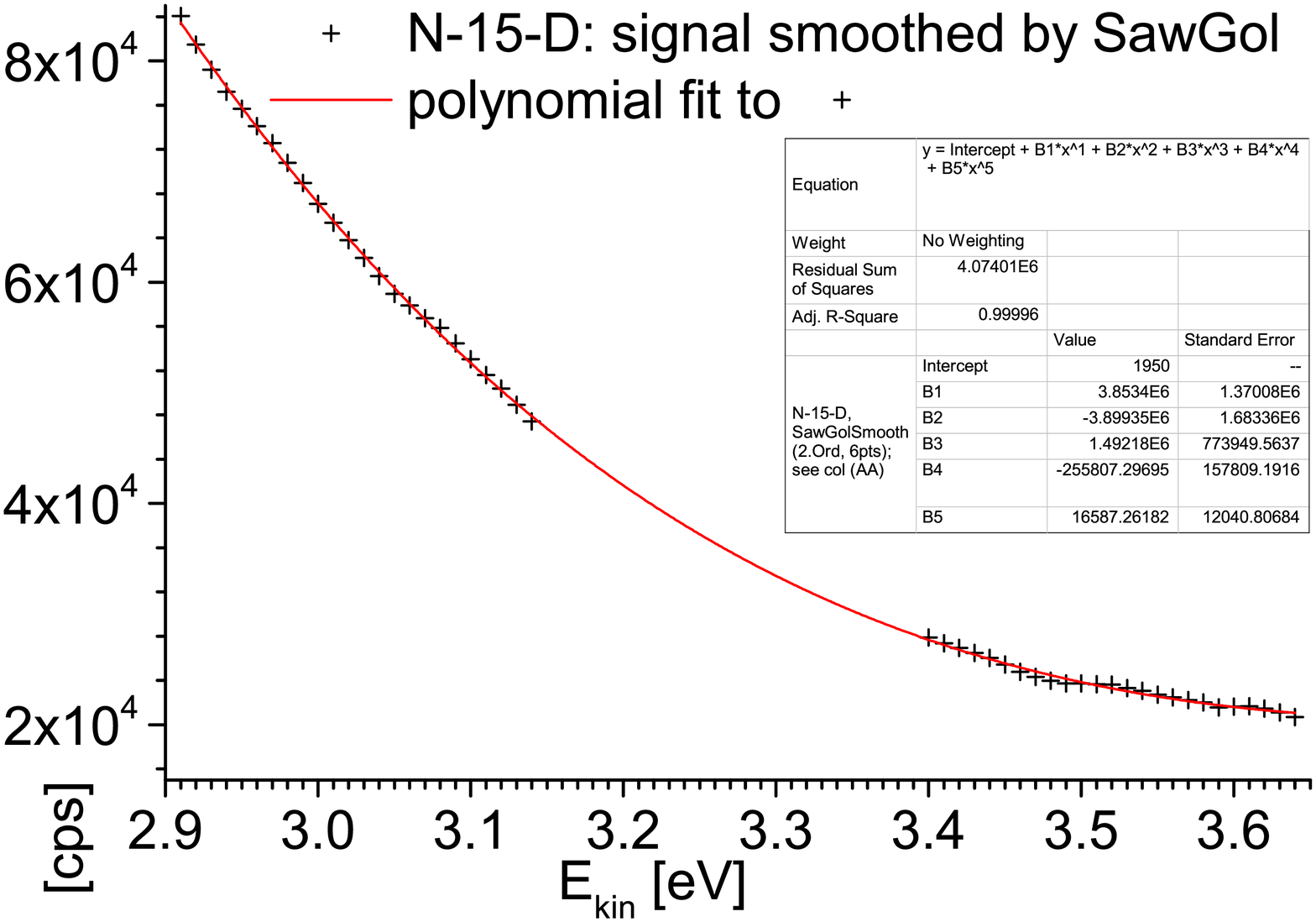}\hfill
\includegraphics[width=4.299cm,keepaspectratio]{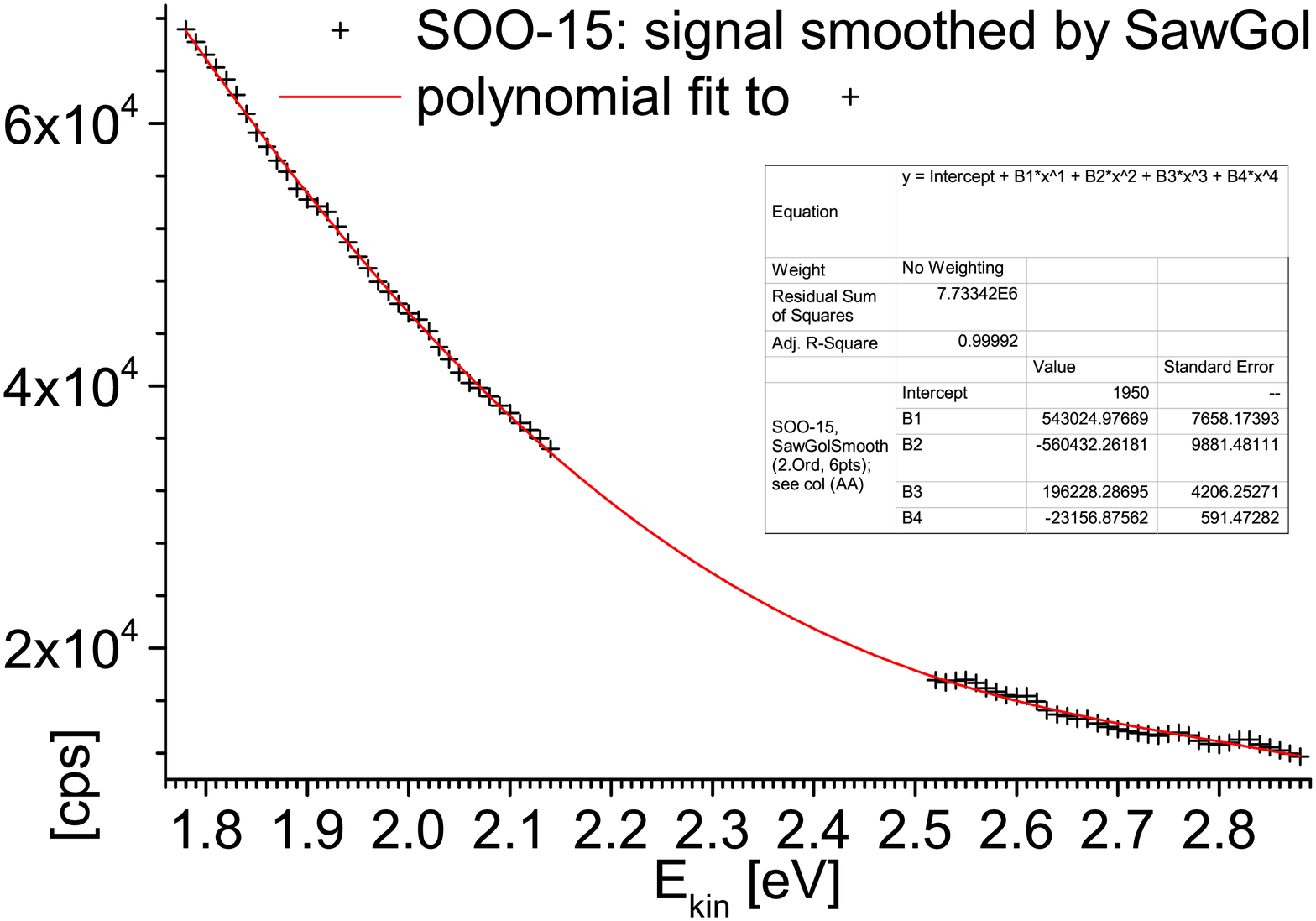}\\
\begin{center}
\vspace*{-0.25cm}
{\bf(a)\hspace{4.3cm}(b)}\\
\vspace*{-0.5cm}
\end{center}
\caption{\label{fig15} Sawitzky-Golay smoothed data of UPS scan in $\pm 1$ $E_{\rm{kin}}(\rm{edge})$ interval, shown for sample N-15-D (a) and SOO-15 (b), see text for details. The red line shows the polynomial fit to the smoothed UPS data which present the local surrounding of the edge region, serving as local background with lower SNR.} 
\end{figure}

Since the algorithm on its own may yield a couple of possible solutions, we carry out additional plausibility tests. One of them is a combination of the smoothed UPS signal, an associated background fit and the interpolation of the resulting differential signal. First, we take the UPS data of the edge region plus one energy range as defined by the edge region above and below the edge (e.g. N-15-D has an edge from $E_{\rm{kin}}=3.15$ to 3.39 eV [25 pts], we now pick UPS data from 2.90 to 3.64 eV [$25+25+25=75$ pts]). To this data set, we apply a Sawitzky-Golay smoothing \cite{Savi64} over 6 points with a $2^{\rm{nd}}$ order polynomial as an effective means to \emph{remove single point runaways without altering the spectral information}, \emph{cf.} Fig. \ref{fig14}. Thereby, we get less noisy UPS data in the range of interest. This smoothed UPS data is used for a background fit, whereby we exclude the actual edge region ($E_{\rm{kin}}=3.15$ to 3.39 eV [25 pts] for sample N-15-D). Since higher order polynomials can be interpreted as the finite Taylor series of an exponential function, $\exp(x)=\sum_i^\infty[x^i/i!]$, we choose polynomials to fit the background and use the signal calibration intercept of 1950 cps as the coefficient for $i=0$, see Fig. \ref{fig15}. While the coefficients of the $x^i$ decrease from $i>1$, qualitatively agreeing with the Taylor series of an exponential function, there is still some degree of freedom to adapt to local deviations. These are evident even after Sawitzky-Golay smoothing, \emph{cf.} Figs. \ref{fig12} and \ref{fig15}. An exponential behavior of the signal background originates from the Urbach tails of the amorphous dielectric \cite{BoerI}, with additional contributions due to inelastic electron scattering.

\begin{figure}[h!]
\includegraphics[width=4.299cm,keepaspectratio]{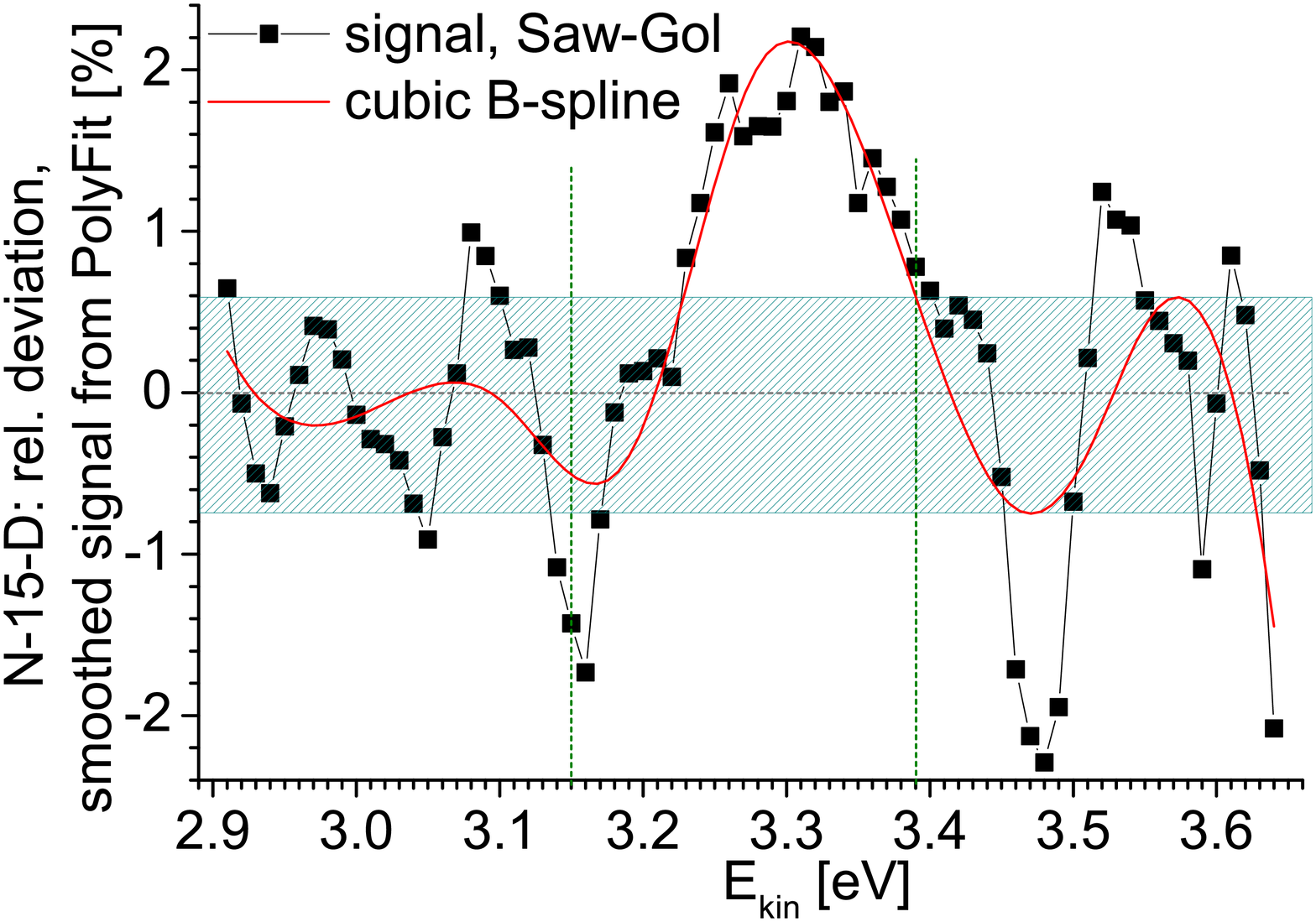}\hfill
\includegraphics[width=4.299cm,keepaspectratio]{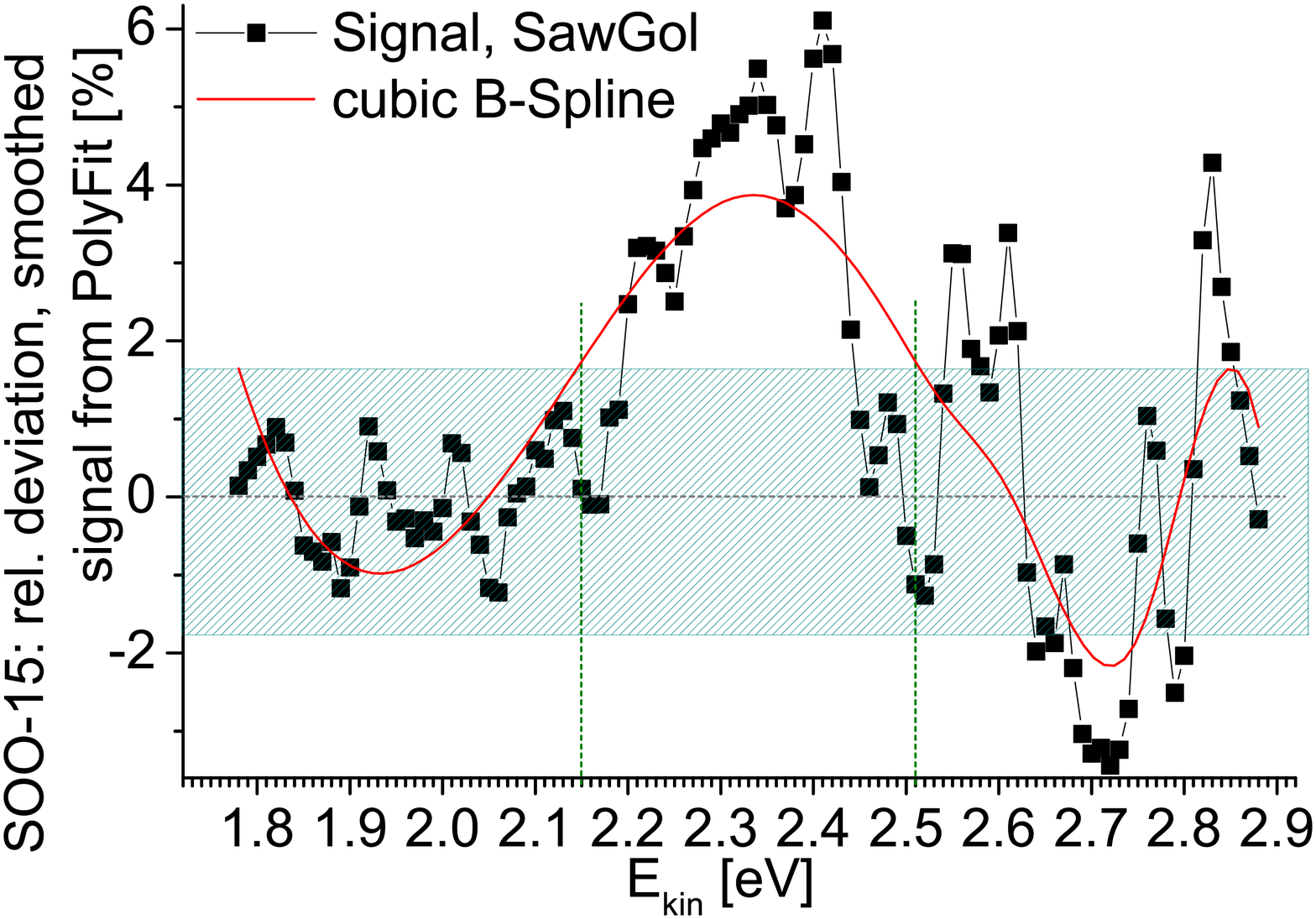}
\begin{center}
\vspace*{-0.25cm}
{\bf(a)\hspace{4.3cm}(b)}\\
\vspace*{-0.5cm}
\end{center}
\caption{\label{fig16} Difference between Sawitzky-Golay smoothed UPS scan and background fit (\emph{cf.} Fig. \ref{fig15}) of edge region $\pm 1$ $E_{\rm{kin}}(\rm{edge})$ interval around it, shown together with a cubic B-Spline interpolation to minimize noise for sample N-15-D (a) and SOO-15 (b). Green dashed lines show $E_{\rm{kin}}$ limits of edge region. Hashed dark cyan area is a guide to noise limit.} 
\end{figure}

We now can calculate the difference between the smoothed UPS signal and the background fit where the edge region shows up above the noise, \emph{cf.} Fig. \ref{fig16}. Due to the finite error in background fitting, there are local deviations of the fit from data points in the local environment of the edge region which cannot be accounted for. As a consequence, the first few points outside the edge region tend to be further away from the background fit which can cause a dumping of the first few data points within the edge region. We also note that the fit does not account for an irregular, rather rapid low amplitude oscillation of the UPS signal. Hence, we have to look for positive signals -- the edge signal occurs on top of the background -- on a prolonged energy scale, ruling out rapid oscillations. To this end, we run a cubic B-Spline interpolation \cite{OxMathUsrGde04} on the differential signal which minimizes the oscillatory behavior, \emph{cf.} Fig. \ref{fig16}.

\begin{figure}[h!]
\includegraphics[width=4.299cm,keepaspectratio]{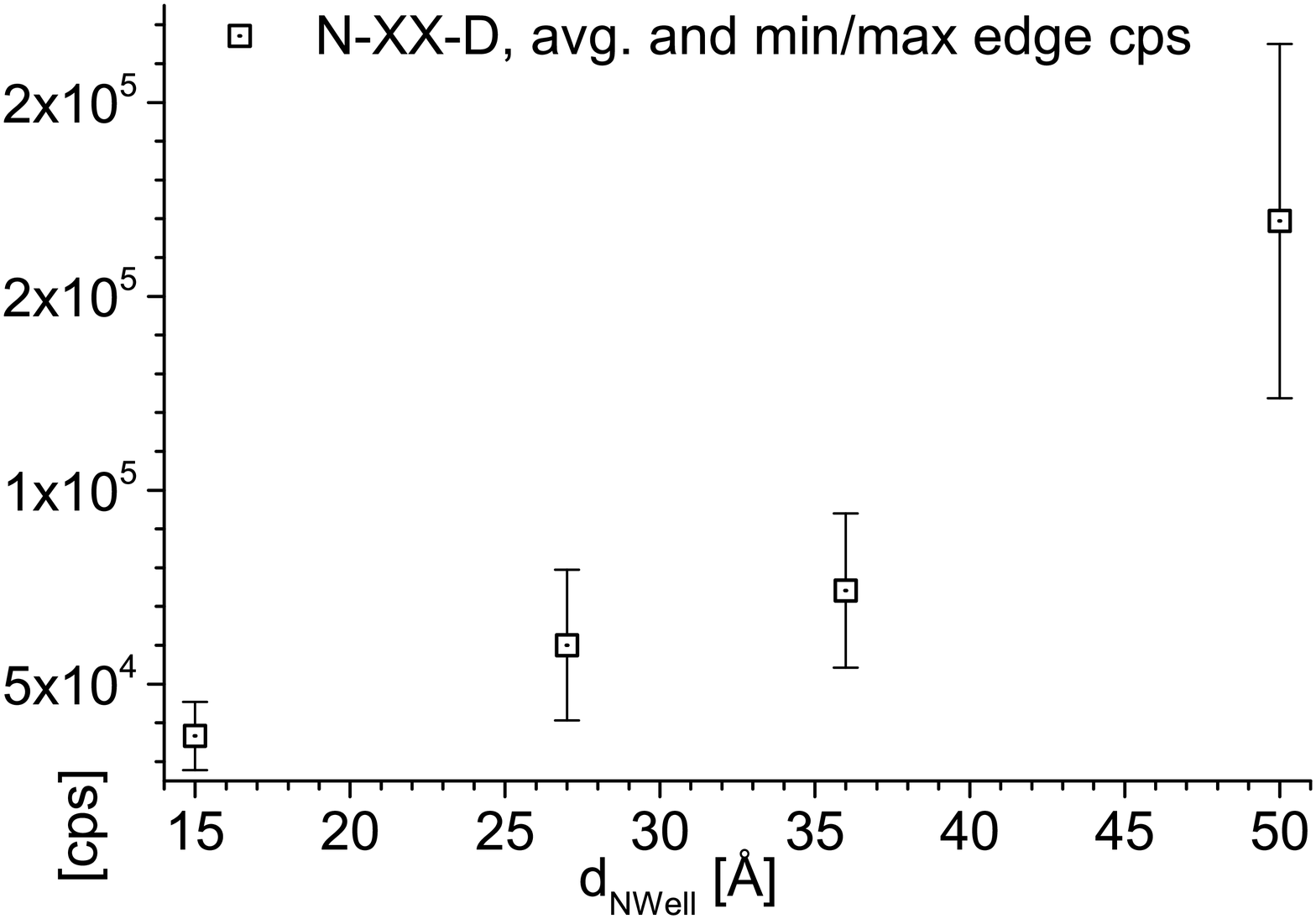}\hfill
\includegraphics[width=4.299cm,keepaspectratio]{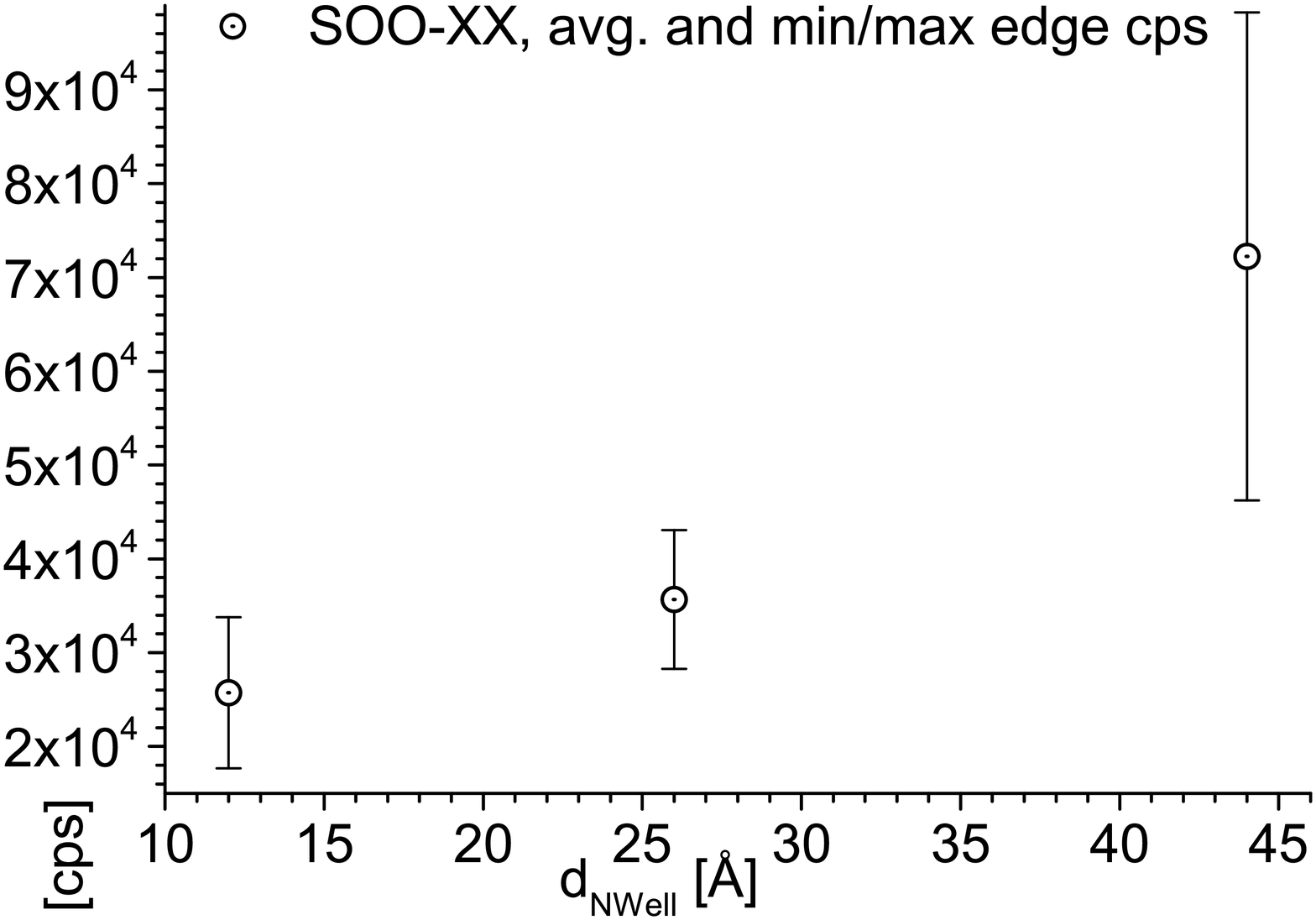}
\begin{center}
\vspace*{-0.25cm}
{\bf(a)\hspace{4.3cm}(b)}\\
\vspace*{-0.5cm}
\end{center}
\caption{\label{fig17} Average value of cps of edge region per sample of  beamtime 1, \emph{cf.} \ref{fig07} for results, for Si-NWells embedded in Si$_3$N$_4$ (a) and in SiO$_2$ (b). Error bars show mini- and maximum cps values per edge region.} 
\end{figure}

Another plausibility test is given by measuring NWell samples of different $d_{\rm{NWell}}$ with a dielectric top layer of possibly constant thickness. Assuming that the dielectric top layer and its interface to the NWell have exactly the same thickness for all samples per dielectric type and all measurement conditions are constant, the signal intensity (cps) of the NWell is related to the layer thickness of the top dielectric layer  $d_{\rm{diel}}^{\rm{top}}$ and $d_{\rm{NWell}}$ via $\rm{cps}(d_{\rm{NWell}})\propto \exp(-d_{\rm{diel}}^{\rm{top}}/\lambda_{\rm{imfp}})\big[1-\exp(-d_{\rm{NWell}}/\lambda_{\rm{imfp}})\big]$, reducing to $\rm{cps}(d_{\rm{NWell}})\propto 1-\exp(-d_{\rm{NWell}}/\lambda_{\rm{imfp}})$ for $d_{\rm{diel}}^{\rm{top}}=\rm{const}\,$. The average inelastic mean free path of excited electrons is presented by $\lambda_{\rm{imfp}}$. For brevity, we introduce one value for Si-NWell and the respective dielectric. It is interesting to note that $\lambda_{\rm{imfp}}$ has been overestimated in UPS for several compounds \cite{Offi08,Iaco19} since its widely used empirical description by Seah and Dench \cite{Seah79} was derived for many chemical elements and compounds for $h\nu \geq 150$ eV but only for three compound materials using $h\nu \leq 40$ eV. In a perfect world, we should see a cps increasing asymptotically  with $d_{\rm{NWell}}$. In reality, minor thickness deviations in particular of the top dielectric layer will soften this asymptotic behavior into a cps monotonically increasing with $d_{\rm{NWell}}$, \emph{cf.} Fig. \ref{fig17}.

As for UPS signal quality, we can state that the SNR is higher for Si-NWells embedded in SiO$_2$ vs. Si$_3$N$_4$. This finding may come as a surprise as the average cps per edge region is higher for NWells embedded in Si$_3$N$_4$ vs. SiO$_2$. However, the difference in valence band top energies is $E_{\rm{V}}(\mbox{bulk-Si})-E_{\rm{V}}(\rm{Si_3N_4})= 1.78$ to 1.88 eV \cite{Keis99,Koe18aa} and $E_{\rm{V}}(\mbox{bulk-Si})-E_{\rm{V}}(\rm{SiO_2})= 4.53$ to 4.54 eV \cite{Keis99,Koe18aa}. As a consequence, the Urbach tails of the amorphous Si$_3$N$_4$ layers deliver a significantly higher contribution to the UPS signal as opposed to SiO$_2$, \emph{cf.} Fig. \ref{fig11}. To add, the packing fraction of Si$_3$N$_4$ and the density of electronic defects is notably higher as compared to SiO$_2$. This finding is also reflected in the bigger range of the valence band offset from Si$_3$N$_4$ to bulk-Si of ca. 0.1 eV as opposed to the value from SiO$_2$ to bulk-Si (0.01 eV).

\end{appendix}

\bibliography{RevTeX_manuscript_v01}

\end{document}